\documentclass[aps,showpacs,twocolumn,preprintnumbers,amsmath,amssymb]{revtex4}

\newcommand{\bra}[1]{\langle #1|}
\newcommand{\ket}[1]{|#1\rangle}

\newcommand{\mean}[1]{\langle #1 \rangle}
\newcommand{\trace}{{\rm Tr}}
\newcommand{\shat}[1]{\breve{#1}}
\newcommand{\sbra}[1]{\ll \hspace{-0.1cm} #1|}
\newcommand{\sket}[1]{| #1 \hspace{-0.1cm} \gg}
\newcommand{\sbraket}[2]{\ll \hspace{-0.1cm} #1|#2 \hspace{-0.1cm} \gg}
\newcommand{\ic}{{\rm i}}
\newcommand{\e}{{\rm e}}

\usepackage{graphicx}
\usepackage{dcolumn}
\usepackage{bm}
\usepackage{amssymb}
\usepackage{amsmath}
\usepackage{epsf}
\usepackage{subfigure}
\usepackage{epstopdf}

\def\xp{{{x^\prime}}}
\def\sp{{{s^\prime}}}

\def\tp{{t^\prime}}

\def\nup{{\nu^{\prime}}}
\def\cz{{\cal Z}}

\begin{document}

\title{Nonequilibrium fluctuations, fluctuation theorems,
and counting statistics in quantum systems}
\author{Massimiliano Esposito}
\affiliation{Department of Chemistry, University of California, San Diego\\
and Center for Nonlinear Phenomena and Complex Systems, Universit\'e Libre de Bruxelles, Brussels.}
\author{Upendra Harbola and Shaul Mukamel}
\affiliation{Department of Chemistry, University of California, Irvine}
\date{\today}

\begin{abstract}
Fluctuation theorems (FTs), which describe some universal properties of nonequilibrium fluctuations, are examined from a quantum perspective and derived by introducing a two-point measurement on the system. FTs for closed and open systems driven out of equilibrium by an external time-dependent force, and for open systems maintained in a nonequilibrium steady-state by nonequilibrium boundary conditions, are derived from a unified approach. Applications to fermion and boson transport in quantum junctions are discussed. Quantum master equations and Green's functions techniques for computing the energy and particle statistics are presented.
\end{abstract}

\maketitle

\tableofcontents

\section{Introduction}

Small fluctuations of systems at equilibrium or weakly driven
near equilibrium satisfy a universal relation known as the
fluctuation-dissipation (FD) theorem
\cite{Callen51,Kubo57,GrootMazur,KuboB98b,zwanzig,StratonovichI}.
This relation that connects spontaneous fluctuations to the
linear response holds for classical and quantum systems alike.
The search for similar relations for systems driven far from
equilibrium has been an active area of research for many decades.
A major breakthrough in this regard had taken place over the
past fifteen years with the discovery of exact fluctuation
relations which hold for classical systems far from equilibrium.
These are collectively referred to as fluctuation theorems (FTs).
In order to introduce these theorems we will adopt the following terminology.
A system that follows a Hamiltonian dynamics is called isolated.
By default, we assume that the Hamiltonian is time independent.
Otherwise, it means that some work is performed on the system
and we denote it driven isolated system.
A system that can only exchange energy with a reservoir
will be denoted closed.
If particles are exchanged as well, we say that the system is open.\\

The first class of FTs (and the earliest discovered) deal with
irreversible work fluctuations in isolated driven systems described
by an Hamiltonian dynamics where the Hamiltonian is time-dependent
\cite{BochkovKuzovlev1,BochkovKuzovlev2,BochkovKuzovlev3,BochkovKuzovlev4,
Stratonovich,Jarzynski97,Jarzynski97b,Cohen04,JarzynskiReply04,VandenBroeck06b,
Jarzynski07b,Jarzynski07,VandenBroeck07,VandenBroeck08}.
An example is the Crooks relation which states that the nonequilibrium
probability $p(W)$, that a certain work $w=W$ is performed by an external
time-dependent driving force acting on a system initially at equilibrium
with temperature $\beta^{-1}$, divided by the probability $\tilde{p}(-W)$,
that a work $w=-W$ is performed by the time-reversed external driving
force acting on the system which is again initially at equilibrium, satisfies
$p(W)/\tilde{p}(-W)=\exp{[\beta (W-\Delta F)]}$, where $\Delta F$ is the
free energy difference between the initial (no driving force)
and final (finite driving force) equilibrium state.
The Jarzynski relation $\mean{\exp{[-\beta W]}}=\exp{[-\beta \Delta F]}$
follows immediately from $\int dW \tilde{p}(-W)=1$.
A second class of FTs is concerned with entropy fluctuations
in closed systems described by deterministic thermostatted equations
of motions \cite{Evans93,Gallavotti95,Gallavotti95b,Evans94,Evans94b,Evans94c,Gallavotti99,Dellago06}
and a third class treats the fluctuations of entropy (or related
quantities such as irreversible work, heat and matter currents)
in closed or open systems described by a stochastic dynamics
\cite{Crooks98,Kurchan98,Crooks99,Lebowitz99,Searles99,Crooks00,
HatanoSasa01,Gaspard04,Seifert05,ChernyakJarzynski06,Ross88,
TaniguchiCohen07,AndrieuxGaspard07a,EspositoHarbola07PRE,Chetrite}.
As an example for the last two classes, we give the steady-state
FT for the entropy production. We consider a trajectory quantity
$s$ whose ensemble average $\mean{s}$ can be associated with
an entropy production (the specific form of $s$ depends on the
underlying dynamics). If $p(S)$ denotes the probability that
$s=S$ when the system is in a nonequilibrium steady-state,
then for long times the FT reads $p(S)/p(-S)=\exp{[S]}$.
FTs valid at any time such as the work FTs are called transient FTs
while those who require a long time limit are called steady-state FTs.\\

The FTs are all intimately connected to time-reversal symmetry and the
relations between probabilities of forward and backward classical trajectories.
Close to equilibrium the FTs reduce to the known fluctuation-dissipation
relations such as the Green-Kubo relation for transport coefficients
\cite{Gallavotti96,Gallavotti96b,Lebowitz99,AndrieuxGaspard04,AndrieuxGaspard07b}.
These classical fluctuation relations have been reviewed in
Refs. \cite{Maes03,Gaspard05rev,Gallavotti06rev,Harris07,Gallavotti08}.
Some of these relations were verified experimentally in mesoscopic
systems where fluctuations are sufficiently large to be measurable.
Work fluctuations have been studied in macromolecule pulling
experiments \cite{Bustamante02,Bustamante05} and in optically
driven microspheres \cite{Bustamante04}, entropy fluctuations
have also been measured in a similar system \cite{Evans05}
and in spectroscopic experiments on a defect center in diamond
\cite{Seifert05exp,Seifert06exp}.
When decreasing system sizes, quantum effects may become significant.
Applying the standard trajectory-based derivations of FTs to quantum
regime is complicated by the lack of a classical trajectory picture
when coherences are taken into account and by the essential role of
measurements, which can be safely ignored in ideal classical systems.
We show that the FTs follow from fundamental dynamical symmetries
that apply equally to classical and quantum systems.\\

Earlier derivations of the Jarzynski relation were done
for quantum systems by defining a work operator
\cite{BochkovKuzovlev1,BochkovKuzovlev2,BochkovKuzovlev3,BochkovKuzovlev4,
Stratonovich,Yukawa00,MonnaiTasaki03,
ChernyakMukamel04,Allahverdyan05,Engel06,Kosov08}.
Since work is not in general an ordinary quantum "observable" (the final 
Hamiltonian does not commute with the initial Hamiltonian) \cite{TalknerLutzHanggi07}, 
attempts to define such an operator had led to quantum corrections to the classical Jarzynski result.
However, the Jarzynski relation in a closed driven quantum
system may be derived without quantum corrections by introducing
an initial and final projective measurment of the system
energy in accordance with the quantum mechanical
measurement postulate.
This has been done (not always in a explicit way) in Refs.
\cite{HTasaki00,Kurchan00,Mukamel03,Monnai05,TalknerLutzHanggi07,
TalknerHanggi07,TalknerHanggi07b}. The work is then a two-point
quantity obtained by calculating the difference between the initial
and final energy of the system.
When the reservoir is explicitly taken into account, the Jarzynski
relation has often been derived using a master equation approach
\cite{Maes04a,EspositoMukamel06,Crooks07a,Crooks07b}.
Alternative derivations can be found in Refs. \cite{Monnai05,TalknerCampisi08}.\\

The derivation of a steady-state FT for quantum systems
has been considered as well \cite{Maes04b,GaspardAndrieuxMeso,Maes07,
DeRoeck07,EspositoMukamel06,EspositoHarbola07,HarbolaEspositoBoson07,
NazarovTobiska05,SaitoUtsumi07,SaitoDhar07,JarzynskiQ,VandenBroeck07EPL,AndrieuxGaspardTasaki}.
Because of the need to describe nonequilibrium fluctuations in
closed or open quantum systems exchanging energy or matter with
their reservoir, many similarities exist with the rapidly
developing field of electron counting statistics
\cite{Nazarov99,Belzig01a,Belzig01b,Belzig03,KindermannNazarov03,NazarovKindermann03,
KindermannPilgram04,Nazarov07,NazarovBagrets03,Rammer03,Levitov93,LevitovLeeJMathPhys96,
Levitov04,Schonhammer07,Utsumi06,Buttiker03,Buttiker07,SnymanNazarov08,Gurvitz97,Rammer04,
Rammer05,Jauho05,Scholl06,AguadoBrandes07,Welack08,Braggio06,Braggio08,Belzig08},
where small nano-scale electronic devices exchange electrons.
Fluctuations in such systems can nowadays be experimentally
resolved at the single electron level
\cite{Lu03,Fujisawa04,Bylander05,Gustavsson06,Hirayama06}.
Similarities also exist with the more established field of photon
counting statistics where photons emitted by a molecule or an atom
driven out of equilibrium by a laser, are individually detected
\cite{Glauber,Kelley,Mandel,MandelWolf,Gardiner00,
Brown,Brownb,Mukamel03b,Mukamel05,BarkaiRev,Orrit}.\\

Different types of approaches have been used to derive
these FTs and describe these counting experiments.
The first is based on the quantum master equation (QME)
\cite{Maes04b,EspositoMukamel06,EspositoHarbola07,
HarbolaEspositoBoson07,Gurvitz97,Rammer04,Rammer05,Jauho05,
Scholl06,AguadoBrandes07,Welack08,Braggio06,Braggio08}.
Here one starts with an isolated system containing
the system and the reservoir in weak interaction.
By tracing the reservoir degrees of freedom, taking the infinite
reservoir limit and using perturbation theory, one can derive a
closed evolution equation for the reduced density matrix of the system.
The information about the reservoir evolution is discarded.
However, the evolution of a quantum system described by a QME can
be seen as resulting from a continuous projective measurement on
the reservoir leading to a continuous positive operator-valued
measurement on the system.
Such interpretation allows to construct a trajectory picture of
the system dynamics, where each realization of the continuous
measurement leads to a given system trajectory
\cite{Gardiner00,Nielsen,Breuer02,Brun00,Brun02}.
The QME is recovered by ensemble averaging over all possible trajectories.
This {\it unraveling} of the QME into trajectories has been originally
developed in the description of photon counting statistics
\cite{Gardiner00,Breuer02,Wiseman93a,Wiseman93b,PlenioKnight98}.
Another approach is based on a modified propagator defined on a
Keldysh loop which, under certain circumstances, can be interpreted
as the generating function of the electron counting probability distribution
\cite{Nazarov99,Belzig01a,Belzig01b,Belzig03,KindermannNazarov03,
NazarovKindermann03,KindermannPilgram04,Nazarov07}.
Using a path integral formalism, the propagator of the density matrix of
a ``detector" with Hamiltonian $p^2/2m$ interacting with a system, can be
expressed in term of the influence functional that only depends
on the system degrees of freedom \cite{Feynman63}.
The modified propagator is the influence functional when the system
is linearly coupled to the detector (with coupling term $x A$, where
$x$ is the position of the detector and $A$ a system observable) in
the limit of very large detector inertia $m \to \infty$.
It is only under some specific assumptions (such as a classical
detector where the detector density matrix is assumed diagonal)
that the modified propagator becomes the generating function associated
with the probability distribution that the detector momentum changes
from a given amount which can be interpreted as the probability
to measure the time average of the system observable $A$:
$\int_0^t d\tau A(\tau)$.
If $A$ is an electric current, then the integral
gives the number of electrons transfered.
An early quantum FT for electronic junctions has been derived
in this context in Ref. \cite{NazarovTobiska05} based on the
time-reversal invariance of the Hamiltonian quantum dynamics.
Different derivations of quantum FTs relying on this approach
have been considered in Ref. \cite{SaitoUtsumi07,SaitoDhar07}.
A third, semiclassical scattering, approach is often used in
electron counting statistics \cite{Buttiker03,Pilgram04,
NagaevButtiker04,PilgramButtiker04,JordanPilgram04}.
This can be recovered from the modified propagator
approach as recently shown in \cite{SnymanNazarov08},
but will not be addressed here.\\

We consider fluctuations in the output of a two-point 
projective measurement (of energy, particle, charge, etc.).
This allows us to avoid the detailed modeling of detectors and their dynamics.
The projective measurement can be viewed as an effective modeling
of the effect of the system-detector interaction on the system or as
resulting in a fundamental way from the quantum measurement postulate.
The three other approaches (unraveling of the QME, modified propagator
on Keldysh loop and the scattering approach) can be recovered in some
limits of the two-point measurement approach.
This provides a unified framework from
which the different types of FTs previously
derived for quantum systems can be obtained.\\

In section \ref{2pointFT}, we give the general expression for the
probability of the output of a two-point measurement at different
times on a quantum system described by the quantum Liouville equation.
The calculation is repeated for a system described by the time-reversed dynamics.
In section \ref{FT}, we start by discussing the basic ingredients required for FTs to hold.
We use these results to derive three transient FTs, the Jarzynski and Crooks
relation in isolated and closed driven systems and a FT for matter and heat
exchange between two systems in direct contact.
We also show that a steady-state FT can be derived for matter and heat
exchange between two reservoirs through an embedded system.
In section \ref{MEGF}, we consider a small quantum
system weakly interacting with multiple reservoirs.
We develop a projection superoperator formalism to derive equations of
motion for the generating function associated with the system reduced
density matrix conditional of the output of a two-point measurement of
the energy or number of particles in the reservoirs.
We apply this generalized quantum master equation (GQME)
formalism to calculate the statistics of particles
or heat transfer in different models of general interest in
nanosciences in order to verify the validity of the steady-state FT.
In section \ref{GreenFun}, we present a nonequilibrium Green's functions
formalism in Liouville space which provides a powerful tool to calculate
the particle statistics of many body quantum systems.
In section \ref{NLcoef}, we show that the FTs can be used
to derive generalized fluctuation-dissipation relations.
Conclusions and perspectives will be drawn in section \ref{ConcPersp}.

\section{Two-point measurement statistics}
\label{2pointFT}

We consider an isolated, possibly driven, quantum system described by a
density matrix $\hat{\rho}(t)$ which obeys the von Neumann (quantum
Liouville) equation
\begin{eqnarray}
\frac{d}{dt}\hat{\rho}(t)= -\frac{\ic}{\hbar} [\hat{H}(t) , \hat{\rho}(t)] \;.
\label{quantumliouville}
\end{eqnarray}
Its formal solution reads
\begin{eqnarray}
\hat{\rho}(t)=\hat{U}(t,0) \hat{\rho}_0 \hat{U}^{\dagger}(t,0) \;.
\label{unitary_evolution}
\end{eqnarray}
The propagator
\begin{eqnarray}
\hat{U}(t,0) &=& \exp_{+}{\{-\frac{\ic}{\hbar} \int_{0}^{t}d\tau \hat{H}(\tau) \}} \\
&&\hspace{-2cm} \equiv 1+ \sum_{n=1}^{\infty}
\big(-\frac{\ic}{\hbar}\big)^n
\int_{0}^{t}dt_1 \int_{0}^{t_1}dt_2 \hdots \int_{0}^{t_{n-1}}dt_n \nonumber\\
&&\hspace{2cm} \hat{H}(t_1) \hat{H}(t_2) \hdots \hat{H}(t_n) \;.\nonumber
\label{explicitform}
\end{eqnarray}
is unitary $\hat{U}^{\dagger}(t,0)=\hat{U}^{-1}(t,0)$ and satisfies
$\hat{U}^{\dagger}(t,0)=\hat{U}(0,t)$ and $\hat{U}(t,t_1) \hat{U}(t_1,0)=\hat{U}(t,0)$.
We use the subscript $+$ ($-$) to denotes a antichronological
(chronological) time ordering from left to right.
We call (\ref{unitary_evolution}) the forward evolution to distinguish
it from the the time-reversed evolution that will be defined below.

\subsection{The forward probability}

We consider an observable $\hat{A}(t)$ in the Schr\"odinger picture whose
explicit time dependence solely comes from an external driving.
For non-driven systems $\hat{A}(t)=\hat{A}$.
In the applications considered below, $\hat{A}(t)$ will be either
an energy operator $\hat{H}$ or a particle number operator $\hat{N}$.
The eigenvalues (eigenvectors) of $\hat{A}(t)$ are
denoted by $a_t$ ($\ket{a_t}$): $\hat{A}(t)=\sum_{a_t} \ket{a_t} a_t \bra{a_t}$.\\

The basic quantity in the following discussion will be the joint
probability to measure $a_0$ at time $0$ and $a_t$ at time $t$
\begin{eqnarray}
P[a_t,a_0] &\equiv& \trace \left\{ \hat{P}_{a_t} \hat{U}(t,0) \hat{P}_{a_0} \hat{\rho}_0
\hat{P}_{a_0} \hat{U}^{\dagger}(t,0) \hat{P}_{a_t} \right\} \nonumber\\
&=& P^*[a_t,a_0] \;, \label{Aaaaa}
\end{eqnarray}
where the projection operators are given by
\begin{eqnarray}
\hat{P}_{a_t} = \ket{a_t} \bra{a_t} \;.
\label{projector}
\end{eqnarray}
Using the properties $\hat{P}_{a_t}=\hat{P}_{a_t}^2$ and $\sum_{a_t} \hat{P}_{a_t}=\hat{1}$,
we can verify the normalization $\sum_{a_ta_0} P[a_t,a_0]=1$.
Consider two complete Hilbert space basis sets $\{\ket{i,a_0}\}$
and $\{\ket{j,a_t}\}$, where $i$ ($j$) are used to differentiate
between the states with same $a_0$ ($a_t$).
The basis $\{\ket{i,a_0}\}$ is chosen such that it diagonalizes
$\hat{\rho}_0$ (this is always possible since $\hat{\rho}_0$ is hermitian).
We can also write (\ref{Aaaaa}) as
\begin{eqnarray}
P[a_t,a_0] = \sum_{i,j} P[j,a_t;i,a_0] \;, \label{Aaaad_bis}
\end{eqnarray}
where
\begin{eqnarray}
P[j,a_t;i,a_0] \equiv \vert \bra{j,a_t} \hat{U}(t,0) \ket{i,a_0} \vert^2
\bra{i,a_0} \hat{\rho}_0 \ket{i,a_0} \label{Aaaad_bisB} \;.
\end{eqnarray}

The probability distribution for the difference $\Delta a = a_t-a_0$
between the output of the two measurements is given by
\begin{eqnarray}
\label{Aaaab}
p(\Delta a) = \sum_{a_t a_0} \delta\big(\Delta a-(a_t-a_0)\big) P[a_t,a_0] \;,
\end{eqnarray}
where $\delta(a)$ denotes the Dirac distribution.
It is often more convenient to calculate the generating
function (GF) associated with this probability
\begin{eqnarray}
G(\lambda) &\equiv&
\int_{-\infty}^{\infty} d\Delta a \; \e^{\ic \lambda \Delta a} p(\Delta a)
=G^*(-\lambda) \nonumber\\
&=& \sum_{a_t a_0} \e^{\ic \lambda (a_t-a_0)} P[a_t,a_0] \;.
\label{Aaaac}
\end{eqnarray}
The $n$'th moment, $\langle \Delta a^n\rangle$, of $p(\Delta a)$ is
obtained by taking $n$'th derivative of the GF with respect to
$\lambda$ evaluated at $\lambda=0$
\begin{eqnarray}
\langle \Delta a^n \rangle = (-\ic)^n \left. \frac{\partial^n}
{\partial \lambda^n} G(\lambda) \right|_{\lambda=0} \;.
\label{Generalmoments}
\end{eqnarray}
We further define the cumulant GF
\begin{eqnarray}
\label{c-gf}
{\cal Z}(\lambda) = \ln G(\lambda) \;.
\end{eqnarray}
The $n$'th cumulant, $K_n$, of $p(\Delta a)$ is
obtained by taking $n$'th derivative of the cumulant GF
with respect to $\lambda$ evaluated at $\lambda=0$
\begin{eqnarray}
\label{GeneralCumulants}
K_n = (-\ic)^n \left. \frac{\partial^n}{\partial \lambda^n}
{\cal Z}(\lambda)\right|_{\lambda=0}.
\end{eqnarray}
The first cumulant coincides with the first moment
which gives the average $K_1=\langle \Delta a \rangle$.
Higher order cumulants can be expressed in term of the moments.
The variance, $K_2=\langle \Delta a^2\rangle -\langle \Delta a\rangle^2$,
gives the fluctuations around the average, and the skewness
$K_3=\langle(\Delta a-\langle \Delta a \rangle)^3$ gives the leading
order deviation of $p(\Delta a)$ from a Gaussian.
When measuring the statistics of quantities associated to nonequilibrium
fluxes, in most cases (but not always \cite{EspositoLindenberg}) the
cumulants grow linearly with time and it becomes convenient to define
the long time limit of the cumulant GF
\begin{eqnarray}
\label{c-gfLimit}
{\cal S}(\lambda) =\lim_{t \to \infty} \frac{1}{t} {\cal Z}(\lambda)
\end{eqnarray}
which measures the deviations to the central limit theorem \cite{Sornette}.

We next turn to computing the GF.
The initial density matrix can be expressed as
\begin{eqnarray}
\hat{\rho}_0 = \bar{\hat{\rho}}_0 + \bar{\bar{\hat{\rho}}}_0 \;, \label{densityMatSplitt}
\end{eqnarray}
where
\begin{eqnarray}
\bar{\hat{\rho}}_0 = \sum_{a_0} \hat{P}_{a_0} \hat{\rho}_0 \hat{P}_{a_0} \ \ \;, \ \
\bar{\bar{\hat{\rho}}}_0 = \sum_{a_0 \neq a_0'} \hat{P}_{a_0} \hat{\rho}_0 \hat{P}_{a_0'}
\;.\label{densityMatSplittB}
\end{eqnarray}
$\bar{\hat{\rho}}_0$ commutes with $\hat{A}(0)$.
Using the fact that $f(\hat{A})= \sum_a \hat{P}_a f(a)$ where
$f$ is an arbitrary function, and using also
\begin{eqnarray}
\sum_{a_0} \e^{-\ic \lambda a_0} \hat{P}_{a_0} \hat{\rho}_0 \hat{P}_{a_0}
= \e^{-\ic \frac{\lambda}{2} \hat{A}(0)} \bar{\hat{\rho}}_0
\e^{-\ic \frac{\lambda}{2} \hat{A}(0)} \;,
\end{eqnarray}
we find, by substituting (\ref{Aaaaa}) in (\ref{Aaaac}), that
\begin{eqnarray}
G(\lambda) = \trace \; \hat{\rho}(\lambda,t) \label{Aaaacx1} \;,
\end{eqnarray}
where we have defined
\begin{eqnarray}
\hat{\rho}(\lambda,t) \equiv \hat{U}_{\frac{\lambda}{2}}(t,0) \bar{\hat{\rho}}_0
\hat{U}_{-\frac{\lambda}{2}}^{\dagger}(t,0) \label{Aaaacx2}
\end{eqnarray}
and the modified evolution operator
\begin{eqnarray}
\hat{U}_{\lambda}(t,0) \equiv \e^{\ic \lambda \hat{A}(t)} \hat{U}(t,0) \e^{-\ic \lambda \hat{A}(0)} \;.
\label{Aaaacx3}
\end{eqnarray}
For $\lambda=0$, $\hat{\rho}(\lambda,t)$ reduces to the system density
matrix and $\hat{U}_{\lambda}(t,0)$ to the standard evolution operator.
Defining the modified Hamiltonian
\begin{eqnarray}
\hat{H}_{\lambda}(t) \equiv \e^{\ic \lambda \hat{A}(t)} \hat{H}(t) \e^{-\ic \lambda \hat{A}(t)}
- \hbar \lambda \partial_t \hat{A}(t) \;, \label{Aaaacx5}
\end{eqnarray}
we find that $\hat{U}_{\lambda}(t,0)$ satisfies the equation of motion
\begin{eqnarray}
\frac{d}{dt} \hat{U}_{\lambda}(t,0) = - \frac{\ic}{\hbar} \hat{H}_{\lambda}(t)
\hat{U}_{\lambda}(t,0) \;. \label{Aaaacx4}
\end{eqnarray}
Since $\hat{U}_{\lambda}(0,0)=\hat{1}$, we get
\begin{eqnarray}
\hat{U}_{\frac{\lambda}{2}}(t,0) &=&
\exp_{+}{\{-\frac{\ic}{\hbar} \int_{0}^{t} d\tau
\hat{H}_{\frac{\lambda}{2}}(\tau)\}} \label{Aaaacx6a}\\
\hat{U}_{-\frac{\lambda}{2}}^{\dagger}(t,0) &=&
\exp_{-}{\{ \frac{\ic}{\hbar} \int_{0}^{t} d\tau
\hat{H}_{-\frac{\lambda}{2}}(\tau)\}} \label{Aaaacx6b} \;.
\end{eqnarray}
Equations (\ref{Aaaacx1}) and (\ref{Aaaacx2}) together with (\ref{Aaaacx6a})
and (\ref{Aaaacx6b}) provide an exact formal expression for the statistics
of changes in $\hat{A}(t)$ derived from the two-point measurements.

We note that if and only if the eigenvalues of $\hat{A}$ are integers (as in
electron counting where one considers the number operator), using the integral
representation of the Kronecker Delta
\begin{eqnarray}
\delta_{K}(a-a')=\int_{0}^{2 \pi} \frac{d\Lambda}{2 \pi} \e^{-\ic \Lambda (a-a')} \;,
\end{eqnarray}
(\ref{Aaaacx2}) can be written as
\begin{eqnarray}
\hat{\rho}(\lambda,t) = \int_{0}^{2 \pi} \frac{d\Lambda}{2 \pi} \;
\; \hat{\rho}(\lambda,\Lambda,t) \label{Aaaacx2BIS} \;,
\end{eqnarray}
where
\begin{eqnarray}
\hat{\rho}(\lambda,\Lambda,t) \equiv \hat{U}_{\Lambda+\frac{\lambda}{2}}(t,0) \hat{\rho}_0
\hat{U}_{\Lambda-\frac{\lambda}{2}}^{\dagger}(t,0) \label{Aaaacx2TRIS} \;.
\end{eqnarray}
We see that by introducing an additional $\Lambda$ dependence, we where able
to keep the initial density matrix $\hat{\rho}_0$ in (\ref{Aaaacx2TRIS}) instead
of $\bar{\hat{\rho}}_0$ as in (\ref{Aaaacx2}).

The current operator associated with $\hat{A}(t)$ is given by
\begin{eqnarray}
\hat{I}(t) \equiv
\frac{\ic}{\hbar} [ \hat{H}(t),\hat{A}(t) ] + \partial_t \hat{A}(t) \;.
\label{DefCurrentOperator}
\end{eqnarray}
As a result,
\begin{eqnarray}
\hat{I}^{(h)}(t) = \frac{d}{dt} \hat{A}^{(h)}(t) \;,
\label{DefCurrentOperatorh}
\end{eqnarray}
where the subscript $(h)$ denotes the Heisenberg representation
$\hat{A}^{(h)}(t) \equiv \hat{U}^{\dagger}(t,0) \hat{A}(t) \hat{U}(t,0)$.
We can write (\ref{Aaaacx5}) as
\begin{eqnarray}
\hat{H}_{\lambda}(t) = \hat{H}(t) - \lambda \hbar \hat{I}(t) + {\cal O}(\lambda^2 \hbar^2)
\;. \label{ModifHamiltSemiclass}
\end{eqnarray}
In the semiclassical approximation where terms ${\cal O}(\lambda^2 \hbar^2)$
are disregarded, the GF (\ref{Aaaacx1}) [with (\ref{Aaaacx2}), (\ref{Aaaacx6a}) and
(\ref{Aaaacx6b})], after going to the interaction representation, becomes
\begin{eqnarray}
G(\lambda) = \trace \big\{ \e_{+}^{ \ic \frac{\lambda}{2} \int_{0}^{t}
d\tau \hat{I}^{(h)}(\tau) } \bar{\hat{\rho}}_0 \e_{-}^{ \ic \frac{\lambda}{2}
\int_{0}^{t} d\tau \hat{I}^{(h)}(\tau)} \big\} \label{GFSemiclass} \; .
\end{eqnarray}
This form is commonly found in the modified propagator
approach (described in the introduction) to counting statistics
\cite{KindermannNazarov03,NazarovKindermann03,KindermannPilgram04}.
Notice that in these Refs. the full initial density matrix $\hat{\rho}_0$
is used in (\ref{GFSemiclass}) instead of $\bar{\hat{\rho}}_0$.\\

In most applications considered in this review, we will consider initial
density matrices with no initial coherences in $\hat{A}(0)$ space
\begin{eqnarray}
[\hat{A}(0),\hat{\rho}_0]=0 \;. \label{Condforw}
\end{eqnarray}
This is equivalent to say that $[\hat{P}_{a_0},\hat{\rho}_0]=0$
or that $\hat{\rho}_0 = \bar{\hat{\rho}}_0$.
In this case, Eq. (\ref{Aaaaa}) can be written as
\begin{eqnarray}
P[a_t,a_0]=\trace \left\{ \hat{U}^{\dagger}(t,0) \hat{P}_{a_t}
\hat{U}(t,0) \hat{P}_{a_0} \hat{\rho}_0 \right\}
\label{Aaaad}
\end{eqnarray}
and using (\ref{Aaaad}) in (\ref{Aaaac}), the GF simplifies to
\begin{eqnarray}
G(\lambda) = \trace \left\{ \e^{\ic \lambda \hat{U}^{\dagger}(t,0)
\hat{A}(t) \hat{U}(t,0)} \e^{-\ic \lambda \hat{A}(0)} \hat{\rho}_0 \right\} \;.
\label{Aaaae}
\end{eqnarray}

\subsection{The time-reversed probability}\label{TRunitary}

The time-reversed evolution brings the final density matrix
of the forward quantum evolution (\ref{unitary_evolution})
back to its initial density matrix.
This means that if the initial condition of the time-reversed
evolution is $\hat{\rho}^{\rm tr}_0=\hat{\rho}(t)=\hat{U}(t,0)
\hat{\rho}_0 \hat{U}^{\dagger}(t,0)$, the time-reversed evolution
must be defined as $\hat{\rho}^{\rm tr}(t) =
\hat{U}^{\dagger}(t,0) \hat{\rho}^{\rm tr}_0 \hat{U}(t,0)$,
so that $\hat{\rho}^{\rm tr}(t) =\hat{\rho}_0$.
The time-reversed expression of the two-point probability (\ref{Aaaaa})
is therefore
\begin{eqnarray}
P^{{\rm tr}}[a_0,a_t]
\equiv \trace \left\{ \hat{P}_{a_0} \hat{U}^{\dagger}(t,0) \hat{P}_{a_t}
\hat{\rho}^{{\rm tr}}_0 \hat{P}_{a_t} \hat{U}(t,0) \hat{P}_{a_0} \right\} \;.
\label{Aaaaf}
\end{eqnarray}
A more systematic discussion on time-reversal operation in quantum mechanics
and its relation to the definition (\ref{Aaaaf}) is given in appendix \ref{appA}.
Without loss of generality, we choose a basis set $\{\ket{j,a_t}\}$ that
diagonalizes $\hat{\rho}^{\rm tr}_0$, to show that (\ref{Aaaaf}) can be rewritten as
\begin{eqnarray}
P^{{\rm tr}}[a_0,a_t] = \sum_{i,j} P^{{\rm tr}}[i,a_0;j,a_t] \;, \label{Aaaah_bis}
\end{eqnarray}
where
\begin{eqnarray}
P^{{\rm tr}}[i,a_0;j,a_t] \equiv \vert \bra{j,a_t} \hat{U}(t,0) \ket{i,a_0} \vert^2
\bra{j,a_t} \hat{\rho}^{\rm tr}_0 \ket{j,a_t} \;.
\label{Aaaah_bisB}
\end{eqnarray}

The probability to measure the difference $\Delta a=a_0-a_t$ between
the two measurements is given by
\begin{eqnarray}
p^{{\rm tr}}(\Delta a) &\equiv& \sum_{a_ta_0} \delta \big(\Delta a-(a_0-a_t)\big)
P^{{\rm tr}}[a_0,a_t] \;. \label{Aaaag}
\end{eqnarray}
The associated GF reads
\begin{eqnarray}
G^{{\rm tr}}(\lambda) &\equiv& \int_{-\infty}^{\infty} d\Delta a \;
\e^{\ic \lambda \Delta a} p^{{\rm tr}}(\Delta a) \nonumber \\
&=& \sum_{a_t,a_0} \e^{-\ic \lambda (a_t-a_0)} P^{{\rm tr}}[a_0,a_t] \;.
\label{AaaaiBis}
\end{eqnarray}
Let us note that for a non-driven system with $\hat{\rho}^{\rm tr}_0
=\hat{\rho}_0$, using (\ref{Aaaaa}) and (\ref{Aaaaf}), we find
that $P[a_t,a_0]=P^{{\rm tr}}[a_t,a_0]$.
This means, using (\ref{Aaaad_bis}) and (\ref{Aaaah_bis}), that
\begin{eqnarray}
p^{{\rm tr}}(\Delta a)=p(\Delta a)  \label{nondrivenforwrevprob}
\end{eqnarray}
and
\begin{eqnarray}
G^{{\rm tr}}(\lambda)=G(\lambda)  \;. \label{nondrivenforwrevGF}
\end{eqnarray}

Using again the partitioning
\begin{eqnarray}
\hat{\rho}_0^{{\rm tr}} = \bar{\hat{\rho}}_0^{{\rm tr}}
+ \bar{\bar{\hat{\rho}}}_0^{{\rm tr}} \;, \label{densityMatSplittTR}
\end{eqnarray}
where
\begin{eqnarray}
\bar{\hat{\rho}}_0^{{\rm tr}} = \sum_{a_t} \hat{P}_{a_t} \hat{\rho}_0^{{\rm tr}} \hat{P}_{a_t} \ \ \;, \ \
\bar{\bar{\hat{\rho}}}_0^{{\rm tr}} = \sum_{a_t \neq a_t'} \hat{P}_{a_t} \hat{\rho}_0^{{\rm tr}} \hat{P}_{a_t'}
\label{densityMatSplittBTR}
\end{eqnarray}
and following the same procedure as for the forward GF, we obtain
\begin{eqnarray}
G^{{\rm tr}}(\lambda) = \trace \; \hat{\rho}^{{\rm tr}}(\lambda,t)
\;, \label{AaaaiBis2}
\end{eqnarray}
where
\begin{eqnarray}
\hat{\rho}^{{\rm tr}}(\lambda,t)
\equiv \hat{U}_{\frac{\lambda}{2}}^{\dagger}(t,0) \bar{\hat{\rho}}_0^{\rm tr}
\hat{U}_{-\frac{\lambda}{2}}(t,0) \label{AaaaiBis3} \;.
\end{eqnarray}
As for (\ref{GFSemiclass}), in the semiclassical limit we find
\begin{eqnarray}
G^{{\rm tr}}(\lambda) = \trace \big\{ \e_{-}^{ \ic \frac{\lambda}{2} \int_{0}^{t}
d\tau \hat{I}^{(h)}(\tau) } \bar{\hat{\rho}}_0^{\rm tr} \e_{+}^{ \ic \frac{\lambda}{2}
\int_{0}^{t} d\tau \hat{I}^{(h)}(\tau)} \big\} \label{GFSemiclassTR} \;.
\end{eqnarray}

We again note that if the initial density matrix of the time-reversed 
evolution contains no initial coherences in $\hat{A}(t)$ space
\begin{eqnarray}
[\hat{A}(t),\hat{\rho}^{\rm tr}_0]=0 \;, \label{Condrev}
\end{eqnarray}
or equivalently if $[\hat{P}_{a_t},\hat{\rho}^{\rm tr}_0]=0$ or
$\hat{\rho}_0^{{\rm tr}} = \bar{\hat{\rho}}_0^{{\rm tr}}$,
(\ref{Aaaaf}) becomes
\begin{eqnarray}
P^{{\rm tr}}[a_0,a_t] = \trace \left\{ \hat{U}(t,0) \hat{P}_{a_0}
\hat{U}^{\dagger}(t,0) \hat{P}_{a_t} \hat{\rho}^{\rm tr}_0 \right\} \;,
\label{Aaaah}
\end{eqnarray}
and
\begin{eqnarray}
G^{{\rm tr}}(\lambda) = \trace \left\{ \e^{\ic \lambda \hat{U}(t,0) \hat{A}(0) \hat{U}^{\dagger}(t,0)}
\e^{-\ic \lambda \hat{A}(t)} \hat{\rho}^{\rm tr}_0 \right\} \;. \label{Aaaai}
\end{eqnarray}

\section{The fluctuation theorem} \label{FT}

\subsection{General derivation and connection to entropy} \label{FTgen}

We define the log of the ratio of the forward and time-reversed probabilities
defined in section \ref{2pointFT}, which in the classical theory of FTs 
is associated with the irreversible contribution to an entropy change 
\begin{eqnarray}
R[j,a_t;i,a_0] \equiv \ln \frac{P[j,a_t;i,a_0]}
{P^{{\rm tr}}[i,a_0;j,a_t]} \label{defRtildef} \;.
\end{eqnarray}
It follows from (\ref{Aaaad_bisB}) and (\ref{Aaaah_bisB}) that
\begin{eqnarray}
R[j,a_t;i,a_0] = \ln \frac{\bra{i,a_0} \hat{\rho}_0 \ket{i,a_0}}
{\bra{j,a_t} \hat{\rho}^{\rm tr}_0 \ket{j,a_t}} \;. \label{defRtil}
\end{eqnarray}
An {\it integral FT} immediately follows from
the normalization of $P^{{\rm tr}}[i,a_0;j,a_t]$
\begin{eqnarray}
\mean{\e^{-R}} &\equiv& \sum_{j,a_t,i,a_0} P[j,a_t;i,a_0] \e^{-R[j,a_t;i,a_0]}
\nonumber \\ &=& \sum_{j,a_t,i,a_0} P^{{\rm tr}}[i,a_0;j,a_t] = 1 \;.
\label{integralFTtil}
\end{eqnarray}
Using Jensen's inequality $\langle\mbox{e}^{X}\rangle \geq
\mbox{e}^{\langle X\rangle}$, (\ref{integralFTtil}) implies
\begin{eqnarray}
\mean{R} = \sum_{j,a_t,i,a_0} P[j,a_t;i,a_0] \; R[j,a_t;i,a_0] \geq 0 \;.
\label{integralFTtilinegal}
\end{eqnarray}
Using (\ref{defRtildef}), we see that $\mean{R}$ resembles a
Kullback-Leibler (or relative) entropy \cite{Kullback,Nielsen}.\\

We define the probability distributions
\begin{eqnarray}
&&\hspace{-0.8cm} p(R) \equiv \sum_{j,a_t,i,a_0} P[j,a_t;i,a_0] \delta(R-R[j,a_t;i,a_0])
\label{detailedFTtildef2a} \\
&&\hspace{-0.8cm} p^{{\rm tr}}(R) \equiv \sum_{j,a_t,i,a_0}
P^{{\rm tr}}[i,a_0;j,a_t] \delta(R-R^{{\rm tr}}[i,a_0;j,a_t]) \;.\label{detailedFTtildef2b}
\end{eqnarray}
where
\begin{eqnarray}
R^{{\rm tr}}[i,a_0;j,a_t] \equiv \ln
\frac{P^{{\rm tr}}[i,a_0;j,a_t]}{P[j,a_t;i,a_0]}
\;. \label{defRtilBis}
\end{eqnarray}
Using (\ref{defRtil}) and (\ref{defRtilBis}), we see that
\begin{eqnarray}
R^{{\rm tr}}[i,a_0;j,a_t] = - R[j,a_t;i,a_0] \;.
\label{essentialproptil}
\end{eqnarray}
It then follows that
\begin{eqnarray}
&&p(R)=\nonumber\\
&&\sum_{j,a_t,i,a_0} \e^{R[j,a_t;i,a_0]} P^{{\rm tr}}[i,a_0;j,a_t]
\delta(R-R[j,a_t;i,a_0]) \nonumber\\
&&= \e^{R} \sum_{j,a_t,i,a_0} P^{{\rm tr}}[i,a_0;j,a_t]
\delta(R-R[j,a_t;i,a_0]) \nonumber\\
&&= \e^{R} \sum_{j,a_t,i,a_0} P^{{\rm tr}}[i,a_0;j,a_t]
\delta(R+R^{{\rm tr}}[i,a_0;j,a_t]) \nonumber\\
&&=\e^{R}p^{{\rm tr}}(-R)\;,
\end{eqnarray}
which gives the {\em detailed} FT
\begin{eqnarray}
\ln \frac{p(R)}{p^{{\rm tr}}(-R)}= R  \;.
\label{detailedFTtil}
\end{eqnarray}
The FTs (\ref{integralFTtil}) and (\ref{detailedFTtil}) are completely general
but only useful when $R$ can be exclusively expressed in terms of physical
and measurable quantities (the eigenvalues of $\hat{A}(0)$ and $\hat{A}(t)$).
In sections \ref{WorkFTsection} and \ref{SSFT2point}, we will see that
the $i$ and $j$ dependence of $R$, that labels states which cannot
be differentiated by a projective measurement of the physical
observable $\hat{A}(t)$, can be eliminated by making specific 
choices of $\hat{\rho}$ and $\hat{\rho}^{\rm tr}$.\\

If the assumptions (\ref{Condforw}) and (\ref{Condrev}) are satisfied
(this will be the case in most of the following applications),
(\ref{integralFTtilinegal}) can be expressed in term of quantum entropies.
Using (\ref{defRtil}), the general property
\begin{eqnarray}
&&\sum_{j,a_t} P[j,a_t;i,a_0] =\bra{i,a_0} \hat{\rho}_0 \ket{i,a_0} \;,
\end{eqnarray}
and the fact that [using assumption (\ref{Condforw})]
\begin{eqnarray}
&&\sum_{i,a_0} P[j,a_t;i,a_0] = \bra{j,a_t} \hat{\rho}(t) \ket{j,a_t} \;,
\end{eqnarray}
(\ref{integralFTtilinegal}) can be rewritten as a quantum relative 
entropy \cite{Breuer02,Nielsen} between $\hat{\rho}(t)$ and $\hat{\rho}_0^{\rm tr}$
\begin{eqnarray}
\mean{R}= \bar{S}-S = \trace \hat{\rho}(t) \big( \ln \hat{\rho}(t)
-\ln \hat{\rho}_0^{\rm tr} \big) \geq 0 \label{2law} \;,
\label{entropyprodqentropy}
\end{eqnarray}
where
\begin{eqnarray}
S &\equiv& -\trace \hat{\rho}(t) \ln \hat{\rho}(t)
= -\trace \hat{\rho}_0 \ln \hat{\rho}_0 \nonumber \\
&=&-\sum_{i,a_0} \bra{i,a_0} \hat{\rho}_0 \ket{i,a_0}
\ln \bra{i,a_0} \hat{\rho}_0 \ket{i,a_0} \label{defproblocalB}
\end{eqnarray}
and
\begin{eqnarray}
\bar{S} &\equiv& - \trace \hat{\rho}(t) \ln \hat{\rho}_0^{\rm tr}
= - \trace \hat{\rho}_0 \ln \hat{\rho}^{\rm tr}(t) \nonumber \\
&=& -\sum_{j,a_t} \bra{j,a_t} \hat{\rho}(t) \ket{j,a_t}
\ln \bra{j,a_t} \hat{\rho}_0^{\rm tr} \ket{j,a_t} \;.
\label{modifiedentropy}
\end{eqnarray}
The second line of (\ref{defproblocalB}) [(\ref{modifiedentropy})]
is obtained using the assumption (\ref{Condforw}) [(\ref{Condrev})].
$S$ is a von Neumann entropy but $\bar{S}$ is not.
It can be compared to the von Neumann entropy
\begin{eqnarray}
S^{\rm tr} &\equiv& -\trace \hat{\rho}^{\rm tr}(t) \ln \hat{\rho}^{\rm tr}(t)
=-\trace \hat{\rho}^{\rm tr}_0 \ln \hat{\rho}^{\rm tr}_0 \nonumber \\
&=&-\sum_{j,a_t} \bra{j,a_t} \hat{\rho}^{\rm tr}_0 \ket{j,a_t}
\ln \bra{j,a_t} \hat{\rho}^{\rm tr}_0 \ket{j,a_t} \label{defproblocalA}
\end{eqnarray}
which is obtained using the general property
\begin{eqnarray}
&&\sum_{i,a_0} P^{\rm tr}[i,a_0;j,a_t]
= \bra{j,a_t} \hat{\rho}_0^{\rm tr} \ket{j,a_t}
\end{eqnarray}
together with [using assumption (\ref{Condrev})]
\begin{eqnarray}
&&\sum_{j,a_t} P^{\rm tr}[i,a_0;j,a_t]
=\bra{i,a_0} \hat{\rho}^{\rm tr}(t) \ket{i,a_0} \;.
\end{eqnarray}
We will see in the following applications that $\mean{R}$ is always 
associated to the irreversible contribution of an entropy change.
Eq. (\ref{entropyprodqentropy}) is therefore the quantum analog of the 
classical relation derived in Refs. \cite{VandenBroeck07,VandenBroeck08} 
and of the stochastic relation of Refs. \cite{Gaspard04b,Gaspard07exp}.\\

In appendix \ref{CoarsegrainedFT}, following Refs. \cite{Maes04c,Maes05},
we show that if one allows for a coarse-graining of $\hat{\rho}_0$ and
$\hat{\rho}^{\rm tr}_0$ in their measured subspaces, one can derive FTs for
$R$'s which can be expressed exclusively in terms of measurable
probabilities (no $i$ and $j$ index) and such that $\mean{R}$
is the difference between the Gibbs-von Neumann entropy
associated to the coarse-grained $\hat{\rho}_0^{\rm tr}$ and $\hat{\rho}_0$.\\

We now examine the detailed FT from the GF perspective.
We define the GFs associated with $p(R)$ and $p^{{\rm tr}}(R)$
\begin{eqnarray}
G(\lambda) &\equiv&
\int_{-\infty}^{\infty} dR \; \e^{\ic \lambda R} p(R) \nonumber \\
G^{{\rm tr}}(\lambda) &\equiv&
\int_{-\infty}^{\infty} dR \; \e^{\ic \lambda R} p^{{\rm tr}}(R) \;.
\label{defGFentropy}
\end{eqnarray}
By combining (\ref{detailedFTtil}) with (\ref{defGFentropy}), we get
\begin{eqnarray}
G(\lambda) = G^{{\rm tr}}(\ic-\lambda) \label{GFFTsymm-1} \;.
\end{eqnarray}
For a non-driven system with $\hat{\rho}^{\rm tr}_0=\hat{\rho}_0$,
we have seen that (\ref{nondrivenforwrevGF}) is satisfied.
Combining this with (\ref{GFFTsymm-1}), the detailed FT (\ref{detailedFTtil})
implies the fundamental symmetry $G(\lambda) = G(\ic-\lambda)$ on the GF.
This type of symmetry will be used in section \ref{NLcoef}
to derive generalized fluctuation-dissipation relations.

\subsection{Transient fluctuation theorems} \label{WorkFTsection}

In this section, we show that the FT (\ref{detailedFTtil}) can be used to
derive the Crooks \cite{Crooks99,Crooks00,Jarzynski07} and the
Jarzynski relations \cite{Jarzynski97,Jarzynski97b,JarzynskiReply04}
in either isolated or closed driven quantum systems as well as a
FT for for heat and particles exchange between two finite systems.

\subsubsection{Work fluctuation theorem for isolated driven systems}\label{WorkFTsectionIsol}

We consider an isolated system initially described by the
Hamiltonian $\hat{H}(0)$ and at equilibrium $\e^{-\beta \hat{H}(0)}/Z(0)$,
where $Z(0)=\trace \e^{-\beta \hat{H}(0)}$ is the partition function.
We can imagine that the system was in contact
with a reservoir at temperature $\beta^{-1}$ for $t<0$. 
At $t=0$ the reservoir is removed and the system 
energy is measured for the first time.
After the first measurement, the system is then subjected to an 
external and arbitrary driving (the Hamiltonian is time-dependent).
The second energy measurement occurs at time $t$,
where the Hamiltonian is $\hat{H}(t)$.
From the two measurements of this forward
process we can calculate $P[E_t,E_0]$.

In the backward process, the isolated system is initially described
by the Hamiltonian $\hat{H}(t)$ and at equilibrium $\e^{-\beta \hat{H}(t)}/Z(t)$,
where $Z(t)=\trace \e^{-\beta \hat{H}(t)}$.
We can imagine that at the end of the forward process, the system
described by the Hamiltonian $\hat{H}(t)$ is put in contact with a
reservoir at temperature $\beta^{-1}$ until it thermalizes, and
that the reservoir is then removed at time zero when the energy of
the system is measured for the first time in the backward process.
After this first measurement, an external driving, which is the time
reversed driving of the forward process, is applied.
The second energy measurement occurs at time $t$,
where the Hamiltonian is $\hat{H}(0)$.

In appendix \ref{appA}, we show that the time-reversed evolution
(as defined in section \ref{TRunitary}) of an isolated system driven
externally according to a given protocol, corresponds to the forward
evolution of the isolated system externally driven according
to the time-reversed protocol.
This means that the backward process just described is identical
to the time-reversal of our forward process, so that the two
measurements occurring during the backward process can be used
to calculate $P^{{\rm tr}}[E_0,E_t]$.

To make the connection with the results of section
\ref{2pointFT}, we define the initial density matrices
for the forward and backward process
\begin{eqnarray}
\hat{\rho}_0=\frac{\e^{-\beta \hat{H}(0)}}{Z(0)}  \  \  \;, \  \
\hat{\rho}_0^{\rm tr}=\frac{\e^{-\beta \hat{H}(t)}}{Z(t)}  \;.
\end{eqnarray}
We further set $a_t=E_t$ ($a_0=E_0$), where
$\hat{H}(t) \ket{E_t,j} = E_t \ket{E_t,j}$ ($\hat{H}(0) \ket{E_0,i} = E_0 \ket{E_0,i}$).
The index $j$ ($i$) distinguish between degenerate eigenstates so that
$\{ \ket{E_t,j} \}$ ($\{ \ket{E_0,i} \}$) constitute a complete
basis in Hilbert space.
We also define the free-energy difference $\Delta F(t)=F(t)-F(0)$
between the initial and final state, where $F(t)=-\beta^{-1} \ln Z(t)$.
Since the system is isolated, no heat exchange occurs and the change
in the system energy can be interpreted as the work done by the
driving force on the system
\begin{eqnarray}
w=\Delta a=E_t-E_0 \label{Workdefisol} \;.
\end{eqnarray}
Eq. (\ref{Aaaad_bisB}) and (\ref{Aaaah_bisB}) become
\begin{eqnarray}
P[j,E_t;i,E_0] &=& \vert \bra{j,E_t} \hat{U}(t,0) \ket{i,E_0} \vert^2
\e^{-\beta  \big(E_0-F(0) \big)} \nonumber \\
P^{{\rm tr}}[i,E_0;j,E_t] &=& \vert \bra{j,E_t} \hat{U}(t,0) \ket{i,E_0}
\vert^2 \e^{-\beta \big( E_t-F(t) \big)} \nonumber \;,
\end{eqnarray}
so that Eq. (\ref{defRtil}) becomes
\begin{eqnarray}
R[j,E_t;i,E_0]=
\beta \big( w - \Delta F (t)\big) = R[E_t,E_0] \;. \label{Risoldriv}
\end{eqnarray}
The essential property that $R$ is independent of $i$ and $j$ and only
expressed in terms of observable quantities is therefore satisfied.

(\ref{defproblocalB}) and (\ref{modifiedentropy}) become
\begin{eqnarray}
\bar{S} &=& \beta \big( \trace \hat{H}(t) \hat{\rho}(t) - F(t) \big) \nonumber \\
S &=& \beta \big( \trace \hat{H}(0) \hat{\rho}_0 - F(0) \big) \;,
\label{entropyWorkisolA}
\end{eqnarray}
and
\begin{eqnarray}
\mean{R} = \bar{S}-S = \beta \big( \mean{w} - \Delta F \big) \geq 0 \;,
\label{secondlawWorkisol}
\end{eqnarray}
where
\begin{eqnarray}
\mean{w} = \trace \hat{H}(t) \hat{\rho}(t) - \trace \hat{H}(0) \hat{\rho}_0 \label{entropyWorkisol} \;.
\end{eqnarray}
$\mean{w}$ is the average work, so that $\beta^{-1} \mean{R}$ is the 
irreversible work (the irreversible contribution to the entropy change). 
Using (\ref{detailedFTtil}), we get the Crooks relation
\begin{eqnarray}
\frac{p(w)}{p^{{\rm tr}}(-w)} = \e^{\beta (w-\Delta F)} \;.
\label{QCrooks}
\end{eqnarray}
The Jarzynski relation follows immediately from (\ref{QCrooks})
[by integrating $p^{{\rm tr}}(-w)$ over $w$ which is equal
to one because of normalization]
\begin{eqnarray}
\mean{\e^{-\beta w}}
= \e^{-\beta \Delta F} \;. \label{QJarzynski}
\end{eqnarray}

Equations (\ref{QJarzynski}) and (\ref{QCrooks}) have been first
derived in Ref. \cite{Kurchan00} for a periodic driving (where
$\Delta F=0$) and in Ref. \cite{HTasaki00} for finite $\Delta F$.
Further studies of (\ref{QJarzynski}) have been done in
Refs. \cite{Mukamel03,TalknerLutzHanggi07,TalknerBurada08} and of (\ref{QCrooks}) 
in Refs. \cite{TalknerHanggi07}. It was generalized to the microcanonical
ensemble in Refs. \cite{TalknerHanggi07b,Campisi08}.

\subsubsection{Work fluctuation theorem for closed driven systems}\label{WorkFTsectionRes}

We consider the same forward and backward process as described above,
except that during the driving the system now remains in weak contact
with a reservoir at equilibrium.
The total Hamiltonian is therefore of the form $\hat{H}(t)=\hat{H}_S(t)+\hat{H}_B+\hat{V}$,
where $\hat{H}_S(t)$ ($\hat{H}_B$) is the system (reservoir) Hamiltonian
and $\hat{V}$ the weak interaction between the two.
The work done by the driving force on the system is now given by
the difference between the system and the reservoir energy change
(this last one represents heat) according to the first
law of thermodynamics.

In this case, the connection with the results
of section \ref{2pointFT} is done using
\begin{eqnarray}
\hspace{-0.6cm} \hat{\rho}_0=\frac{\e^{-\beta \hat{H}_S(0)}}{Z_S(0)}
\frac{\e^{-\beta \hat{H}_B}}{Z_B} \  \  \;, \  \
\hat{\rho}_0^{\rm tr}&=&\frac{\e^{-\beta \hat{H}_S(t)}}{Z_S(t)} \frac{\e^{-\beta \hat{H}_B}}{Z_B} \;,
\end{eqnarray}
as well as $a_0=E_s(0)+E_b$ and $a_t=E_{s'}(t)+E_{b'}$, where $E_s(0)$ ($E_s(t)$)
are the eigenvalues of $\hat{H}_S(0)$ ($\hat{H}_S(t)$) and $E_b$ the eigenvalues of $\hat{H}_B$.
We define $i=(i_s,i_b)$ and $j=(j_s,j_b)$, where $i_s$ and $j_s$ are used
to distinguish between degenerate eigenstates of $\hat{H}_S(0)$ and $\hat{H}_S(t)$
and $i_b$ and $j_b$ between degenerate eigenstates of $\hat{H}_B$.
The work is therefore
\begin{eqnarray}
w = \Delta a = U_{s's}+Q_{b'b}
\end{eqnarray}
where $U_{s's}=E_{s'}(t)-E_s(0)$ is the change in the system energy
and $Q_{b'b}=E_{b'}-E_b$ is the heat transferred from the system to
the reservoir.
Since the eigenstates of the Hamiltonian constitute a complete basis
set, (\ref{Aaaad_bisB}) and (\ref{Aaaah_bisB}) become
\begin{eqnarray}
\label{alter1}
&&P[j,E_{s'}(t)+E_{b'};i,E_s(0)+E_b] \\
&&\hspace{2.4cm}=\vert \bra{j s' b'} \hat{U}(t,0) \ket{i s b} \vert^2
\bra{s b} \hat{\rho}_0 \ket{s b} \nonumber \\
&&P^{\rm tr}[i,E_s(0)+E_b;j,E_{s'}(t)+E_{b'}] \\
&&\hspace{2.4cm}= \vert \bra{j s' b'} \hat{U}(t,0) \ket{i s b} \vert^2
\bra{s'b'} \hat{\rho}_0^{\rm tr} \ket{s'b'} \;.\nonumber
\end{eqnarray}
Eq. (\ref{defRtil}) therefore gives
\begin{eqnarray}
&&R[j,E_{s'}(t)+E_{b'};i,E_s(0)+E_b] =\beta \big( w - \Delta F \big)
\nonumber \\ &&\hspace{2.2cm}=
R[E_{s'}(t)+E_{b'},E_s(0)+E_b]\;, \label{Workopeninirho}
\end{eqnarray}
where $\Delta F(t)=F(t)-F(0)$ is the free-energy difference between
the initial and final system state ($F(t)=-\beta^{-1} \ln Z_S(t)$).
The essential property that $R$ is independent of $i$ and $j$ and
expressed solely in terms of observable quantities is therefore again satisfied.
Using (\ref{detailedFTtil}), we get the same Crooks (\ref{QCrooks})
and Jarzynski (\ref{QJarzynski}) relation as in the isolated case.
The two relations were derived for quantum open driven
systems in many different ways in Refs.
\cite{Kurchan00,Maes04a,Monnai05,EspositoMukamel06,Crooks07a,Crooks07b,TalknerCampisi08}.
Using (\ref{defproblocalB}) and (\ref{modifiedentropy}),
we also find that (\ref{secondlawWorkisol}) still holds with
\begin{eqnarray}
\mean{w} = \trace \big( \hat{H}_S(t)+\hat{H}_B \big) \hat{\rho}(t)
- \trace \big( \hat{H}_S(0)+\hat{H}_B \big) \hat{\rho}_0 \label{entropyWorkOpen} \;.
\end{eqnarray}

\subsubsection{Fluctuation theorem for direct heat and matter exchange between two systems}
\label{TransientFTHM}

We consider two finite systems $A$ and $B$ with Hamiltonians
$\hat{H}_A$ and $\hat{H}_B$, each initially at equilibrium with
its own temperature and chemical potential.
The two systems are weakly interacting, allowing
heat and matter exchange between them.
The total Hamiltonian is of the form $\hat{H}_{tot}=\hat{H}_A+\hat{H}_B+\hat{V}$,
where $\hat{V}$ is the coupling term between $A$ and $B$.
The joint Hilbert space is ${\cal H}_A \times {\cal H}_B$.
The energy $E_A$ and the number of particles $n_A$ of
system $A$ is measured at time zero and again at time $t$.
We assume
\begin{eqnarray}
\hat{\rho}_0=\hat{\rho}_0^{\rm tr}=
\hat{\rho}_A^{eq}(\beta_A,\mu_A) \hat{\rho}_B^{eq}(\beta_B,\mu_B) \;,
\label{exchangeDM1}
\end{eqnarray}
where
\begin{eqnarray}
\hat{\rho}_X^{eq}(\beta_X,\mu_X)=\e^{-\beta_X (\hat{H}_X-\mu_X \hat{N}_X)}/\Xi_X
\label{exchangeDM2}
\end{eqnarray}
and $X=A,B$. $\Xi_X$ is the grand canonical partition function.
The index $i_X$ is used to distinguish between eigenstates
of $\hat{H}_X$ with same energy $E_X$ and number of particles $n_X$.
We define $i=(i_A,i_B)$ and $\alpha=(E_A,n_A,E_B,n_B)$.
Using (\ref{Aaaad_bisB}) and (\ref{Aaaah_bisB}), we find
\begin{eqnarray}
P[i',\alpha';i,\alpha] &=& \vert \bra{i',\alpha'} \hat{U}_t \ket{i,\alpha} \vert^2
\bra{\alpha} \hat{\rho}_0 \ket{\alpha} \label{ProbforwExch}\\
P^{\rm tr}[i,\alpha;i',\alpha']&=&
\vert \bra{i',\alpha'} \hat{U}_t \ket{i,\alpha} \vert^2
\bra{\alpha'} \hat{\rho}_0 \ket{\alpha'} \;. \label{ProbbackExch}
\end{eqnarray}
Eq. (\ref{defRtil}) with (\ref{exchangeDM1}) give
\begin{eqnarray}
R[\alpha',\alpha]&=&-\beta_A \big( (E_A-E_A')-\mu_A (n_{A}-n_{A}') \big)
\label{alterrationprob1ini}\\
&&-\beta_B \big( (E_B-E_B')-\mu_B (n_{B}-n_{B}') \big) \nonumber \;.
\end{eqnarray}
Conservation laws imply that changes in matter and energy in one system
are accompanied by the opposite changes in the other system so that
\begin{eqnarray}
E_A-E_A' &\approx& -(E_B-E_B') \label{alterConsEnerini} \\
n_{A}-n_{A}' &=& n_{B}-n_{B}' \label{alterConsPartini} \;.
\end{eqnarray}
The weak-interaction assumption is required for (\ref{alterConsEnerini}) to hold.
Using (\ref{alterConsEnerini}) and (\ref{alterConsPartini})
and defining the heat and matter nonequilibrium constraints
\begin{eqnarray}
&&{\cal A}_h \equiv - \beta_A + \beta_B \nonumber \\
&&{\cal A}_m \equiv \beta_A \mu_A - \beta_B \mu_B \;, \label{affinitiesDef}
\end{eqnarray}
we find that (\ref{alterrationprob1ini}) can be expressed
exclusively in terms of measured quantities $E_A$ and $n_{A}$
\begin{eqnarray}
R[E_A',n_{A}';E_A,n_{A}]
\approx - {\cal A}_h (E_A'-E_A) - {\cal A}_m (n_{A}'-n_{A}) .
\nonumber \\  \label{alterrationprob2b}
\end{eqnarray}
Using (\ref{defproblocalB}) and (\ref{modifiedentropy}), we find
\begin{eqnarray}
S &=& - \sum_{X=A,B} \beta_X \big( \trace \hat{H}_X \hat{\rho}_0
- \mu_X \trace \hat{N}_X \hat{\rho}_0 \big) \\
\bar{S} &=& - \sum_{X=A,B} \beta_X \big( \trace \hat{H}_X \hat{\rho}(t)
- \mu_X \trace \hat{N}_X \hat{\rho}(t) \big).
\end{eqnarray}
From (\ref{2law}), the ensemble average of (\ref{alterrationprob2b})
is the time-integrated entropy production which has the familiar force-flux 
form of nonequilibrium thermodynamics \cite{GrootMazur,Prigogine,Andrei08}
\begin{eqnarray}
\mean{R} &\approx& - {\cal A}_h \big( \trace \hat{H}_A \hat{\rho}(t) - \trace \hat{H}_A \hat{\rho}(0) \big)
\nonumber\\&&- {\cal A}_m \big( \trace \hat{N}_A \hat{\rho}(t) - \trace \hat{N}_A \hat{\rho}(0) \big).
\end{eqnarray}
The detailed FT follows from (\ref{detailedFTtil}) and (\ref{alterrationprob2b})
\begin{eqnarray}
\frac{p(\Delta E_A,\Delta n_A)}{p(-\Delta E_A,-\Delta n_A)}
\approx e^{-({\cal A}_h \Delta E_A + {\cal A}_m \Delta n_A)}  \;.
\label{FTonProbforExchange}
\end{eqnarray}
Positive ${\cal A}_h$ [${\cal A}_m$] means that $T_A>T_B$
[$\beta_A \mu_A > \beta_B \mu_B$] so that the probability for an
energy transfer $\Delta E_A$ [of a particle transfer $\Delta n_A$]
from $A$ to $B$ is exponentially more likely than from $B$ to $A$.\\

Such a FT for heat has been derived in Ref. \cite{JarzynskiQ}.
A similar FT for exchange of bosons has been derived in \cite{VandenBroeck07}.
This FT for particles can also be derived from the GF of Ref. \cite{Rammer03,Levitov04}.
Derivations of this detailed FT for specific models are presented
in section \ref{IsolTunnelJunct} and \ref{ThermoLimitRes}.

\subsection{Steady-state fluctuation theorems} \label{SSFT2point}

We give simple qualitative and general arguments to show that the FT
(\ref{detailedFTtil}) can be used to obtain a quantum steady-state FT for
heat and matter exchange between two reservoirs through an embedded system.\\

We consider two reservoirs $A$ and $B$ (with Hamiltonians $\hat{H}_A$
and $\hat{H}_B$) each initially at equilibrium with its own temperature
and chemical potential. A heat and matter exchange occurs between the two
reservoirs through a weakly coupled embedded system (e.g. a molecule or a quantum dot).
The total Hamiltonian is $\hat{H}_{tot}=\hat{H}_A+\hat{H}_B+\hat{V}$,
where $\hat{V}=\hat{H}_S+\hat{V}_{AS}+\hat{V}_{BS}$ contains the free
Hamiltonian of the system $\hat{H}_S$ and the coupling term between each
of the reservoirs and the system $\hat{V}_{AS}$ and $\hat{V}_{BS}$.
The total Hilbert space is ${\cal H}_A \times {\cal H}_B \times {\cal H}_S$.
We use the index $i_X$ to distinguish between eigenstates
of $\hat{H}_X$ with same energy $E_X$ and number of particles $n_X$,
where $X=A,B,S$. We define the abbreviated notation $i=(i_A,i_B,i_S)$
and $\alpha=(E_A,n_A,E_B,n_B,E_S,n_S)$.
The energy $E_A$ and the number of particles $n_A$ is measured
in reservoirs $A$ at time zero and again at time $t$. We assume
\begin{eqnarray}
\hat{\rho}_0=\hat{\rho}_0^{\rm tr}
=\hat{\rho}_A^{eq}(\beta_A,\mu_A) \hat{\rho}_B^{eq}(\beta_B,\mu_B)
\hat{\rho}_S^{eq}(\beta_S,\mu_S) , \label{transportDM1a}
\end{eqnarray}
where $\hat{\rho}_S^{eq}$ is the equilibrium system reduced density matrix.
Since
\begin{eqnarray}
P[i',\alpha';i,\alpha] &=&
\vert \bra{i',\alpha'} \hat{U}_t \ket{i,\alpha} \vert^2
\bra{i,\alpha} \hat{\rho}_0 \ket{i,\alpha} \label{alter1mod}\\
P^{\rm tr}[i,\alpha;i',\alpha'] &=&
\vert \bra{i',\alpha'} \hat{U}_t \ket{i,\alpha} \vert^2
\bra{i',\alpha'} \hat{\rho}_0 \ket{i',\alpha'} \;,\label{alter2mod}
\end{eqnarray}
Eq. (\ref{defRtil}) reads
\begin{eqnarray}
\label{alterrationprob1}
R[\alpha',\alpha]
&=&-\beta_A \big( (E_A-E_A')-\mu_A (n_{A}-n_{A}') \big)  \\
&&-\beta_B \big( (E_B-E_B')-\mu_B (n_{B}-n_{B}') \big) \nonumber \\
&&-\beta_S \big( (E_S-E_S')-\mu_S (n_{S}-n_{S}') \big) \nonumber \;.
\end{eqnarray}
Since the system-reservoir couplings are weak, conservation
laws of the total unperturbed system ($\hat{H}_{tot}$ with
$\hat{V}_{AS}+\hat{V}_{BS}=0$) implies that
\begin{eqnarray}
E_B-E_B' &\approx&-(E_A-E_A')-(E_{S}-E_{S}') \label{alterConsEner}\\
n_{B}-n_{B}' &=& n_{A}'-n_{A}+n_{S}'-n_{S} \label{alterConsPart}\;.
\end{eqnarray}
This means that (\ref{alterrationprob1}) is equal to
\begin{eqnarray}
R[E_A',n_{A}';E_A,n_{A}] &\approx&
-{\cal A}_h (E_A'-E_A)-{\cal A}_m (n_{A}'-n_{A}) \nonumber\\
&&\hspace{-0.4cm}
+ {\cal O}(E_S'-E_S) + {\cal O}(n_S'-n_S).\label{alterrationprob2}
\end{eqnarray}
Since $A$ and $B$ are assumed macroscopic (i.e. reservoirs), the change
in energy $E_A'-E_A$ and matter $n_A-n_A'$ in reservoir $A$ is not bounded.
However, because system $S$ is assumed small and finite,
$E_S'-E_S$ and $n_S'-n_S$ are always bounded and finite.
This means that in the long time limit, these contribution
to $R$ will become negligible in (\ref{alterrationprob2}).
For long times, the FT (\ref{detailedFTtil}) with
(\ref{alterrationprob2}) becomes a universal (independent of
system quantities) steady-state FT for the heat and matter currents
\begin{eqnarray}
\lim_{t \to \infty} \frac{1}{t} \ln
\frac{p(\Delta E_A,\Delta N_A)}{p(-\Delta E_A,-\Delta N_A)}
= {\cal A}_h I_h+ {\cal A}_m I_m \;, \label{currentFTembeded}
\end{eqnarray}
where $I_h= \Delta E_A/t$ and $I_m=\Delta N_A/t$ are the heat
and matter current between the system and the reservoir $A$.
The r.h.s. of (\ref{currentFTembeded}) can thus be 
interpreted as an entropy production.
A rigorous proof of (\ref{currentFTembeded}) has been 
recently given in Ref. \cite{AndrieuxGaspardTasaki}.
In the long time limit, the steady-state FT (\ref{currentFTembeded})
is similar to the detailed FT (\ref{FTonProbforExchange}).
We note that the long time limit is related to the
existence of a large deviation function (see appendix \ref{largedev}).
We also note that when the system $S$ is not finite,
${\cal O}(E_S'-E_S)$ and ${\cal O}(n_S'-n_S)$ terms in
(\ref{alterrationprob2}) may not be negligible in the
long time limit, as observed in Ref. \cite{Zon,Zonb}.
Similar problems are expected if $A$ and $B$ are not "good" reservoirs.
A "good" reservoirs should allow the system to reach a steady-state.
Since it is known that such reservoirs cannot be properly described within
the Hamiltonian formalism, it should be no surprise that more systematic
derivations of quantum steady-state FT (\ref{currentFTembeded}) require to use
some effective (and irreversible) description of the embedded system dynamics.
A common way to do this is the quantum master equation approach which consists in
deriving an approximate equation of motion for the system reduced density
matrix containing the effects of reservoir through its correlation functions.
As required for a "true" reservoir, the back-action
of the system on the reservoir is neglected (Born approximation).
Such a derivation of the steady-state FTs will be presented in
section \ref{MEGF} [see (\ref{4april-7}) and (\ref{7may-1})].
Another approach, is based on a system Greens functions description.
Here, the effect of the reservoirs appear through the self-energies.
These derivations will be presented in section \ref{GreenTransportJunction}.
It has been recently suggested that finite thermostats (commonly used to model
thermostatted classical dynamics) could also be used to describe thermostatted
quantum dynamics \cite{Gallavotti08}.

\section{Heat and matter transfer statistics in weakly-coupled open systems} \label{MEGF}

We now consider a small quantum system weakly interacting with a reservoir.
Heat and matter exchanges are measured by a projective measurement in the reservoir.
We will derive a generalized quantum master equation (GQME) for the GF associated
to the system density matrix conditional to a given transfer with the reservoir.
The statistics is therefore obtained from the solution of the GQME.
When summing the GQME over all possible transfer processes,
one recovers the standard quantum master equation (QME).

\subsection{Generalized quantum master equation}

We consider a single reservoir, but the extension
to multiple reservoirs is straightforward.
The total Hamiltonian is the sum of the system $S$ Hamiltonian,
$\hat{H}_S$, the reservoir $R$ Hamiltonian, $\hat{H}_R$, and the weak interaction
between the two, $\hat{V}$.
\begin{eqnarray}
\hat{H} = \hat{H}_0 + \hat{V} = \hat{H}_S + \hat{H}_R + \hat{V} \;. \label{Hamiltonian}
\end{eqnarray}
We use the index $s$ ($r$) to label the eigenstates
of the Hamiltonian of system $S$ ($R$).
The reservoir is initially assumed to be at equilibrium
$\hat{\rho}^{eq}_R=\e^{-\beta (\hat{H}_R-\mu \hat{N}_R)}/\Xi_R$.
The measured observable is the energy $\hat{H}_R$ and number
of particle $\hat{N}_R$ in the reservoir.
Since the measured observables commutes with the
initial density matrix $\hat{\rho}_0=\hat{\rho}_S(0) \hat{\rho}^{eq}_R$,
using (\ref{Aaaacx1}), we get
\begin{eqnarray}
G(\lambda,t) = \trace  \hat{\rho}(\lambda,t) \;, \label{GFtottot}
\end{eqnarray}
where $\lambda=\{\lambda_h,\lambda_m\}$,
\begin{eqnarray}
\hat{\rho}(\lambda,t)
\equiv \e^{-\frac{\ic}{\hbar} \hat{H}_{\lambda} t} \hat{\rho}_0
\e^{\frac{\ic}{\hbar} \hat{H}_{-\lambda} t} \label{GFtot}
\end{eqnarray}
and
\begin{eqnarray}
\hat{H}_{\lambda} &=& \e^{\frac{\ic}{2} ( \lambda_h \hat{H}_R + \lambda_m \hat{N}_R )}
\hat{H} \e^{-\frac{\ic}{2} ( \lambda_h \hat{H}_R + \lambda_m \hat{N}_R )} \label{defHgamma}\\
&=& \hat{H}_0 + \hat{V}_{\lambda} \;. \nonumber
\end{eqnarray}
Obviously, $\hat{\rho}(t)=\hat{\rho}(\lambda=0,t)$.

We define the system GF
\begin{eqnarray}
\hat{\rho}_S(\lambda,t) \equiv \trace_R \hat{\rho}(\lambda,t) \;,
\label{GFIdef}
\end{eqnarray}
which is an operator in the system space.
Since $\hat{\rho}_S(t)=\hat{\rho}_S(\lambda=0,t)$ is the reduced density matrix
of the system, $\hat{\rho}_S(\lambda,t)$ is a reduced density matrix
of the system conditional to a certain energy and
matter transfer between $S$ and $R$.
We can now rewrite (\ref{GFtottot}) as
\begin{eqnarray}
G(\lambda,t) = \trace_S \hat{\rho}_S(\lambda,t) \;.
\label{GFtottotastrace}
\end{eqnarray}
We will derive a closed evolution equation for $\hat{\rho}_S(\lambda,t)$
by using projection operator technique and second order perturbation
theory in $\hat{V}$ on $\hat{\rho}(\lambda,t)$.
By solving this equation one can get $G(\lambda,t)$.
Details are given in appendix \ref{GFMEweak}.
The final result reads
\begin{eqnarray}
\dot{\hat{\rho}}_S(\lambda,t)&=& -\frac{\ic}{\hbar} [\hat{H}_S ,\hat{\rho}_S(\lambda,t)]
+ \frac{1}{\hbar^2}
\sum_{\kappa\kappa'} \int_{0}^{t} d\tau \label{GFevolofRedGen} \\
&&\hspace{0cm}\Big\{- \trace_R \{ \hat{V}^{\kappa}_{\lambda} \hat{V}^{\kappa'}_{\lambda}(-\tau)
\hat{\rho}^{eq}_R \hat{\rho}_S(\lambda,t) \} \nonumber \\ &&\hspace{0.4cm}
- \trace_R \{ \hat{\rho}^{eq}_R \hat{\rho}_S(\lambda,t) \hat{V}^{\kappa}_{-\lambda}(-\tau)
\hat{V}^{\kappa'}_{-\lambda} \} \nonumber \\ &&\hspace{0.4cm}
+ \trace_R \{ \hat{V}^{\kappa}_{\lambda} \hat{\rho}^{eq}_R \hat{\rho}_S(\lambda,t)
\hat{V}^{\kappa'}_{-\lambda}(-\tau) \} \nonumber \\ &&\hspace{0.4cm}
+ \trace_R \{ \hat{V}^{\kappa}_{\lambda}(-\tau) \hat{\rho}^{eq}_R
\hat{\rho}_S(\lambda,t) \hat{V}^{\kappa'}_{-\lambda} \} \; \Big\} \; , \nonumber
\end{eqnarray}
where
\begin{eqnarray}
\hat{V}_{\lambda}^{\kappa}(t)=\e^{\frac{\ic}{\hbar} \hat{H}_0 t}
\hat{V}_{\lambda}^{\kappa} \e^{-\frac{\ic}{\hbar} \hat{H}_0 t} . \label{Megacoupling}
\end{eqnarray}

\subsubsection{Generalized reservoir correlation functions}
\label{mark-gencorr}

We now consider an interaction of the form
\begin{eqnarray}
\hat{V} = \sum_{\kappa} \hat{S}^{\kappa} \hat{R}^{\kappa} \;, \label{HamilCouplBis}
\end{eqnarray}
where $\hat{S}^{\kappa}$ ($\hat{R}^{\kappa}$) is a coupling operator of system $S$ ($B$).
It follows from (\ref{defHgamma}) that
$\hat{V}_{\lambda} \equiv \sum_{\kappa} \hat{S}^{\kappa} \hat{R}^{\kappa}_{\lambda}$, where
\begin{eqnarray}
\hat{R}^{\kappa}_{\lambda} \equiv \e^{\frac{\ic}{2}(\lambda_h \hat{H}_R +
\lambda_m \hat{N}_R )} \hat{R}^{\kappa} \e^{-\frac{\ic}{2} ( \lambda_h \hat{H}_R +
\lambda_m \hat{N}_R )}. \label{defVgamma}
\end{eqnarray}
For such interaction, (\ref{GFevolofRedGen}) becomes
\begin{eqnarray}
\dot{\hat{\rho}}_S(\lambda,t)&=& -\frac{\ic}{\hbar} [\hat{H}_S ,\hat{\rho}_S(\lambda,t)]
+ \frac{1}{\hbar^2}
\sum_{\kappa\kappa'} \int_{0}^{t} d\tau  \label{GFevolofRed} \\
&&\hspace{-0cm}\Big\{ -\alpha_{\kappa \kappa'}(\tau) \hat{S}^{\kappa}
\hat{S}^{\kappa'}(-\tau) \hat{\rho}_S(\lambda,t) \nonumber \\ &&\hspace{0.4cm}
- \alpha_{\kappa \kappa'}(-\tau)
\hat{\rho}_S(\lambda,t) \hat{S}^{\kappa}(-\tau) \hat{S}^{\kappa'} \nonumber \\
&&\hspace{0.4cm}+\alpha_{\kappa' \kappa}(-\lambda,-\tau) \hat{S}^{\kappa}
\hat{\rho}_S(\lambda,t) \hat{S}^{\kappa'}(-\tau) \nonumber \\ &&\hspace{0.4cm}
+ \alpha_{\kappa' \kappa}(-\lambda,\tau)
\hat{S}^{\kappa}(-\tau) \hat{\rho}_S(\lambda,t) \hat{S}^{\kappa'} \; \Big\} \; . \nonumber
\end{eqnarray}
Here we have defined the generalized reservoir correlation functions
\begin{eqnarray}
\alpha_{\kappa \kappa'}(\lambda,t)
&\equiv& \trace_R \hat{\rho}^{eq} \hat{R}^{\kappa}_{2\lambda}(t) \hat{R}^{\kappa'} \label{DefTimecorrel} \\
&=& \sum_{rr'} \frac{\e^{-\beta (E_r-\mu N_r)}}{Z_G}
\e^{\frac{i}{\hbar} (E_r-E_{r'}) t} \nonumber \\
&&\hspace{0.5cm}\e^{\ic \{ \lambda_h (E_r-E_{r'}) + \lambda_m (N_r-N_{r'}) \}}
R^{\kappa}_{rr'} R^{\kappa'}_{r'r} \nonumber
\end{eqnarray}
where $\hat{R}^{\kappa}_{\lambda}(t)=\e^{\frac{i}{\hbar} \hat{H}_R t}
\hat{R}^{\kappa}_{\lambda} \e^{-\frac{i}{\hbar} \hat{H}_R t}$.
The reservoir correlation functions are given by
$\alpha_{\kappa \kappa'}(t) \equiv \alpha_{\kappa \kappa'}(\lambda=0,t)$.
For $\lambda=0$, (\ref{GFevolofRed}) therefore reduces to the
non-Markovian Redfield QME of Ref. \cite{GaspNaga99}.

The ordinary reservoir correlation functions satisfy the standard
Kubo-Martin-Schwinger (KMS) condition \cite{KuboB98b}
\begin{eqnarray}
\alpha_{\kappa\kappa'}(t) =
\alpha_{\kappa'\kappa}(-t-\ic \hbar \beta) \label{KMStime} \;.
\end{eqnarray}
In the frequency domain
\begin{eqnarray}
\label{fourier} \tilde{\alpha}_{\kappa
\kappa'}(\lambda,\omega)&=&\int_{-\infty}^{\infty} \frac{dt}{2\pi}
\e^{i \omega t} \alpha_{\kappa \kappa'}(\lambda,t)
\end{eqnarray}
the KMS relation reads
\begin{eqnarray}
\tilde{\alpha}_{\kappa\kappa'}(\omega) = \e^{\beta \hbar \omega}
\tilde{\alpha}_{\kappa'\kappa}(-\omega). \label{KMSfreq}
\end{eqnarray}
The generalized reservoir correlation functions satisfy the symmetry
\begin{eqnarray}
&&\alpha_{\kappa\kappa'}(\{\lambda_h,\lambda_m\},t) = \nonumber \\
&&\hspace{1.3cm}
\alpha_{\kappa'\kappa}(\{-\lambda_h-\ic \beta,-\lambda_m+\ic \beta \mu\},-t) \; .
\label{KMStimeGen}
\end{eqnarray}
We note also that if $\hat{R}^{\kappa}$ and $\hat{S}^{\kappa}$ are Hermitian,
we further have
\begin{eqnarray}
\alpha_{\kappa \kappa'}(\lambda,t) = \alpha_{\kappa' \kappa}^*(-\lambda,-t) \;.
\end{eqnarray}

\subsubsection{The Markovian and the rotating wave approximation}
\label{mark-rwa}

Two approximations commonly used to simplify the QME may also be used on the GQME.
The Markovian approximation consist of setting the upper bound
of the time integral in (\ref{GFevolofRed}) to infinity.
The rotating wave approximation (RWA) \cite{Gardiner00,Breuer02}
(also known as secular approximation \cite{CohenTann96,Brandes08} or
Davis procedure \cite{Spohn80,KampenB97}) is often used to impose a
Lindblad form \cite{Lindblad76,Spohn80,Breuer02} to the Markovian
QME generator in order to guaranty the complete positivity
of the subsystem density matrix time evolution.
Without RWA, the Markovian QME generator can lead to a positivity breakdown
for certain set of initial conditions due to small errors introduced
on the initial short-time dynamics by the Markovian approximation
\cite{PechukasPRL94,Silbey92,Tannor97,GaspNaga99,Silbey05,Sudarshan08}.
One has to note however that the use of the RWA is not always
physically justified and might miss important effects
\cite{Silbey92,Tannor97,GaspNaga99,Silbey05}.
The RWA is equivalent to define a coarse-grained time derivative
of the system density matrix on times long compared to
the free system evolution \cite{CohenTann96,Brandes08}.
One easy way to perform the RWA consist in time averaging
$\lim_{T\to\infty}\frac{1}{2T}\int_{-T}^{T}dt$ the generator of the
QME in the interaction picture and in the system eigenbasis, using
\begin{eqnarray}
\int_{0}^{\infty} d\tau \e^{\pm i \omega \tau} =
\pi \delta(\omega) \pm i {\rm P} \frac{1}{\omega}= \lim_{\eta \to 0^+} \frac{1}{\eta \mp \ic \omega}.
\label{RelDistribTheo}
\end{eqnarray}
Using these two approximation on the GQME (\ref{GFevolofRed}),
we find that coherences, $\rho_{ss'}(t) \equiv \bra{s}\hat{\rho}_S(t)\ket{s'}$
with $s \neq s'$, follow the dynamics
\begin{eqnarray}
\dot{\rho}_{ss'}(\lambda,t) =
(- \Gamma_{ss'} - i \Omega_{ss'}) \rho_{ss'}(\lambda,t)
\label{B11114} \; ,
\end{eqnarray}
where the relaxation rates are given by
\begin{eqnarray}
\Gamma_{ss'} &=& \frac{1}{\hbar^2} \sum_{\kappa \kappa'} \Big\lbrace
- 2 \pi \tilde{\alpha}_{\kappa\kappa'}(0)
S^{\kappa'}_{ss} S^{\kappa}_{s's'} \label{B11115} \\
&&\hspace{-0.6cm} +\pi \sum_{\bar{s}}
\left[ \tilde{\alpha}_{\kappa\kappa'}
(\omega_{s\bar{s}}) S^{\kappa}_{s\bar{s}}
S^{\kappa'}_{\bar{s}s} + \tilde{\alpha}_{\kappa'\kappa}
(\omega_{s'\bar{s}}) S^{\kappa'}_{s'\bar{s}}
S^{\kappa}_{\bar{s}s'} \right] \Big\rbrace \nonumber
\end{eqnarray}
and the modified system frequencies are
\begin{eqnarray}
\Omega_{ss'} &=& \omega_{ss'} - \frac{1}{\hbar^2} \sum_{\kappa \kappa'}
\sum_{\bar{s}} \left[ \int_{-\infty}^{\infty} d\omega {\rm P}
\frac{\tilde{\alpha}_{\kappa\kappa'}(\omega)}
{\omega+\omega_{\bar{s}s}} S^{\kappa}_{s\bar{s}} S^{\kappa'}_{\bar{s}s}
\right. \nonumber \\
&&\hspace{1.5cm} \left. -\int_{-\infty}^{\infty} d\omega {\rm P}
\frac{\tilde{\alpha}_{\kappa\kappa'}(\omega)}
{\omega+\omega_{\bar{s}s'}} S^{\kappa}_{s'\bar{s}}
S^{\kappa'}_{\bar{s}s'} \right] \label{B11116}.
\end{eqnarray}
The coherences evolve independently from the populations
[diagonal elements $\rho_{ss}(t)$] and also independently
from of each other. They simply undergo an exponentially
damped oscillations which are independent of $\lambda$.
Populations, on the other hand, evolve according to the equation
\begin{eqnarray}
\dot{\rho}_{ss}(\lambda,t) &=&
\frac{1}{\hbar^2} \sum_{\kappa \kappa'} \sum_{\bar{s}}
\Big\lbrace \label{B11117} \\
&&\hspace{0cm}-2 \pi
\tilde{\alpha}_{\kappa\kappa'}(-\omega_{\bar{s}s})
S^{\kappa}_{s\bar{s}} S^{\kappa'}_{\bar{s}s} \rho_{ss}(\lambda,t)
\nonumber \\ &&\hspace{0cm}+2 \pi
\tilde{\alpha}_{\kappa'\kappa}(\lambda,\omega_{\bar{s}s}) S^{\kappa}_{s\bar{s}}
S^{\kappa'}_{\bar{s}s} \rho_{\bar{s}\bar{s}}(\lambda,t) \Big\rbrace \nonumber \; .
\end{eqnarray}
The population dynamics depends on $\lambda$.

\subsection{Applications to particle counting statistics} \label{ManyBodyapplic}

We now calculate the particle statistics for different
models and derive various steady-state FTs using the GQME.

\subsubsection{Fermion transport}\label{fermion-transport}

We consider a many electron quantum system attached to
two metal leads which act as particle reservoirs.
We shall denote the singe-particle eigenstates of the
system and leads by indices $s$ and $i$, respectively.
The total Hamiltonian is
$\hat{H}=\hat{H}_A+\hat{H}_B+\hat{H}_S+\hat{V}$, where
\begin{eqnarray}
\label{fermi-hamil-2a}
\hat{H}_X= \sum_{i\in X=A,B} \epsilon_i \hat{c}_i^\dag \hat{c}_i,
~~~  \hat{H}_S= \sum_s \epsilon_s \hat{c}_s^\dag \hat{c}_s.
\end{eqnarray}
The coupling between the lead $X=A,B$ and the system is $\hat{V}_X=\hat{J}_{X}+\hat{J}_{X}^\dag$
where $\hat{J}_{X} = \sum_{s,i\in X} J^X_{si} \hat{c}_s^\dag \hat{c}_i$ and $J_{si}^X$ are the
coupling elements between the system and the leads $X$. The total coupling is then
\begin{eqnarray}
\label{fermi-hamil-2b}
\hat{V}= \hat{J}_A + \hat{J}_B + \hat{J}_A^\dag +\hat{J}_B^\dag.
\end{eqnarray}
There is no direct coupling between the two leads, and an electron transfer is only
possible by charging or discharging the quantum system.
The operators $\hat{c}(\hat{c}^\dag)$ represent the annihilation
(creation) operators which satisfy the Fermi anticommutation relations
\begin{eqnarray}
\hat{c}_s\hat{c}_{s^\prime}^\dag+\hat{c}_{s^\prime}^\dag \hat{c}_s &=&\delta_{ss^\prime},\nonumber\\
\hat{c}_s^\dag \hat{c}_{s^\prime}^\dag+\hat{c}_{s^\prime}^\dag \hat{c}_s^\dag &=&
\hat{c}_s\hat{c}_{s^\prime}+\hat{c}_{s^\prime}\hat{c}_s=0.
\end{eqnarray}
To connect with the notation of the Hamiltonian (\ref{Hamiltonian}),
we have $\hat{H}_R=\hat{H}_A+\hat{H}_B$ and $\hat{V}=\hat{V}_A+\hat{V}_B$.
Apart from the difference in chemical potentials $\mu_A$ and $\mu_B$
with $eV=\mu_A-\mu_B$, the two leads are assumed be identical.

To count the change in the number of electrons in
the lead $A$, the projection is done on $A$.
Therefore (\ref{Megacoupling}) for this model reads
\begin{eqnarray}
\label{4april-1}
\hat{V}_{\lambda} &=& \mbox{e}^{\frac{\ic}{2}\lambda \hat{N}_A}\left(\hat{J}_A+\hat{J}_A^\dag\right)
\mbox{e}^{-\frac{\ic}{2}\lambda \hat{N}_A}+\hat{V}_B\nonumber\\
&=& \mbox{e}^{-\frac{\ic}{2}\lambda}\hat{J}_A+ \mbox{e}^{\frac{\ic}{2}\lambda}\hat{J}_A^\dag +\hat{V}_B\;.
\end{eqnarray}
To get the second line, we used the relation $\hat{J}_A\hat{N}_A=(\hat{N}_A+1)\hat{J}_A$.
Substituting Eq. (\ref{4april-1}) in (\ref{GFevolofRedGen}), the GQME becomes
\begin{eqnarray}
\label{4april-2}
&&\hspace{-0.5cm}\dot{\hat{\rho}}_S(\lambda,t)
= -\frac{\ic}{\hbar}[\hat{H}_S,\hat{\rho}_S(\lambda,t)]+\frac{1}{\hbar^2}\sum_{ss^\prime}\int_0^td\tau\nonumber\\
&&\hspace{0cm} \left[ \frac{}{}\; \; \{\mbox{e}^{\ic \lambda}\alpha^A_{ss^\prime}(-\tau)+\alpha^B_{ss^\prime}(-\tau)\}
\hat{c}_{s^\prime}\hat{\rho}_{S}(\lambda,t)\hat{c}_s^\dag(-\tau)\right.\nonumber\\
&&\hspace{0.2cm} +\left. \{\mbox{e}^{\ic \lambda}\alpha^A_{ss^\prime}(\tau)+\alpha^B_{ss^\prime}(\tau)\}
\hat{c}_{s^\prime}(-\tau)\hat{\rho}_{S}(\lambda,t)\hat{c}_s^\dag\right.\nonumber\\
&&\hspace{0.2cm} + \left. \{\mbox{e}^{-\ic \lambda}\beta^A_{ss^\prime}(-\tau)+\beta^B_{ss^\prime}(-\tau)\}
\hat{c}_{s}^\dag\hat{\rho}_{S}(\lambda,t)\hat{c}_{s^\prime}(-\tau)\right.\nonumber\\
&&\hspace{0.2cm} +\left. \{\mbox{e}^{-\ic \lambda}\beta^A_{ss^\prime}(\tau)+\beta^B_{ss^\prime}(\tau)\}
\hat{c}_{s}^\dag(-\tau)\hat{\rho}_{S}(\lambda,t)\hat{c}_{s^\prime}\right.\nonumber\\
&&\hspace{0.2cm} -\left. \alpha_{ss^\prime}(\tau)\hat{c}_{s}^\dag \hat{c}_{s^\prime}(-\tau)\hat{\rho}_{S}(\lambda,t)\right.\nonumber\\
&&\hspace{0.2cm} -\left. \beta_{ss^\prime}(\tau)\hat{c}_{s^\prime}\hat{c}_{s}^\dag(-\tau)\hat{\rho}_{S}(\lambda,t)\right.\nonumber\\
&&\hspace{0.2cm} -\left. \alpha_{ss^\prime}(-\tau)\hat{\rho}_{S}(\lambda,t)\hat{c}_{s}^\dag(-\tau) \hat{c}_{s^\prime}\right.\nonumber\\
&&\hspace{0.2cm} -\left. \beta_{ss^\prime}(-\tau)\hat{\rho}_{S}(\lambda,t)\hat{c}_{s^\prime}(-\tau) \hat{c}_s^\dag \frac{}{} \; \; \right]
\end{eqnarray}
where
\begin{eqnarray}
\label{4april-3}
\alpha^X_{ss^\prime}(\tau)=\sum_{ii^\prime\in X}J^X_{si}(J^X_{s^\prime i^\prime})^*
\mbox{Tr}\{\hat{c}_i(\tau)\hat{c}_{i^\prime}^\dag\hat{\rho}_B\}\nonumber\\
\beta^X_{ss^\prime }(\tau)=\sum_{ii^\prime\in X}J^X_{si}(J^X_{s^\prime i^\prime})^*
\mbox{Tr}\{\hat{c}_i^\dag(\tau)\hat{c}_i\hat{\rho}_B\}
\end{eqnarray}
are the equilibrium correlation functions for leads $X$ and where
$\alpha_{ss^\prime}(\tau)=\alpha_{ss^\prime}^A(\tau)+\alpha_{ss^\prime}^B(\tau)$
and $\beta_{ss^\prime}(\tau)=\beta_{ss^\prime}^A(\tau)+\beta_{ss^\prime}^B(\tau)$.

For $\lambda=0$, Eq. (\ref{4april-2}) reduces
to the QME derived in Ref. \cite{HarbolaEsposito06}.
After applying the Markovian approximation described in section \ref{mark-rwa}
(the upper limit of the time integral in Eq. (\ref{4april-2}) is extended to
infinity), we perform the RWA approximation which is equivalent to assume that the
lead correlation functions are diagonal in $s$ \cite{HarbolaEsposito06}.
Eq. (\ref{4april-2}) then becomes
\begin{eqnarray}
\label{4april-4}
\dot{\hat{\rho}}_S(\lambda,t) &=& -\frac{\ic}{\hbar}[\hat{H}_S,\hat{\rho}_S(\lambda,t)] \\
&&\hspace{-0.8cm} + \sum_{s}\left[ \; \; \frac{}{}
\{\mbox{e}^{-\ic \lambda}\alpha^A_{ss}+\alpha^B_{ss}\}\hat{c}_s^\dag\hat{\rho}_S(\lambda,t)\hat{c}_s \right.\nonumber\\
&&\hspace{0.4cm}+ \left.\{\mbox{e}^{\ic \lambda}\beta^A_{ss}+\beta^B_{ss}\}\hat{c}_s\hat{\rho}_S(\lambda,t)\hat{c}_s^\dag\right.\nonumber\\
 &&\hspace{0.4cm}-\left.\alpha_{ss}\hat{c}_s\hat{c}_s^\dag \hat{\rho}_S(\lambda,t)-\beta_{ss}\hat{c}_s^\dag \hat{c}_s \hat{\rho}_S(\lambda,t)
\frac{}{}\right] \nonumber .
\end{eqnarray}
The rates $\alpha^X_{ss}$ and $\beta^X_{ss}$ are calculated by assuming a constant
density of states $\sigma$ for the leads over the energy range around the Fermi level
\begin{eqnarray}
\label{4april-5}
\alpha^X_{ss} &=& \frac{2\pi}{\hbar^2} \sigma |J_s^X|^2 (1-f_X(\epsilon_s))\nonumber\\
\beta^X_{ss} &=& \frac{2\pi}{\hbar^2} \sigma |J_s^X|^2 f_X(\epsilon_s) ,
\end{eqnarray}
where $f_X(\epsilon)=[1+\mbox{e}^{-\beta(\epsilon-\mu_X)}]^{-1}$ is the
Fermi function of lead $X$, and $\beta=1/k_B T$.
These rates satisfy the relation
\begin{eqnarray}
\label{4april-6}
\frac{\alpha^A_{ss}\beta^B_{ss}}{\alpha^B_{ss}\beta^A_{ss}} = \mbox{e}^{\beta eV}.
\end{eqnarray}
The solution of (\ref{4april-4}) allows to compute the time-dependent
electron statistics between lead $A$ and the system at any time.
For $\lambda=0$, (\ref{4april-4}) is the Lindblad QME derived in \cite{HarbolaEsposito06}.
Equation (\ref{4april-4}) was first derived in Ref. \cite{EspositoHarbola07} by unraveling this QME.
This means that the QME is interpreted as resulting from a continuous positive
operator-valued measurement \cite{Nielsen,Breuer02} on the system by the leads.
This allows to construct probabilities for histories of electron transfers,
and to use them to derive equations of motion for the GF associated with the
probability distribution of a net transfer of electrons during a given time
interval, which are identical to (\ref{4april-4}).
We thus find that the two-point projection method and the positive
operator-valued measurement lead to the same electron statistics
result in the weak coupling regime (with Markovian and RWA).
A similar conclusion was reached in Refs. \cite{Maes07,DeRoeck07}.

In (\ref{4april-4}), the GF factorizes in terms of single orbital GF of the system,
$\hat{\rho}_S(\lambda,t)=\prod_{s=1}^M \hat{\rho}_s(\lambda,t)$, where $M$ is the total
number of orbital and $\hat{\rho}_s(\lambda,t)$ is the single orbital GF, so that
\begin{eqnarray}
\label{22april-1}
\dot{\hat{\rho}}_s(\lambda,t) &=&  -\frac{\ic}{\hbar}\epsilon_s[\hat{c}_s^\dag \hat{c}_s,\hat{\rho}_s]\\
&&+ \left[ \; \frac{}{} \{\mbox{e}^{-\ic \lambda}\alpha^A_{ss}+\alpha^B_{ss}\}
\hat{c}_s^\dag\hat{\rho}_s(\lambda,t)\hat{c}_s \right.\nonumber\\
&&\hspace{0.4cm}+\left.\{\mbox{e}^{\ic \lambda}\beta^A_{ss}+\beta^B_{ss}\}\hat{c}_s
\hat{\rho}_s(\lambda,t)\hat{c}_s^\dag\right.\nonumber\\
&&\hspace{0.4cm}-\left.\alpha_{ss}\hat{c}_s\hat{c}_s^\dag \hat{\rho}_s(\lambda,t)
-\beta_{ss}\hat{c}_s^\dag \hat{c}_s \hat{\rho}_s(\lambda,t)
\frac{}{} \; \right].\nonumber
\end{eqnarray}
As discussed in Sec. (\ref{mark-rwa}), the GQME (\ref{4april-4}), when expressed in the
eigenbasis of the system describes an independent dynamics for coherences and populations.
The coherences simply decay in time following damped
oscillations while populations follow a classical rate equation.
If the eigenstates of each orbital are denoted by $|n_s \rangle$, where $n_s=0,1$,
the vector made of the population of $\hat{\rho}_s(\lambda,t)$ in this basis
denoted by $\tilde{\rho}_s(\lambda,t) \equiv \{\langle 0|\hat{\rho}_s(\lambda,t)|
0 \rangle,\langle 1|\hat{\rho}_s(\lambda,t)|1 \rangle\}$ evolves according to
\begin{eqnarray}
\label{22april-2}
\dot{\tilde{\rho}}_s(\lambda,t)= \Gamma_s(\lambda) \tilde{\rho}_s(\lambda,t)
\end{eqnarray}
where $\Gamma_s(\lambda)$ is a $2\times 2$ matrix
\begin{eqnarray}
\label{23april-1}
\Gamma_s(\lambda) = \left(
\begin{array}{cc}
-\alpha_{ss}& \mbox{e}^{\ic \lambda}\beta^A_{ss}+\beta^B_{ss}\\
\mbox{e}^{-\ic \lambda}\alpha^A_{ss}+\alpha^B_{ss}& -\beta_{ss}
\end{array}
\right).
\end{eqnarray}
The eigenvalues of this matrix are given by
\begin{eqnarray}
\label{23april-2}
\gamma_{s\pm}(\lambda) = -\frac{\alpha_{ss}+\beta_{ss}}{2}\pm \sqrt{f(\lambda)}
\end{eqnarray}
where
\begin{eqnarray}
\label{9may-1}
f(\lambda)=(\mbox{e}^{\ic \lambda}\beta^A_{ss}+\beta^B_{ss})
(\mbox{e}^{-\ic \lambda}\alpha^A_{ss}+\alpha^B_{ss})
+ \frac{1}{4}(\alpha_{ss}-\beta_{ss})^2. \nonumber
\end{eqnarray}
Since $G(\lambda,t)=\prod_s G_s(\lambda,t)$, where
$G_s(\lambda,t)=\langle 0|\hat{\rho}_s(\lambda,t)|0 \rangle
+\langle 1|\hat{\rho}_s(\lambda,t)|1 \rangle$, the long time
limit of the cumulant GF is given by the dominant eigenvalue
\begin{eqnarray}
{\cal S}(\lambda) = \lim_{t \to \infty} \frac{1}{t} \ln G(\lambda,t)
= \sum_{s} \gamma_{s+}(\lambda). \label{9may-2}
\end{eqnarray}
Using (\ref{4april-6}) and (\ref{23april-2}), we find that
$\gamma_{s\pm}(\lambda)=\gamma_{s\pm}(-\ic\beta eV-\lambda)$,
which implies that
\begin{eqnarray}
\label{23april-3}
{\cal S}(\lambda)={\cal S}(-\ic\beta eV-\lambda).
\end{eqnarray}
In appendix \ref{largedev}, we show that this symmetry
implies the steady-state fluctuation-theorem
\begin{eqnarray}
\label{4april-7}
\lim_{t\to\infty}\frac{p(k,t)}{p(-k,t)}= e^{\beta eVk},
\end{eqnarray}
where $p(k,t)$ is the probability of transferring a net number
$k$ of electrons in time $t$ from lead $A$ to the system.
Similar FTs have been derived in Refs. \cite{NazarovTobiska05,
GaspardAndrieuxMeso,EspositoHarbola07,SaitoUtsumi07}.

\subsubsection{Boson transport}\label{boson-transport}

We consider a single oscillator mode at frequency $\epsilon_0/\hbar$
$\hat{H}_S = \epsilon_0 \hat{a}^\dag_0 \hat{a}_0$
coupled to two baths $X=A,B$ at different temperatures
$\beta_A^{-1}$ and $\beta_B^{-1}$ ($k_B=1$) that consist in a
collection of noninteracting bosons (e.g. phonons) $\hat{H}_R = \hat{H}_A+\hat{H}_B$,
where $\hat{H}_X = \sum_{i\in X} \epsilon_i \hat{a}^\dag_i \hat{a}_i$.
The coupling is taken of the form $\hat{V} = \hat{V}_A + \hat{V}_B$, where
$\hat{V}_X = \sum_{i\in X} J_{i0}^X(\hat{a}_0 + \hat{a}^\dag_0)(\hat{a}_i^\dag + \hat{a}_i)$.
The subscript $0$ denotes the system oscillator and  $i$ is
for the $i$'th oscillator in the bath.
$J_{i0}^X$ is the coupling between the system and the
$i$'th bath oscillator from $X$.
All operators satisfy the boson commutation relations
\begin{eqnarray}
\label{comm-boson}
\hat{a}_s \hat{a}_{s^\prime}^\dag-\hat{a}_{s^\prime}^\dag \hat{a}_s &=& \delta_{ss^\prime},\nonumber\\
\hat{a}_s^\dag \hat{a}_{s^\prime}^\dag - \hat{a}_{s^\prime}^\dag \hat{a}_s^\dag &=& \hat{a}_s
\hat{a}_{s^\prime}-\hat{a}_{s^\prime}\hat{a}_s = 0.
\end{eqnarray}
The system eigenstates have an energy $N_S \epsilon_0$ where $N_S=1,2,\cdots $.
We are interested in the statistics of the energy transfers between the
system and the $A$ reservoir, so that the two energy measurements
are performed on system $A$.
It can be shown that performing the RWA on the GQME is equivalent to
assume from the beginning that the coupling term is of the simplified form
$\hat{V}_X = \sum_{i\in X} J_{i0}^X(\hat{a}^\dag_i\hat{a}_0+ \hat{a}_0^\dag \hat{a}_i)$.
We thus have
\begin{eqnarray}
\label{7april-2}
\hat{V}_\lambda &=& \mbox{e}^{\frac{\ic}{2}\lambda \hat{H}_A}
\left(\hat{V}_A+\hat{V}_B \right)\mbox{e}^{-\frac{\ic}{2}\lambda \hat{H}_A}\nonumber\\
&=& \hat{J}_{A}(\lambda) +\hat{J}_A^\dag(\lambda)+\hat{V}_B \;,
\end{eqnarray}
where
\begin{eqnarray}
\label{7april-3}
\hat{J}_A(\lambda) = \sum_{i} J_{i0}^A \hat{a}_0^\dag \hat{a}_i \mbox{e}^{\frac{i}{2}\lambda\epsilon_i}.
\end{eqnarray}
We have used $\hat{a}_i \hat{H}_A = (\epsilon_i+\hat{H}_A) \hat{a}_i$.
Note that unlike fermions, Eq. (\ref{4april-1}), in this case we have a factor
$\epsilon_i$ in the exponential in the coupling, because we now measure energy.
However, in the present model the energy change is directly proportional to
particle change, i.e. their statistics is the same.

Substituting Eq. (\ref{7april-2}) in Eq. (\ref{GFevolofRedGen})
, we get
\begin{eqnarray}
\label{7april-4}
\dot{\hat{\rho}}_S(\lambda,t) &=& -\frac{\ic}{\hbar}[\hat{H}_S,\hat{\rho}_S(\lambda,t)] \\
&&-\alpha_d \hat{a}_0\hat{a}_0^\dag \hat{\rho}_S(\lambda,t)-\alpha_u \hat{a}_0^\dag \hat{a}_0 \hat{\rho}_S(\lambda,t)\nonumber\\
&&+(\alpha_u^A\mbox{e}^{\ic\lambda\epsilon_0}+\alpha_u^B)\hat{a}_0\hat{\rho}_S(\lambda,t)\hat{a}_0^\dag \nonumber\\
&&+(\alpha_d^A\mbox{e}^{-\ic\lambda\epsilon_0}+\alpha_d^B)\hat{a}_0^\dag\hat{\rho}_S(\lambda,t)\hat{a}_0 \nonumber ,
\end{eqnarray}
where the rates $\alpha_u$ and $\alpha_d$ correspond to the
"up" and "down" jumps between the system states
\begin{eqnarray}
\label{7april-5}
\alpha_u^X &=& \frac{2\pi\sigma}{\hbar^2}|J_{0}^X|^2 (1+n_X(\epsilon_0))\nonumber\\
\alpha_d^X &=& \frac{2\pi\sigma}{\hbar^2}|J_{0}^X|^2 n_X(\epsilon_0).
\end{eqnarray}
$n_X(\epsilon_0)= [\mbox{e}^{\beta_X \epsilon_0}-1]^{-1}$ is the Bose
distribution function and $\alpha_{d(u)}=\alpha_{d(u)}^A+\alpha_{d(u)}^B$.
The rates satisfy,
\begin{eqnarray}
\label{7april-6}
\frac{\alpha_u^A\alpha_d^B}{\alpha_d^A\alpha_u^B} = \mbox{e}^{\epsilon_0(\beta_A-\beta_B)}.
\end{eqnarray}
For $\lambda=0$, (\ref{7april-4}) is the Lindblad form QME
derived in \cite{Nitzan05,HarbolaEspositoBoson07}.
In the system eigenbasis $\{\ket{N_S}\}$, Eq. (\ref{7april-4}) describes
a populations dynamics which follows the equation
\begin{eqnarray}
\label{13april-1}
\dot{\rho}_{N_S}(\lambda,t) &=&
\left( \alpha_u^A \mbox{e}^{\ic\lambda\epsilon_0}+\alpha_u^B \right)
(N_{S}+1) \rho_{N_S+1}(\lambda,t) \nonumber\\
&& - \left\{\alpha_d (N_{S}+1) +\alpha_u N_{S} \right\} \rho_{N_S}(\lambda,t) \nonumber\\
&&+\left( \alpha_d^A \mbox{e}^{-\ic\lambda\epsilon_0}+\alpha_d^B \right)
N_{S} \rho_{N_S-1}(\lambda,t) ,
\end{eqnarray}
where $\rho_{N_S}(\lambda,t) \equiv \bra{N_S} \hat{\rho}_{S}(\lambda,t) \ket{N_S}$.
Like Eq. (\ref{22april-2}) , (\ref{13april-1}) may also be recast into a matrix form.
However, unlike fermions, in this case since the matrix is infinite. $\tilde{\rho}$ is
an infinite dimensional vector and $\Gamma(\lambda)$ is a tridiagonal infinite dimensional matrix.
The determinant of a tridiagonal matrix can be expressed as a sum of terms
where the nondiagonal terms always appear in pair with its
symmetric nondiagonal term with respect to the diagonal.
With the help of Eq. (\ref{7april-6}), this pair is symmetric with respect
to $\lambda \to -\ic \epsilon_0 (\beta_A-\beta_B) - \lambda$, so that
$\mbox{det}\{\Gamma(\lambda)\}=\mbox{det}\{\Gamma(-\ic\epsilon_0 (\beta_A-\beta_B)-\lambda)\}$.
This implies that the eigenvalues have the same symmetry and therefore that
the following steady-state FT hold
\begin{eqnarray}
\lim_{t\to\infty}\frac{p(k,t)}{p(-k,t)} =
e^{\epsilon_0(\beta_A-\beta_B) k} \label{7may-1}.
\end{eqnarray}
$p(k,t)$ is the probability that a net number of bosons are
transferred from the reservoir $A$ to the system in a time $t$.
Similar FTs have been derived in Refs.
\cite{HarbolaEspositoBoson07,Maes07,SaitoDhar07}.
The transport statistics of bosons and fermions is different
and was compared in Ref. \cite{HarbolaEspositoBoson07}.
However, both satisfy the same type of FT [(\ref{4april-7}) and (\ref{7may-1})].

\subsubsection{Modulated-tunneling} \label{ModTunn}

In the above, fermion and bosons are transferred from one
lead to another by charging or discharging an embedded system.
We now consider electron tunneling between two coupled
leads, where the tunneling elements are modulated by
the state of an embedded system.
Contrary to the model of section \ref{fermion-transport}, the system never
gets charged, however it affects the electron tunneling between the leads.
This can happen for example if an impurity at the leads interface
interacts with the spin of the tunneling electrons.
The effect of this interaction is to modulate the
tunneling elements between the two leads.
This model of electron transfer was proposed in Ref. \cite{Rammer04}.
Here, we treat this model using the GQME approach.

The Hamiltonian of the junction is of the form (\ref{Hamiltonian}),
where $\hat{H}_S$ is the system Hamiltonian and $\hat{H}_R=\hat{H}_A+\hat{H}_B$
with $\hat{H}_X = \sum_{i\in X}\epsilon_i\hat{c}_i^\dag \hat{c}_i$ ($X=A,B$)
are the two leads Hamiltonian.
The coupling between the two leads is of the form $\hat{V} = \hat{J}+\hat{J}^\dag$,
where $\hat{J}= \sum_{i\in A,j\in B} \hat{J}_{ij} \hat{c}_i^\dag \hat{c}_j$.
The tunneling elements between the leads $\hat{J}_{ij}^\dag=\hat{J}_{ji}$
are now operators in the system space.
We measure the number of particles in the lead $A$. We then have
\begin{eqnarray}
\label{8april-4}
\hat{V}_\lambda = \mbox{e}^{\frac{\ic}{2}\lambda \hat{N}_A} \hat{V} \mbox{e}^{-\frac{\ic}{2}\lambda \hat{N}_A}
= \mbox{e}^{\frac{\ic}{2}\lambda}\hat{J} +  \mbox{e}^{-\frac{\ic}{2}\lambda}\hat{J}^\dag.
\end{eqnarray}
Substituting this in Eq. (\ref{GFevolofRedGen}), we obtain
\begin{eqnarray}
\label{8april-5}
&&\hspace{-0.3cm}\dot{\hat{\rho}}_S (\lambda,t)
= -\frac{\ic}{\hbar}[\hat{H}_S,\hat{\rho}_S(\lambda,t)]- \sum_{i\in A, j\in B } \\
&&\left[\; f_A(\epsilon_i)(1-f_B(\epsilon_j)) \hat{J}_{ij}\{\hat{J}_{ij}^\dag(t)\}
\hat{\rho}_S(\lambda,t) +h.c.\right.\nonumber\\
&&+ \left. f_B(\epsilon_j)(1-f_A(\epsilon_i)) \hat{J}_{ij}^\dag\{\hat{J}_{ij}(t)\}
\hat{\rho}_S(\lambda,t)+h.c.\right.\nonumber\\
&&- \left. f_B(\epsilon_j)(1-f_A(\epsilon_i))\mbox{e}^{\ic \lambda}
\left(\hat{J}_{ij}\hat{\rho}_S(\lambda,t)\{\hat{J}_{ij}^\dag(t)\}+h.c.\right).\right.\nonumber\\
&&-\left. f_A(\epsilon_i)(1-f_B(\epsilon_j))\mbox{e}^{-\ic \lambda}
\left(\{\hat{J}_{ij}^\dag(t)\}\hat{\rho}_S(\lambda,t)\hat{J}_{ij}+h.c.\right)
\right] \nonumber
\end{eqnarray}
where
\begin{eqnarray}
\label{8april-6}
\{\hat{J}_{ij}(t)\} = \frac{1}{\hbar^2}\int_0^t d\tau \mbox{e}^{\ic \epsilon_{ij}\tau}
\mbox{e}^{-\ic \hat{H}_S\tau}\hat{J}_{ij}\mbox{e}^{\ic \hat{H}_S\tau}.
\end{eqnarray}
For $\lambda=0$, Eq. (\ref{8april-5}) reduces to a Redfield
equation for the reduced density-matrix of the system.
A QME for the charge specific reduced density-matrix of the system was derived in Ref. \cite{Rammer04}.
(\ref{8april-5}) is the evolution equation for the GF associated to it.

When applying the Markovian approximation and the RWA to (\ref{8april-5}) in the system
eigenbasis $\{\ket{s}\}$, the populations
$\rho_{ss}(\lambda,t)=\bra{s}\hat{\rho}_S(\lambda,t) \ket{s}$
evolve independently from the exponentially damped coherences according to
\begin{eqnarray}
&&\dot{\rho}_{ss}(\lambda,t) \label{13april-2}\\
&&\hspace{0.3cm}= \sum_{s'} \big( \Gamma_{s's}(\lambda) \rho_{s's'}(\lambda,t)
- \Gamma_{s's}(\lambda=0) \rho_{ss}(\lambda,t) \big) .\nonumber
\end{eqnarray}
The rates are given by
\begin{eqnarray}
\label{23april-4}
\Gamma_{ss'}(\lambda) = \e^{-\ic \lambda} \alpha_{s's}
+ \e^{\ic \lambda} e^{\beta (E_{ss'}-e V)}  \alpha_{ss'}
\end{eqnarray}
where
\begin{eqnarray}
&&\hspace{-0.2cm}\alpha_{ss'} \label{13april-3}\\
&&\hspace{0cm}=\frac{2\pi}{\hbar^2} \sum_{ij} f_A(\epsilon_i)(1-f_B(\epsilon_j))
|\langle s|\hat{J}_{ij}|s'\rangle|^2 \delta(\epsilon_{ij}-E_{ss'}) \;.\nonumber
\end{eqnarray}
They satisfy the symmetry
\begin{eqnarray}
\label{23april-5}
\Gamma_{s's}(-\lambda-\ic \beta eV) = \e^{\beta E_{s's}} \Gamma_{ss'}(\lambda).
\end{eqnarray}
We define $\Gamma(\lambda)$ as the matrix generating the dynamics (\ref{13april-2}).
Using Leibniz formula, the determinant reads
\begin{eqnarray}
\label{23april-6}
\mbox{det} \{ \Gamma(\lambda) \}=
\sum_{\sigma}\mbox{sgn}(\sigma) \prod_{s}^N \Gamma_{s\sigma(s)}(\lambda),
\end{eqnarray}
where $N$ is the order of matrix $\Gamma$ and the sum is computed
over all permutations $\sigma$ of the numbers $\{1, 2, ... , N\}$.
$\mbox{sgn}(\sigma)$ denotes the sign of the permutation, $\mbox{sgn}(\sigma) = + 1$
if $\sigma$ is an even permutation and $\mbox{sgn}(\sigma) = -1$ if it is odd.
Using Eq. (\ref{23april-5}), it can be shown that
\begin{eqnarray}
\label{23april-6b}
\mbox{det}\{\Gamma(\lambda)\}&=&
\sum_{\sigma} \mbox{sgn}(\sigma) \prod_{s=1}^N \mbox{e}^{\beta E_{s\sigma(s)}}
\Gamma_{\sigma(s)s}(-\lambda-\ic \beta eV)\nonumber\\
&=& \sum_{\sigma}\mbox{sgn}(\sigma) \prod_{s=1}^N \Gamma_{\sigma(s)s}(-\lambda-\ic \beta eV)\nonumber\\
&=& \sum_{\sigma}\mbox{sgn}(\sigma) \prod_{s=1}^N \Gamma_{s\sigma(s)}(-\lambda-\ic \beta eV)\nonumber\\
&=& \mbox{det} \{\Gamma(-\lambda-\ic \beta eV)\}.
\end{eqnarray}
In going from first to second line, we used the fact that $\prod_{s=1}^N \mbox{e}^{\beta E_{s\sigma(s)}}=1$
due to $\sum_{s=1}^N E_{s\sigma(s)}=0$. This property follows from the bijective nature of
permutations which implies that for a given $E_{s\sigma(s)}$ in the sum such that $\sigma(s)=s'$,
there will always be a $E_{s' \sigma(s')}$ in the sum that cancels the $E_{s'}$.
Since the eigenvalues of $\Gamma(\lambda)$ satisfy the same symmetry property
as the determinant, we get the same steady-state FT as (\ref{4april-7}),
where $p(k,t)$ is the probability for a net number $k$ of electron transfer
from the lead $A$ to the lead $B$.
This shows that the FT (\ref{4april-7}) is not model-specific
but rather a generic property of nonequilibrium distribution
of electron transfers between two leads.

\subsubsection{Direct-tunneling limit} \label{IsolTunnelJunct}

When the system is decoupled from the junction, the tunneling elements
between the two leads are given by $\hat{J}_{ij}=J_{ij} \hat{1}$.
Using the Markov approximation, $t\to \infty$ in Eq. (\ref{8april-6}), we get
\begin{eqnarray}
\label{8april-7}
\{\hat{J}_{ij}\}=\frac{J_{ij} \hat{1}}{\hbar^2}\left(\pi \delta(\epsilon_i-\epsilon_j)
-\ic \mbox{P}\frac{1}{\epsilon_i-\epsilon_j}\right)
\end{eqnarray}
where $\mbox{P}\frac{1}{x}$ is the principal part of $x$ which we shall neglect.
Under these approximations, it is possible to obtain the explicit form
of the GF for the particle transfer statistics between the two leads.
Substituting Eq. (\ref{8april-7}) in (\ref{8april-5}) and tracing over
system degrees of freedom [Eq. (\ref{GFtottotastrace})], we obtain
\begin{eqnarray}
\label{8april-8}
&&\dot{G}(\lambda,t) = \frac{2\pi}{\hbar^2}\sum_{ij} |J_{ij}|^2 \delta(\epsilon_i-\epsilon_j) \\
&&\hspace{0.3cm}\left[\frac{}{} \big\{f_A(\epsilon_i)+f_B(\epsilon_j)-f_A(\epsilon_i)f_B(\epsilon_j)\big\}
(\mbox{cos}\lambda-1)  \right.\nonumber\\ &&\hspace{2.3cm}+\left.\ic \big\{f_B(\epsilon_i)
-f_A(\epsilon_j) \big\}\mbox{sin}\lambda \frac{}{} \right] G(\lambda,t).\nonumber
\end{eqnarray}
The solution of this equation with the initial condition $G(\lambda,0)=1$ is
\begin{eqnarray}
\label{9april-1}
G(\lambda,t)= \mbox{exp}\left\{t\mu_1(\mbox{e}^{\ic \lambda}-1)
+ t\mu_2(\mbox{e}^{-\ic \lambda}-1)\right\},
\end{eqnarray}
where
\begin{eqnarray}
\label{9april-2}
&&\hspace{-0.5cm}\mu_1= \frac{2\pi}{\hbar^2}\sum_{ij} |J_{ij}|^2\delta(\epsilon_i-\epsilon_j)
f_B(\epsilon_j)\{1-f_A(\epsilon_i)\}\nonumber\\
&&\hspace{-0.5cm}\mu_2= \frac{2\pi}{\hbar^2}\sum_{ij} |J_{ij}|^2\delta(\epsilon_i-\epsilon_j)
f_A(\epsilon_i) \{1-f_B(\epsilon_j)\}.
\end{eqnarray}
We show in appendix \ref{bidirpoisson}, that the probability distribution
associated to the GF (\ref{9april-1}) is a bidirectional Poisson process:
the difference of two Poisson processes with moments $\mu_1$ and $\mu_2$.
Since the moments $\mu_1$ and $\mu_2$ satisfy $\mu_1=\mbox{e}^{-\beta eV}\mu_2$,
the GF has the symmetry [see appendix \ref{bidirpoisson}]
\begin{eqnarray}
\label{9april-4}
G(\lambda,t)= G(-\lambda-\ic \beta eV,t).
\end{eqnarray}
This immediately implies the FT
\begin{eqnarray}
\label{9april-3}
\frac{p(k,t)}{p(-k,t)} = \e^{\beta eVk}
\end{eqnarray}
which is satisfied at all times (transient FT) unlike (\ref{4april-7})
which only hold at long times (steady-state FT).
The entire distribution $p(k,t)$ is calculated in appendix \ref{bidirpoisson}.

\section{Many-body approach to particle counting statistics} \label{GreenFun}

In previous sections, we formulated the counting statistics using a kinetic equation approach.
This simple and intuitive approach makes some key assumptions.
It assumes an initially factorized density matrix of the interacting
systems so that initial Fock space coherences are ignored.
Moreover, the approach is valid only in the weak coupling
limit and it is not obvious how to include many-body
interactions such as electron-electron and electron-phonon.
In this section we present a formulation of counting statistics
based on superoperator non-equilibrium Green's functions (SNGF) \cite{HarbolaPhysRep08}
which allows to relax these approximations.

\subsection{Liouville space formulation of particle counting statistics}
\label{GeneralFormulation}

We consider particle transfer between two coupled
systems $A$ and $B$ described by the Hamiltonian
\begin{eqnarray}
\label{tunneling-hamil-1}
\hat{H}=\hat{H}_A + \hat{H}_B + \hat{V},
\end{eqnarray}
where the coupling reads $\hat{V} = \hat{J} + \hat{J}^\dag$.
By choosing suitable form for $\hat{J}$, we can recover the
different models studied in section (\ref{ManyBodyapplic}).
For the present discussion, we do not need to specify the explicit form of $\hat{J}$.

The measurement of the net number of particles transferred from $A$ to $B$ is
performed using a two-point measurement as described in Sec. \ref{2pointFT}.
Here the measured observable is the number of particles in $A$.
A measurement is done at time $t=0$.If right before this measurement
the system is described by a density
matrix $\sket{\rho(0)}$, the measurement destroys all Fock
space coherences and immediately after the measurement the
density-matrix becomes diagonal in the Fock basis.
A second measurement is performed at time $t$.
A difference of the two measurements gives the net number of particles
transferred between $A$ and $B$.
However if the particle transfer between $A$ and $B$ occurs
though an embedded system, the two-point measurement of particle numbers in $A$ measures the net
particle transfer between $A$ and the embedded system rather than between $A$ and $B$.

It will be convenient to work with superoperators in Liouville space
\cite{fano,reuven,MukamelB,up-shaul-JCP,HarbolaPhysRep08}.
These are defined in Appendix \ref{superoperator}.
We shall denote Liouville space-superoperators by a breve
and Hilbert space operators by a hat.
$\breve{H}_\alpha$, $\breve{V}_\alpha$ and $\breve{H}_{0\alpha}$,
where $\alpha=L,R$, are the left and right superoperators corresponding
to $\hat{H}$, $\hat{V}$ and $\hat{H}_0=\hat{H}_A+\hat{H}_B$.
The probability of the net transfer of $k$ electrons from
$A$ to $B$ during the time interval $t$ is [see Eq. (\ref{prob-1})]
\begin{eqnarray}
\label{prob-dist-10} p(k,t)= \sum_n \sbra{I}\shat{P}_{n-k}
\shat{U}(t,0) \shat{P}_n \sket{\rho(0)},
\end{eqnarray}
where $\shat{U}(t,0)= \mbox{e}^{-i \sqrt{2}\shat{H}_-t}$
is the time evolution operator in Liouville space
and $\shat{{P}}_{n}$ is the projection operator associated
with the measurement of $n$ electrons in $A$.
$\sket{\rho(0)}$ is the interacting density matrix when the counting
starts and contains coherences in the number operator basis.
It is constructed by switching on the interaction $\hat{V}$ from the
infinite past, where the density-matrix $\sket{\rho(-\infty)}$ is given by
a direct product of the density-matrices of systems $A$ and $B$, to $t=0$.
\begin{eqnarray}
\label{adiabatic}
\sket{\rho(0)}= \shat{U}_I(0,-\infty)\sket{\rho(-\infty)},
\end{eqnarray}
where
\begin{eqnarray}
\label{uI}
\shat{U}_I(0,-\infty)=
\mbox{exp}_+\left\{-\frac{\ic}{\hbar}\int_{-\infty}^{0}d\tau
\sqrt{2} \shat{V}_-(\tau) \right\}
\end{eqnarray}
with $\sqrt{2}\shat{V}_-(\tau)=\shat{ V}_L(\tau)-\shat{V}_R(\tau)$
[see Eq. (\ref{trans})] and
\begin{eqnarray}
\label{interaction} \shat{V}_\alpha(\tau) = \shat{U}_0^\dag(\tau,0)
\shat{V}_\alpha \shat{U}_0(\tau,0),~~~~\alpha=L,R
\end{eqnarray}
where
\begin{eqnarray}
\label{u0}
\shat{U}_0(\tau,0)=\theta(\tau)e^{-\frac{\ic}{\hbar}\sqrt{2}\shat{H}_{0-}\tau}.
\end{eqnarray}
The GF associated to $p(k,t)$ is defined by
\begin{eqnarray}
\label{gf-1}
G(\lambda,t) = \sum_k \mbox{e}^{\ic\lambda k} p(k,t).
\end{eqnarray}
Substituting Eq. (\ref{gf-1}) in Eq. (\ref{prob-dist-10}),
we get (see Appendix \ref{probdistrib})
\begin{eqnarray}
\label{gf-y} G(\lambda,t)=\int_{0}^{2\pi}\frac{d\Lambda}{(2\pi)^3}
G(\lambda,\Lambda,t)
\end{eqnarray}
with
\begin{eqnarray}
\label{chi-y}
&&G(\lambda,\Lambda,t)=\nonumber\\
&&\sbra{I}\mbox{e}^{-\frac{\ic}{\hbar}\sqrt{2}\breve{H}_{0-}t}
\mbox{e}_+^{- \frac{\ic}{\hbar} \int_{-\infty}^{t} d\tau \sqrt{2}
\shat{{V}}_{-}(\gamma(\tau),\tau)} \sket{\rho(-\infty)},\nonumber\\
\end{eqnarray}
where $\shat{{V}}_{-}(\gamma(t))=\shat{V}_{L}(\gamma_L(t))-\shat{{V}}_{R}(\gamma_R(t))$
with $\gamma_L(t)=\theta(t)(\Lambda+\lambda/2)$ and
$\gamma_R(t)=\theta(t)(\Lambda-\lambda/2)$.
The GF (\ref{chi-y}) includes the initial $t=0$
correlations between systems $A$ and $B$ in the density matrix.
These correlations are built through the
switching of the coupling $\hat{V}$ from $t=-\infty$ and $t=0$.
In the absence of such correlations, the initial density matrix is
diagonal in the number basis and $G(\lambda,t)=G(\lambda,\Lambda=0,t)$
[i.e. $\rho(0)$ commutes with $\hat{N}_A$].
Below, we show how $G(\lambda,\Lambda,t)$ can be computed.

\subsection{Electron counting statistics for direct-tunneling between two systems}
\label{GreenDirTunn}

We next apply Eq. (\ref{chi-y}) to calculate the electron current statistics
for the direct tunneling model of section (\ref{IsolTunnelJunct}).
The Hamiltonian is given by (\ref{tunneling-hamil-1}), where
\begin{eqnarray}
\label{tunnel-hamil-3}
\hat{J}= \sum_{i\in A,j\in B} J_{ij} \hat{c}_{i}^{\dagger} \hat{c}_{j},
\end{eqnarray}
where $J_{ij}^{*}=J_{ji}$.
Hamiltonian $\hat{H}_A$ and $\hat{H}_B$ are general and can include many-body interactions.
The exact form for $\hat{H}_A$ and $\hat{H}_B$ is not necessary in the present discussion. A noninteracting electron model, as studied in Sec. (\ref{IsolTunnelJunct}),
will be considered in the next subsection.

We now define the superoperators $\shat{J}$, $\shat{J}^\dag$ and $\shat{N}$
corresponding to the operators $\hat{J}$, $\hat{J}^\dag$ and the number
operator $N_A$ for the system $A$.
These satisfy commutation relations
\begin{eqnarray}
&&[\shat{{ J}}_{L}, \shat{{ N}}_{L}]  = - \shat{{J}}_{L} \ \ \;,
\ \ [\shat{{ J}}^\dag_{L}, \shat{{ N}}_{L}] = \shat{{J}}^\dag_{L} \\
&&[\shat{{J}}_{R}, \shat{ N}_{R}]  = \shat{J}_{R} \ \ \;,
\ \ [\shat{J}^\dag_{R}, \shat{ N}_{R}] = - \shat{J}^\dag_{R} \nonumber\;.
\end{eqnarray}
Using these commutation relations in Eq. (\ref{new-1}), we can write
\begin{eqnarray}
\label{mm-1}
\shat{V}_\alpha(\gamma_\alpha(t)) = \exp{\{-\ic \gamma_\alpha(t) \}} \shat{J}_{\alpha}
+ \exp{\{\ic \gamma_\alpha(t) \}} \shat{J}_{\alpha}^{\dagger} \;.
\end{eqnarray}

We define
\begin{eqnarray}
\label{z-llt}
{\cal Z}(\lambda,\Lambda,t)\equiv\mbox{ln}G(\lambda,\Lambda,t).
\end{eqnarray}
Expanding the time-ordered exponential in (\ref{chi-y}) we can compute the
GF and the cumulant GF perturbatively in the coupling $J_{ab}$. Since $\sbra{I} \shat{V}_{-} \sket{\rho(-\infty)}=0$, to second order we obtain
\begin{eqnarray}
&&{\cal Z}(\lambda,\Lambda,t) =-\frac{1}{2\hbar^2}  \int_{-\infty}^{t}d\tau_1
\int_{-\infty}^{t}d\tau_2 \label{mm-2}\\
&&\sbra{I} \shat{{\cal T}} \shat{V}_{-}(\gamma(\tau_1),\tau_1)
\shat{V}_{-}(\gamma(\tau_2),\tau_2) \sket{\rho(-\infty)} \nonumber \;.
\end{eqnarray}
Substituting Eq. (\ref{mm-1}) in (\ref{mm-2}) we get
\begin{eqnarray}
\label{mm-3}
{\cal Z}(\lambda,\Lambda,t)
= {\cal Z}^{(0)}(\lambda,t) + {\cal Z}^{(1)}(\lambda,\Lambda,t)
\end{eqnarray}
where
\begin{eqnarray}
\hspace{-1cm}{\cal Z}^{(1)}(\lambda,\Lambda,t)= 2 (\e^{\ic \lambda/2}
-\e^{-\ic \lambda/2}) \mbox{Re} \{ \e^{\ic \Lambda} W(t) \} \label{nmm1b}\\
\end{eqnarray}
and
\begin{eqnarray}
\hspace{-1cm}{\cal Z}^{(0)}(\lambda,t) = (\e^{-\ic \lambda}-1)W^{(0)}_{BA}(t)
+ (\e^{\ic \lambda}-1) W^{(0)}_{AB}(t) \label{nmm1}
\end{eqnarray}
are the contributions coming from time evolution from $t=-\infty$ to $t=0$ and from $t=0$ to time $t$, respectively, and
$W(t)\equiv W^{(1)}_{BA}(t)-W^{(1)}_{AB}(t)$ with
\begin{eqnarray}
W^{(0)}_{AB}(t)&=&\frac{1}{\hbar^2}\int_{0}^{t}dt_1 \int_{0}^{t}dt_2
\sbra{I} \shat{J}_R(t_1) \shat{J}_L^{\dagger}(t_2) \sket{\rho(-\infty)} \nonumber\\
W^{(0)}_{BA}(t)&=&\frac{1}{\hbar^2}\int_{0}^{t}dt_1 \int_{0}^{t}dt_2
\sbra{I} \shat{J}_L(t_1) \shat{J}_R^{\dagger}(t_2) \sket{\rho(-\infty) }\nonumber\\
W^{(1)}_{AB}(t)&=&\frac{1}{\hbar^2}\int_{-\infty}^{0}dt_1 \int_{0}^{t}dt_2
\sbra{I} \shat{J}_R(t_1) \shat{J}_L^{\dagger}(t_2) \sket{\rho(-\infty)} \nonumber\\
W^{(1)}_{BA}(t)&=&\frac{1}{\hbar^2}\int_{-\infty}^{0}dt_1 \int_{0}^{t}dt_2
\sbra{I} \shat{J}_L(t_1) \shat{J}_R^{\dagger}(t_2) \sket{\rho(-\infty)} \nonumber. \\
\label{rates}
\end{eqnarray}

From (\ref{z-llt}) and (\ref{mm-3}) we get
\begin{eqnarray}
\label{10sep-1}
G(\lambda,\Lambda,t) = \mbox{e}^{{\cal Z}^{(0)}(\lambda,t)} \mbox{e}^{{\cal Z}^{(1)}(\lambda,\Lambda,t)}.
\end{eqnarray}
Substituting this in (\ref{gf-y}), the GF is obtained as
\begin{eqnarray}
G(\lambda,t) &=& G^{(0)}(\lambda,t) G^{(1)}(\lambda,t)\label{25}
\end{eqnarray}
where
\begin{eqnarray}
G^{(0)}(\lambda,t) &=& \exp{\{ {\cal Z}^{(0)}(\lambda,t)\}} \label{25a}\\
G^{(1)}(\lambda,t) &=& \int_{0}^{2\pi}\frac{d\Lambda}{2\pi}
\exp{\{ {\cal Z}^{(1)}(\lambda,\Lambda,t) \}} \label{25b}
\end{eqnarray}

The cumulant GF is finally obtained as
\begin{eqnarray}
{\cal Z}(\lambda,t)= {\cal Z}^{(0)}(\lambda,t)
+\mbox{ln}\int_0^{2\pi}\frac{d\Lambda}{2\pi}
\exp{\{ {\cal Z}^{(1)}(\lambda,\Lambda,t)\}} \nonumber. \\
\label{c-gf-11}
\end{eqnarray}
The second term on the rhs of Eq. (\ref{c-gf-11}) is the contribution
due to the initial correlations that exist between systems $A$ and $B$
right before the first measurement.
When these initial correlations are ignored, i.e. initial density
matrix is a direct product of the density matrix of $A$ and $B$
(or equivalently $[\hat{N}_A,\hat{\rho}(0)]$=0), ${\cal Z}^{(1)}=0$.

\subsubsection{Effects of initial correlations} \label{inicorrel}

Here we discuss the corrections to the electron statistics due to
correlations between $A$ and $B$ in the initial density matrix.
We show that these contributions do not affect the
first moment (the current) but only higher moments.\\

Using (\ref{nmm1b}) and expanding in $\lambda$, we find that
\begin{eqnarray}
\label{chi-1}
&&\hspace{-0.7cm}\exp{\{ {\cal Z}^{(1)}(\lambda,\Lambda,t) \}}\nonumber\\
&&= \sum_{n=0}^{\infty}\frac{(2i)^n}{n!}
\sin^n \left(\frac{\lambda}{2}\right)
\left[\e^{-i\Lambda} W(t)+\e^{i\Lambda}W^{*}(t)\right]^n \nonumber\\
&&= \sum_{n(\geq k),k=0}^{\infty}
\frac{(2i)^n}{k!(n-k)!}\sin^n \left(\frac{\lambda}{2}\right)\nonumber\\
&&\hspace{2cm} \times W^{n-k}(t) W^{k*}(t) \e^{-i \Lambda(n-2k)}.
\end{eqnarray}
Integrating over $\Lambda$, (\ref{25b}) becomes
\begin{eqnarray}
\label{chi-1-final}
G^{(1)}(\lambda,t)=1+\sum_{n=1}^\infty \frac{(-4)^n}{(n!)^2}|W(t)|^{2n}
\sin^{2n}\left(\frac{\lambda}{2}\right) \;.
\end{eqnarray}
By differentiating (\ref{25}) with respect to $\lambda$, we can factorize the
moments in two parts, $\langle k^n\rangle_0$ which does not depend on
the initial correlations and $\Delta^{(n)}$ which does:
\begin{eqnarray}
\langle k^n(t) \rangle = \langle k^n(t) \rangle_0 + \Delta^{(n)}(t),
\end{eqnarray}
where
\begin{eqnarray}
\label{nnn1}
\langle k^n(t) \rangle_0 &=& (-\ic)^n\left.\frac{\partial^n}{\partial \lambda^n}
G^{(0)}(\lambda,t)\right|_{\lambda=0}\nonumber\\
\Delta^{(n)}(t) &=& \sum_{k=1}^n\sum_{l=1}^{\infty} \sum_{m=0}^{2l}
\ic^{2l-m+k}(-1)^k \;^{2l}C_m \;^nC_k \nonumber\\
&&\hspace{2cm}\times  \langle k^{n-k}(t) \rangle_0 |W(t)|^{2l} .
\end{eqnarray}
$\;^nC_k=n!/(k!(n-k)!)$ are the binomial coefficients.

We find that $\Delta^{(1)}(t)=0$,
i.e. initial correlations do not contribute to first moment, which is the net
current from $A\to B$.
However, they do contribute to higher moments.
The correction to the second moment is
\begin{eqnarray}
\label{correction-2moment}
\Delta^{(2)}(t) = -32|W(t)|^2 .
\end{eqnarray}
We see that initial correlations always tend to decrease the second moment.

\subsubsection{The thermodynamic limit}\label{ThermoLimitRes}

We consider now the limit where $A$ and $B$ can be assumed to have continuous spectra.
We treat them as non-interacting electron leads and show that initial correlations
do not contribute to the long time statistics.

This corresponds to the model discussed in section \ref{IsolTunnelJunct}.
In this limit, the rates $W_{AB}$ and $W_{BA}$ given in Eq. (\ref{rates})
can be calculated explicitly. The Hamiltonian for two systems ($X=A,B$) is
\begin{eqnarray}
\hat{H}_X = \sum_{i\in X}\epsilon_i \hat{c}_i^\dag \hat{c}_i.
\end{eqnarray}
Using the fact that the density-matrix at $t=-\infty$ is a direct product
$\sket{\rho(-\infty)}=\sket{\rho_A^{eq}} \otimes \sket{\rho_B^{eq}}$, we get
\begin{eqnarray}
&&\sbra{I} \shat{J}_R(\tau_1) \shat{J}_L^{\dagger}(\tau_2) \sket{\rho(-\infty)} \\ \nonumber
&&\hspace{1cm}= \sum_{ij} \vert J_{ij} \vert^2  f_A(\epsilon_i) (1-f_B(\epsilon_j))
\e^{\ic \omega_{ij} (\tau_1-\tau_2)} \\
&&\sbra{I} \shat{J}_L(\tau_1) \shat{J}_R^{\dagger}(\tau_2) \sket{\rho(-\infty)} \\ \nonumber
&&\hspace{1cm}= \sum_{ij} \vert J_{ij} \vert^2  f_B(\epsilon_j) (1-f_A(\epsilon_i))
\e^{\ic \omega_{ij} (\tau_1-\tau_2)} \} \;,
\end{eqnarray}
where $\omega_{ij}=\epsilon_i-\epsilon_j$ and
$f_X(\epsilon)=(\exp{\{\beta(\epsilon-\mu_X)\}}+1)^{-1}$ is the Fermi
function for the system $A (B)$ with $\mu_A$ and $\mu_B$ denoting the
chemical potential of systems $A$ and $B$.

Remembering that
\begin{eqnarray}
\int_{0}^{t}d\tau_1 \int_{0}^{t}d\tau_2 \e^{\pm \ic \omega_{ij} (\tau_1-\tau_2)}
&=& \bigg( \frac{\sin (\omega_{ij}t/2)}{\omega_{ij}/2} \bigg)^2 \nonumber\\
&\stackrel{t \to \infty}{=}& 2 \pi \delta(\omega_{ij}) t \nonumber
\end{eqnarray}
and that
\begin{eqnarray}
&&\int_{-\infty}^{0}d\tau_1 \int_{0}^{t} d\tau_2 \e^{\pm \ic \omega_{ij} (\tau_1-\tau_2)} =
- \frac{ \big( \e^{\mp \ic (\omega_{ij} \mp \ic \eta^+) t} -1 \big) }
{\big(\eta^+ \mp \ic \omega_{ij}\big)^2}  \nonumber \;,
\end{eqnarray}
using (\ref{nmm1}) we find  that
\begin{eqnarray}
{\cal Z}^{(0)}(\lambda,t) =(\e^{-\ic \lambda}-1) \mu_2 t + (\e^{\ic \lambda}-1) \mu_1 t \;,
\label{cumulantasympt}
\end{eqnarray}
where $\mu_1$ and $\mu_2$ are given by (\ref{9april-2}).
$G^{(0)}(\lambda)$ is therefore identical to the
GF for a bidirectional Poisson process obtained in (\ref{9april-1}) within the GQME.

The rate $W(t)$ which appears in the expression for ${\cal Z}^{(1)}$ in Eq. (\ref{nmm1b}) reads
\begin{eqnarray}
W(t)&=&\sum_{ij} \vert J_{ij} \vert^2 \nonumber\\
&\times&
\bigg[ (f_A(\omega_j)-f_B(\omega_i)) \frac{\e^{-\ic(\omega_{ij}-\ic \eta^+)t}-1}
{(\omega_{ij}-\ic \eta^+ \big)^2} \bigg]. \label{wt}
\end{eqnarray}
Taking the continuous limit of the leads' density of states, we find that for
long times $W(t)$ becomes time independent \cite{Rammer03}. Therefore
\begin{eqnarray}
{\cal S}(\lambda)=
\lim_{t \to \infty} \frac{1}{t}{\cal Z}(\lambda,t) =
\lim_{t \to \infty} \frac{1}{t}{\cal Z}^{(0)}(\lambda,t) \;,
\end{eqnarray}
which shows that the long time statistics is not affected
by the initial correlations between $A$ and $B$.

\subsection{Electron counting statistics for transport through a quantum junction}
\label{GreenTransportJunction}

We next apply Eq. (\ref{gf-y}) to calculate the current statistics
in the transport model of section \ref{fermion-transport} where
a quantum system (e.g. a molecule, chain of atoms or quantum dots)
is embedded between two much larger systems $A$ and $B$.
Notice that here the two-point measurement of the particle number in
$A$ does not measure the net particle transfer between $A$ and $B$ as
stated in section \ref{GeneralFormulation} but rather the net particle
transfer between $A$ and the embedded system.
The particle transfer statistics for this model was studied
in section \ref{fermion-transport} using the GQME approach.
Here, we express the transfer statistics in terms of
the SNGF \cite{up-shaul-JCP,HarbolaPhysRep08} of the quantum system.
By connecting this powerful many-body formalism with the
two-point measurement, we can study more complicated models.
The effect of eigenbasis coherences in the quantum system
(which requires to go beyond the RWA in the GQME approach)
and the effect of many-body interactions in the quantum system can be
easily incorporated into the SNGF approach via the self-energy matrix.
In presence of many-body interactions, the SNGF theory involves a self-consistent calculation
for the Green's functions together with their self-energies.
This goes beyond the weak coupling limit of the GQME.
The simple form for the lead-system interactions (\ref{fermi-hamil-2b}) allows us to obtain
analytical results for the corresponding self-energy and hence the GF.
Electron-electron interactions will provide an extra (additive)
self-energy matrix computed in Ref. \cite{up-shaul-JCP}.

The Hamiltonian of the model is given by (\ref{fermi-hamil-2a}) and (\ref{fermi-hamil-2b}).
The superoperators $\shat{H}_{0\alpha}$ and $\shat{V}_{\alpha}$ corresponding
to $\hat{H}_0=\hat{H}_A+\hat{H}_B+\hat{H}_S$ and $\hat{V}_{A(B)}$ can be obtained by using
Eqs. (\ref{super-identity}) in  Eqs. (\ref{fermi-hamil-2a}) and (\ref{fermi-hamil-2b}). We get
\begin{eqnarray}
\shat{H}_{0L}&=& \sum_{x\in A,B,S}\epsilon_{x} \breve{c}_{xL}^\dag \breve{c}_{xL} \nonumber\\
\shat{H}_{0R}&=& \sum_{x\in A,B,S}\epsilon_{x} \breve{c}_{xR} \breve{c}_{xR}^\dag
\end{eqnarray}
and
\begin{eqnarray}
\breve{{V}}_\alpha = \breve{J}_{A,\alpha}
+\breve{J}^\dag_{A,\alpha} + \breve{J}_{B,\alpha} +\breve{J}^\dag_{B,\alpha}
\end{eqnarray}
where
\begin{eqnarray}
\shat{J}_{X,L}&=&\sum_{s,i\in X}J_{si}\breve{c}_{iL}^\dag \breve{c}_{sL}\nonumber\\
\shat{J}_{X,R}&=&\sum_{s,i\in X}J_{si} \breve{c}_{sR}\breve{c}_{iR}^\dag .
\end{eqnarray}
The superoperators $\shat{{J}}_{X,L}$ and $\shat{{J}}_{X,L}^\dag$
satisfy the commutation relations \cite{HarbolaPhysRep08}
\begin{eqnarray}
\label{eq-3n} [\shat{{J}}_{A,L}, \shat{{N}}_{L}] &=&
-\shat{{J}}_{A,L};~~ [\shat{{J}}_{A,R}, \shat{{N}}_{R}]  = \shat{{J}}_{A,R}
\end{eqnarray}
\begin{eqnarray}
\label{eq-30} [\shat{{J}}_{A,L}^\dag, \shat{{N}}_{L}]
&=&  \shat{{J}}_{A,L}^\dag;~~ [\shat{{J}}_{A,R}^\dag,
\shat{{N}}_{R}]  = -\shat{{J}}_{A,R}^\dag
\end{eqnarray}
and $[\shat{{N}}_\alpha, \shat{{J}}_{B,\alpha}]=[\shat{{N}}_\alpha, \shat{J}^\dag_{B,\alpha}]=0$.
Using these in (\ref{new-1}), we obtain
\begin{eqnarray}
\label{new-2}
\hat{{V}}_\alpha(\gamma_\alpha(t),t) &=&
\exp{\{-\ic \gamma_\alpha(t) \}} \shat{{J}}_{A,\alpha}(t)\\
&&+ \exp{\{\ic \gamma_\alpha(t) \}} \shat{{J}}_{A,\alpha}^{\dagger}(t)
+ \shat{{J}}_{B,\alpha}+\shat{{J}}_{B,\alpha}^{\dag},\nonumber
\end{eqnarray}
where $\shat{{J}}_{X,\alpha}=\shat{{J}}_{X,\alpha}(\gamma_\alpha=0)$.
Note that in (\ref{new-2}), exponential
factors are associated only with superoperators of the lead $A$.
This is because the measurement (projection) is done only on $A$.

We can now use (\ref{new-2}) in (\ref{chi-y}) to compute the GF.
$\sket{\rho(-\infty)}$ in (\ref{chi-y}) is given by the direct product of equilibrium density-matrices of the system and the leads,
\begin{eqnarray}
\label{rho0}
\sket{\rho(-\infty)}
&=&|\rho_S\rangle\rangle\otimes|\rho_A\rangle\rangle\otimes|\rho_B\rangle\rangle\\
|\rho_x\rangle\rangle &=&\frac{1}{\Xi_x}|e^{-\beta \hat{H}_x-\mu_x\hat{N}_x}\rangle \rangle
\end{eqnarray}
where $\mu_x$ and $\Xi_x$ are respectively the chemical potential and the partition function for system $x$.

Using Grassmann variables and a path-integral formulation, the GF (\ref{gf-y}) can
be expressed in terms of the Green's functions of the quantum system.
In Appendix \ref{sec-path-int} we present a derivation in terms of Liouville space superoperators.
For a Hilbert space derivation see Ref. \cite{kamenev}. 
Some useful properties of Grassmann variables used in
the derivation are summarized in Appendix \ref{Grassmann}.
The final result for the GF, Eq. (\ref{chi-y}), is
\begin{eqnarray}
\label{gf-final-10}
G(\lambda,\Lambda,t) &=& \e^{{\cal Z}(\lambda,\Lambda,t)},
\end{eqnarray}
with
\begin{eqnarray}
{\cal Z}(\lambda,\Lambda,t)=\int_{-\infty}^t d\tau
\mbox{ln}\mbox{Det}\left[g^{-1}(\tau=0)-\Sigma(\tau,\tau,\gamma(\tau))\right], \nonumber\\
\label{gf-final-10-a}
\end{eqnarray}
where $g(t-\tp)$ and $\Sigma(t,\tp)$ are Green's function and self-energy (due to system-lead 
interaction) matrices in $\nu,\nup = +,-$ representation. The Green's function matrix satisfies
\begin{eqnarray}
\left(\ic\hbar\frac{\partial}{\partial t}-\epsilon_s\frac{}{}\right)
g^{\nu\nup}_{s\sp}(t-\tp)=
\delta(t-\tp)\delta_{\nu\nup}\delta_{s\sp}\label{g-sigma-1}
\end{eqnarray}
and the self-energy matrix is
\begin{eqnarray}
&&\hspace{-0.5cm}\Sigma_{s\sp}^{\nu\nup}(t,\tp,\gamma(t),\gamma(\tp))=\nonumber\\
&&\hspace{0.1cm}\sum_{X}\sum_{ii^\prime\in X}J_{si}^{\nu\nu_1}(\gamma(t))
g_{ii^\prime}^{\nu_1\nu_2}(t-\tp)J_{i^\prime\sp}^{\nu_2\nu^\prime}(\gamma(\tp))\label{g-sigma-2}
\end{eqnarray}
where
\begin{eqnarray}
J^{++}_{is}(\gamma)&=& J^{--}_{is}(\gamma)=J_{is}(e^{i\gamma_L}+e^{\ic\gamma_R})/2\nonumber\\
J^{+-}_{is}(\gamma)&=& J^{-+}_{is}(\gamma)=J_{is}(e^{i\gamma_L}-e^{\ic\gamma_R})/2
\end{eqnarray}
for $i\in A$ and
\begin{eqnarray}
J^{++}_{is}(\gamma) &=& J^{--}_{is}(\gamma)=J_{is}\nonumber\\
J^{+-}_{is}(\gamma) &=& J^{-+}_{is}(\gamma)=0
\end{eqnarray}
for $i\in B$.
One important point to note is that while $g^{+-}(t,t^\prime)$
(zero-order system Green's function without interactions with leads)
is causal and $g^{-+}(t,t^\prime)=0$\cite{up-shaul-JCP,HarbolaPhysRep08}, this is no longer the case for $\Sigma^{+-}$ and $\Sigma^{-+}$ which depend on $\gamma$. This is due to the fact
that when $\gamma_L\neq\gamma_R$, the ket and the bra evolve with a
different Hamiltonian.
The cumulant GF is then given by
\begin{eqnarray}
\label{1-1}
{\cal Z}(\lambda,t)=\mbox{ln}\int_0^{2\pi}
\frac{d\Lambda}{2\pi} G(\lambda,\Lambda,t).
\end{eqnarray}
Equation (\ref{1-1}) with (\ref{gf-final-10}) and (\ref{gf-final-10-a})
give the statistics for the net particle transfer between lead $A$ and
the quantum system embedded between $A$ and $B$.

\subsubsection{Long-time statistics}\label{long-time-statistics}

At steady-state all the two-time functions, such as
$g(t,\tp)$ and $\Sigma(t,\tp)$, depend only on the difference of
their time arguments. We factorize time integration in Eq.
(\ref{gf-final-10-a}) in two regions, one from $-\infty$ to $0$
and other from $0$ to $t$. 
Since $\gamma(t)=0$ for negative times, Eq. (\ref{glr-1}), we obtain
\begin{eqnarray}
\label{x-1}
{\cal Z}(\lambda,\Lambda,t)&=&G_0 \\
&+&\int_{0}^t d\tau
\mbox{ln}\mbox{Det}\left[g^{-1}(\tau=0)-\Sigma(\tau,\tau,\gamma(\tau))\right]. \nonumber
\end{eqnarray}
The term $G_0$, which is independent on time and $\gamma$
comes from integration $t=-\infty$ to $t=0$ and contains all initial
correlations between system and the leads.
Substituting for the self-energy (\ref{g-sigma-2}),
we notice that since the matrix elements $J_{is}^{\nu\nu^\prime}$ and  $J_{si}^{\nu\nu^\prime}$
appear at the same time, the $\Lambda$ dependence drops out.
We can recast (\ref{gf-final-10-a}) for long times as
\begin{eqnarray}
\label{gf-final-1} {\cal Z}(\lambda,t) = t\int \frac{d\omega}{2\pi}~ \mbox{ln}\mbox{Det}
\left[g^{-1}(\omega)-\Sigma(\omega,\lambda)\right]+ G_0.
\end{eqnarray}
At long times the first term in (\ref{gf-final-1})
dominates, and the current GF is given solely by the first term in (\ref{gf-final-1}).
\begin{eqnarray}
\label{11-11} {\cal S}(\lambda) &=&
\lim_{t\to\infty}\frac{1}{t} {\cal Z}(\lambda,t)\nonumber\\
&=& \int \frac{d\omega}{2\pi}~ \mbox{ln}\mbox{Det}
\left[g^{-1}(\omega)-\Sigma(\omega,\lambda)\right]\nonumber\\
&\equiv& \int \frac{d\omega}{2\pi}~ \mbox{ln}\mbox{Det}[\chi^{-1}(\omega)].
\end{eqnarray}
Thus, as in Sec. \ref{ThermoLimitRes}, we can conclude that
contributions coming from the initial correlations between the system and the leads
do not effect the long-time statistics.

We shall compute the self energy in frequency domain. Since the leads are
made of non-interacting electrons, their zeroth order Green's
functions in frequency domain are
\begin{eqnarray}
\label{zero-order-gf}
g^{--}_{ii^\prime}(\omega) &=& \frac{\delta_{ii^\prime}}{\hbar\omega-\epsilon_i+\ic\eta},~~
g^{++}_{ii^\prime}(\omega)={[g^{--}]}^\dag_{ii^\prime}(\omega) \nonumber\\
g^{-+}_{ii^\prime}(\omega)&=& -2\pi \ic
\delta_{ii^\prime}(2f_{i}(\omega)-1)\delta(\hbar\omega-\epsilon_i).
\end{eqnarray}
Substituting this in
(\ref{g-sigma-2}), the self-energy matrix in the wide-band
approximation is obtained as
\begin{eqnarray}
\Sigma_{s\sp}^{++}(\omega,\lambda)&=&\frac{\ic}{2}\Gamma^B_{s\sp} \label{sig++}\\
&+&\frac{\ic}{2}\Gamma^A_{s\sp}\left[e^{\ic\lambda}(1-f_A(\omega))+e^{-\ic\lambda}f_A(\omega)
\right], \nonumber\\
\Sigma_{s\sp}^{--}(\omega,\lambda)&=& -\frac{\ic}{2}\Gamma^B_{s\sp} \label{sig--}\\
&&-\frac{\ic}{2}\Gamma^A_{s\sp}\!\left[e^{\ic\lambda}(1-f_A(\omega))
+e^{-\ic\lambda}f_A(\omega) \right], \nonumber\\
\Sigma_{s\sp}^{+-}(\omega,\lambda) &=&
-\frac{\ic}{2}\Gamma^A_{s\sp}\left[(e^{\ic\lambda}-1)(1-f_A(\omega)\right.\label{sig+-}\\
&&\hspace{2cm}-\left.(e^{-\ic\lambda}-1)f_A(\omega)) \right], \nonumber\\
\Sigma_{s\sp}^{-+}(\omega,\lambda) &=&
-\ic\Gamma_{s\sp}^B(2f_B(\omega)-1) +\frac{\ic}{2}\Gamma_{s\sp}^A \label{sig-+}\\
&\times& \left[(e^{\ic\lambda}+1)(1-f_A(\omega))\right. \nonumber\\
&&\hspace{2cm}\left.-(e^{-\ic\lambda}+1)f_A(\omega)\right] \nonumber,
\end{eqnarray}
where $\Gamma_{s\sp}^X=2\pi\sum_{i\in X}J_{is}J_{\sp i}\delta(\omega-\epsilon_i)$.

Note that when $\lambda=0$, $\Sigma^{+-}=0$ as it should be
(causality)\cite{up-shaul-JCP}, and $\Sigma^{--}$, $\Sigma^{++}$ and
$\Sigma^{-+}$ reduce to usual retarded, advanced and correlation
(Keldysh) self-energies, respectively.
\begin{eqnarray}
\label{keldysh-se}
\Sigma^{--}_{s\sp}(\omega) &=&  -\frac{\ic}{2}\Gamma_{s\sp},~~~
\Sigma^{++}_{s\sp}(\omega)= \frac{\ic}{2}\Gamma_{s\sp}\nonumber\\
\Sigma^{-+}_{s\sp}(\omega) &=& \ic\Gamma_{s\sp}
-2(\Gamma^A_{s\sp}f_A(\omega)+\Gamma^B_{s\sp}f_B(\omega))
\end{eqnarray}
where $\Gamma=\Gamma^A+\Gamma^B$.

The retarded Green's functions for the molecule is then given by
\begin{eqnarray}
\label{ret-green}
R^{--}_{s\sp}(\omega)= \left[(\hbar\omega-\epsilon)\hat{1}
-\ic\frac{\Gamma}{2}\right]^{-1}_{s\sp}
\end{eqnarray}
where $\hat{1}$ is the identity matrix and
$R^{++}=[R^{--}]^\dag$ is the advanced Green's function.

Finally, we transform the self-energy matrix from the $+,-$
($\Sigma$) to the $L,R$ ($\tilde{\Sigma}$) representation.
This can be achieved by the matrix transformation,
$\tilde{\Sigma}=Q^{-1}\Sigma Q$ \cite{up-shaul-JCP}, where
\begin{eqnarray}
\label{q-mat} Q=\frac{1}{\sqrt{2}}\left(
\begin{array}{cc}
-1&-1\\
1&-1
\end{array}
\right).
\end{eqnarray}

This gives the matrix $\tilde{\Sigma}_{\alpha\beta}(\omega,\lambda)$ with elements
\begin{eqnarray}
&&\hspace{-0.4cm}\tilde{\Sigma}_{RR}^{s\sp}(\omega)=
\frac{\ic}{2}\Gamma_{s\sp}-\ic(\Gamma_{s\sp}^Af_A(\omega)
+\Gamma^B_{s\sp}f_B(\omega))\label{sigma-1}\\
&&\hspace{-0.4cm}\tilde{\Sigma}_{LL}^{s\sp}(\omega)=
-\frac{\ic}{2}\Gamma_{s\sp}+\ic(\Gamma^A_{s\sp}f_A(\omega)
+\Gamma^B_{s\sp}f_B(\omega))\label{sigma-2}\\
&&\hspace{-0.4cm}\tilde{\Sigma}_{LR}^{s\sp}(\omega,\lambda)=
\ic\Gamma^B_{s\sp}f_B(\omega)
+\ic\Gamma^A_{s\sp}f_A(\omega)e^{-\ic\lambda}\label{sigma-3}\\
&&\hspace{-0.4cm}\tilde{\Sigma}_{RL}^{s\sp}(\omega,\lambda)=
-\ic\Gamma^B_{s\sp}(f_B(\omega)-1)
-\ic\Gamma^A_{s\sp}(f_A(\omega)-1)e^{i\lambda},
\nonumber\\\label{sigma-4}
\end{eqnarray}
where the $\lambda$ dependence occurs only in $\tilde{\Sigma}_{LR}$ and $\tilde{\Sigma}_{RL}$.

Equation (\ref{11-11}) together with (\ref{zero-order-gf})
and (\ref{sig++})-(\ref{sig-+}) gives the long-time
current statistics within the two-point measurement approach.
It contains the full information about the coherences in the
system eigenbasis through the self-energy matrix $\Sigma$ and can therefore
be used to study effects of coherences on the current statistics.

\subsubsection{Recovering the generalized quantum master equation}\label{GreenQME}

The GF (\ref{11-11}) is different from the GF
obtained using the GQME approach (\ref{9may-2}).
We are now going to show in what limit (\ref{11-11})
reduces to (\ref{9may-2}).\\

Assuming that the $\Sigma$ matrix is diagonal,
$\Sigma_{s\sp}=\delta_{s\sp}\Sigma_{ss}$, the determinant
$|\chi^{-1}|=|g^{-1}-\Sigma|$ in Eq. (\ref{11-11}) factorizes into
a product of determinants corresponding to each orbital $s$,
$|\chi^{-1}|=\prod_s|\chi_{ss}^{-1}|$, and
${\cal Z}(\lambda,t)=\sum_s{\cal Z}_s(\lambda,t)$ becomes the sum of GF for
individual orbitals.
We note that the assumption of a diagonal $\Sigma$ matrix amounts
to ignoring the coherences in the quantum system eigenbasis and is
therefore the analog of the RWA in the GQME approach.
In the following we compute ${\cal Z}_s$. For clarity, we omit the orbital
index $s$ in the self-energies. Since from (\ref{sigma-1})-(\ref{sigma-4})
$\tilde{\Sigma}_{LL}=[\tilde{\Sigma}_{RR}]^*$, we can write
\begin{eqnarray}
\label{determinant-1}
|\chi_{ss}^{-1}|=(\omega-\epsilon_s)^2-|\tilde{\Sigma}_{LL}|^2
-\tilde{\Sigma}_{LR}\tilde{\Sigma}_{RL}.
\end{eqnarray}

Substituting this in (\ref{11-11}), we get for the long time cumulant GF
\begin{eqnarray}
\label{1-2} {\cal S}_s(\lambda)= \int \frac{d\omega}{2\pi}
\mbox{ln}\left[(\omega-\epsilon_s)^2
+|\tilde{\Sigma}_{LL}|^2-\tilde{\Sigma}_{LR}\tilde{\Sigma}_{RL}\right].
\end{eqnarray}
In order to compute the frequency integral we first obtain the $\lambda$-dependent
current by taking the derivative with respect to $\lambda$
\begin{eqnarray}
\label{current} I_s(\lambda)=-\int \frac{d\omega}{2\pi} \frac{\partial_\lambda
(\tilde{\Sigma}_{LR}\tilde{\Sigma}_{RL})}{(\omega-\epsilon_s)^2+
|\tilde{\Sigma}_{LL}|^2-\tilde{\Sigma}_{LR}\tilde{\Sigma}_{RL}}.
\end{eqnarray}
Using (\ref{sigma-1})-(\ref{sigma-4}), we get
\begin{eqnarray}
\label{current-1}
&&\hspace{-0.5cm} I_s(\lambda)= \ic \Gamma^A\Gamma^B\\
&&\times \int \frac{d\omega}{2\pi} \frac{f_B(\omega)[1-f_A(\omega)]e^{\ic\lambda}
-f_A(\omega)[1-f_B(\omega)]e^{-\ic\lambda}}
{(\omega-\epsilon_s)^2+M(\lambda,\omega)} \nonumber
\end{eqnarray}
where
\begin{eqnarray}
\label{m} M(\lambda,\omega)&=&
\frac{1}{4}\Gamma^2+\Gamma^A\Gamma^B\left[f_B(\omega)[1-f_A(\omega)](e^{\ic\lambda}-1)
\right.\nonumber\\
&&+\left.f_A(\omega)[1-f_B(\omega)](e^{-\ic\lambda}-1)\right].
\end{eqnarray}
Assuming that the couplings with the leads are weak $k_B T>>\Gamma^X$
so that resulting broadening is small compared to $\epsilon_s$, the
contribution to the integral comes mainly from the center of the
Lorentzian. This allows us to replace $\omega=\epsilon_s$ in the
Fermi functions inside the integrand. We therefore need to consider
the poles $\omega=\epsilon_s\pm \ic\sqrt{M(\lambda,\epsilon_s)}$.
Computing the residues at the poles, we get
\begin{eqnarray}
\label{current-2}I_s(\lambda)&=&
\frac{-\ic\Gamma^A\Gamma^B}{2\sqrt{M(\lambda,\epsilon_s)}}
\left[\frac{}{}f_B(\epsilon_s)[1-f_A(\epsilon_s)]e^{\ic\lambda}\right.\\
&&\left.\hspace{2.2cm} -
f_A(\epsilon_s)[1-f_B(\epsilon_s)]e^{-\ic\lambda}\frac{}{}\right].\nonumber
\end{eqnarray}
Since $Z_s(\lambda)=\int_0^{\lambda}I(\lambda^\prime)$, we finally get that
\begin{eqnarray}
\label{gf-fnl} {\cal S}_s(\lambda)=
-\frac{\Gamma}{2}+\sqrt{M(\lambda,\epsilon_s)}
\end{eqnarray}
which coincides with the GF obtained from the GQME,
(\ref{9may-1}).

\subsubsection{The Levitov-Lesovik formula}\label{LevLes}

Equation (\ref{11-11}) with (\ref{zero-order-gf})-(\ref{sig-+}) is the most
general formula for the transport statistics at long times for a system of
non-interacting electrons.
It includes the effects of coherences between the various tunneling channels
(system orbitals) available to an electron tunneling between the two lead.
This is due to the non-diagonal structure of the self-energy in the Hilbert
space of the system, Eqs. (\ref{sig++})-(\ref{sig-+}).
Here, we recover Levitov-Lesovik formula \cite{Levitov93,LevitovLeeJMathPhys96}
for the counting statistics.
For that we again assume diagonal self-energies.
As discussed in the previous subsection (\ref{GreenQME}), the cumulant
GF in this case is simply the product of the GFs for each orbital.
Thus all orbitals contribute independently to the electron transport.

Using self-energy expressions (\ref{sigma-1})-(\ref{sigma-4}),
the GF (\ref{1-2}) can be expressed as
\begin{eqnarray}
\label{nmm2}
{\cal S}_s(\lambda)&=& \int \frac{d\omega}{2\pi}
\mbox{ln}\left[(\omega-\epsilon_s)^2+ \frac{\Gamma^2}{4}\right. \\
&&\hspace{1.3cm}+\left.\Gamma^A\Gamma^B[f_B(\omega)
(f_A(\omega)-1)(1-\e^{\ic\lambda})\right.\nonumber\\
&&\hspace{1.3cm}+\left.f_A(\omega)(f_B(\omega)-1)
(1-\e^{-\ic\lambda})] \frac{}{} \right].\nonumber
\end{eqnarray}
Using (\ref{ret-green}), we can write for orbital $s$
\begin{eqnarray}
\label{mod-ret-green}
|R_{ss}^{--}(\omega)|^{-2}= (\omega-\epsilon_s)^2+\frac{\Gamma^2}{4}.
\end{eqnarray}
Substituting (\ref{mod-ret-green}) in (\ref{nmm2}), we obtain
\begin{eqnarray}
\label{nmm3}
&&{\cal S}_s(\lambda)= -2\int \frac{d\omega}{2\pi} \mbox{ln}|R_{ss}^{--}(\omega)| \nonumber\\
&&\hspace{1cm}+ \int \frac{d\omega}{2\pi} \mbox{ln}\left\{\frac{}{}1 + {\cal T}(\omega)[f_B(\omega)
(f_A(\omega)-1)(1-\e^{\ic\lambda})\right.\nonumber\\
&&\hspace{1cm}+ \left.f_A(\omega)(f_B(\omega)-1)(1-\e^{-\ic\lambda})]\frac{}{}\right\}
\end{eqnarray}
where ${\cal T}(\omega)=\Gamma^A\Gamma^B |R_{ss}^{--}(\omega)|^{2}$
is the transmission coefficient for the tunneling region.
The first term on the r.h.s. of (\ref{nmm3}) can be ignored since it does
not contribute to the average current or its fluctuations (independent on $\lambda$).
Therefore
\begin{eqnarray}
\label{nmm4}
{\cal S}_s(\lambda) &=& \int \frac{d\omega}{2\pi} \mbox{ln}\left\{\frac{}{}1 + {\cal T}(\omega)
[f_B(\omega)(f_A(\omega)-1)(1-\e^{\ic\lambda})\right.\nonumber\\
&&\quad\quad+\left.f_A(\omega)(f_B(\omega)-1)(1-\e^{-\ic\lambda})]\frac{}{}\right\}
\end{eqnarray}
which is the Levitov-Lesovik formula \cite{Levitov93,LevitovLeeJMathPhys96,Levitov04}.
It has been recently generalized to a
multi-terminal model for a non-interacting tight-binding model \cite{Schonhammer07}.
Equation (\ref{nmm4}) is valid to all orders of the coupling.
The only approximation required to obtain the Levitov-Lesovik
expression (\ref{nmm4}) is to ignore the coherence effects between
different orbitals in the tunneling junction.
Notice that if ${\cal T}(\omega)$ is small, we can expand the logarithm in Eq. (\ref{nmm4}).
This is equivalent to making a perturbation in the coupling $\hat{V}$.
The leading order in the expansion gives (\ref{cumulantasympt}) with (\ref{9april-2}).

Since $f_A(\omega)[1-f_B(\omega)]=\e^{\beta eV}f_B(\omega)[1-f_A(\omega)]$,
it is straightforward to see that the GF (\ref{nmm4}) satisfy,
${\cal S}(\lambda)={\cal S}(-\lambda-\ic\beta eV)$, and the FT (\ref{4april-7}) follows.

Taking the derivative with respect to $\lambda$ of the GF (\ref{nmm4})
at $\lambda=0$, the average current is
\begin{eqnarray}
\label{nmm5}
I= \int \frac{d\omega}{2\pi} {\cal T}(\omega)\left[f_B(\omega)-f_A(\omega)\right],
\end{eqnarray}
which is the Landauer-Buttiker expression for the average current through
a tunneling junction with transmission coefficient ${\cal T}(\omega)$ \cite{ButtikerRev00}.

\section{Nonlinear coefficients} \label{NLcoef}

As we have seen, the FT implies a specific
symmetry of the GF which depends on the
nonequilibrium constraints imposed on the system.
For weak constraints, i.e. close to equilibrium, this symmetry can be
used to derive fluctuation-dissipation relation as well as Onsager
symmetry relations \cite{Gallavotti96,Gallavotti96b,Lebowitz99,AndrieuxGaspard04}.
A systematic expansion of the GF in the nonequilibrium constrains allows
to derive similar fundamental relations further away from equilibrium.
This has been done for stochastic systems \cite{AndrieuxGaspard07b},
for counting statistics \cite{NazarovTobiska05,Forster,SaitoUtsumi07,UtsumiSaito08}
and for the work FT \cite{GaspardAndrieux08}.
FTs therefore provide a systematic approach for studying
generalized fluctuation-dissipation relations such as previously
considered in Refs. \cite{WangHeinz02,ChouSuHaoYu85,StratonovichI}.

\subsection{Single nonequilibrium constraint}\label{SingleNEC}

We assume that a FT of the form $p(k,{\cal A})= \mbox{e}^{Ak} p(-k,{\cal A})$
holds in a system maintained in a nonequilibrium steady-state by a single
nonequilibrium constraint ${\cal A}$, where $p(k,{\cal A})$ is the probability
distribution that a net amount of energy or matter $k$ crossed the system during
a given time. The cumulant GF defined as
\begin{eqnarray}
\label{eq-2}
\cz(\lambda,{\cal A}) = \ln \left(\sum_k e^{\ic\lambda k} p(k,{\cal A})\right)
\end{eqnarray}
then possesses the symmetry
\begin{eqnarray}
\label{eq-1}
\cz(\lambda,{\cal A}) = \cz(\ic{\cal A}-\lambda,{\cal A}) \;.
\end{eqnarray}
Taking the derivative with respect to ${\cal A}$ of both sides
and using (\ref{eq-2}), we find that in the ${\cal A}\to 0$ limit
\begin{eqnarray}
\label{eq-6} \frac{\partial}{\partial {\cal A}} \left[
\cz(\lambda,0)-\cz(-\lambda,0)\right]
=-\ic\frac{\partial}{\partial \lambda}\cz(\lambda,0).
\end{eqnarray}
The cumulant GF is expressed in terms of cumulants as
\begin{eqnarray}
\label{eq-7}
\cz(\lambda,{\cal A}) =
\sum_{m=1}^{\infty} \frac{(\ic \lambda)^m}{m!} K_m ({\cal A}).
\end{eqnarray}
Using (\ref{eq-7}) in (\ref{eq-6}), we find at each order in $\lambda$, that
\begin{eqnarray}
\label{eq-8}
\left[1 - (-1)^m \right]
\frac{\partial}{\partial {\cal A}} K_m(0) =K_{m+1}(0).
\end{eqnarray}
Equation (\ref{eq-8}) implies that at equilibrium, odd cumulants are zero
and event cumulant are related to the derivative with respect to the
nonequilibrium constraints of the nonequilibrium odd cumulants when
approaching equilibrium
\begin{eqnarray}
K_{2m-1}(0) &=& 0 \label{eq-9} \\
K_{2m}(0) &=& 2 \frac{\partial}{\partial {\cal A}}K_{2m-1}(0).\label{eq-9a}
\end{eqnarray}
Below we show that this leads to the well known fluctuation dissipation relations.

We next consider the second derivative with respect to ${\cal A}$ of both
sides of (\ref{eq-1}). Using (\ref{eq-1}) and (\ref{eq-8}) and after some
algebra, we find in the ${\cal A}\to 0$ limit that
\begin{eqnarray}
&&\frac{\partial^2}{\partial {\cal A}^2}
\left[\cz(\lambda,0)-\cz(-\lambda,0) \right] \label{eq-9b}\\
&&\hspace{1.5cm} = - \ic
\frac{\partial^2}{\partial \lambda \partial {\cal A}}
\left[\cz(\lambda,0)+\cz(-\lambda,0) \right] .\nonumber
\end{eqnarray}
Using (\ref{eq-7}), we find at each order in $\lambda$ that
\begin{eqnarray}
&&\left[1 - (-1)^m \right]
\frac{\partial^2}{\partial {\cal A}^2} K_m(0) \label{eq-9c}\\
&&\hspace{1.6cm}=
\left[1 + (-1)^{m+1} \right] \frac{\partial}{\partial {\cal A}} K_{m+1}(0).
\nonumber
\end{eqnarray}
This relation is only useful for odd $m$ and implies
\begin{eqnarray}
\label{eq-9d}
\frac{\partial^2}{\partial {\cal A}^2} K_{2m-1}(0) =
\frac{\partial}{\partial {\cal A}} K_{2m}(0) .
\end{eqnarray}
This procedure can be continued for higher derivative of
$\cz(i{\cal A}-\lambda,{\cal A})$ with respect to ${\cal A}$.

We can always expand the average process in term of
the nonequilibrium constrain as
\begin{eqnarray}
\label{eq-10}
K_1({\cal A}) = K_1(0) + L^{(1)} {\cal A} + L^{(2)} {\cal A}^2
+ {\cal O}({\cal A}^3) .
\end{eqnarray}
$L^{(1)}$ is the Onsager coefficient.
Using (\ref{eq-9}), (\ref{eq-9a}) and (\ref{eq-9d}) for $m=1$,
we find that $K_1(0) = 0$ and that
\begin{eqnarray}
\label{eq-11}
L^{(1)} &=& \frac{\partial}{\partial {\cal A}} K_1(0) = \frac{K_2(0)}{2} \\
L^{(2)} &=& \frac{1}{2} \frac{\partial^2}{\partial {\cal A}^2} K_1(0) =
\frac{1}{2} \frac{\partial}{\partial {\cal A}} K_2(0) .
\end{eqnarray}
(\ref{eq-11}) is a fluctuation-dissipation relation.
As an illustration, we consider a biased quantum junction such as
in section \ref{fermion-transport}.
$k$ represents the number of electron crossing the junction and the
nonequilibrium constraint is given by ${\cal A}=\beta e V$, where
$V$ is the potential bias across the junction.
In this case, close to equilibrium, $\mean{I} = \beta e^2 V L^{(1)}$ is the
average electrical current through the junction and $e^2 K_2(0)$ is the Fourier
transform of the equilibrium current correlation functions at zero frequency.
(\ref{eq-11}) indicates that the resistance of the junction, which
characterize the dissipation, is related to the current fluctuation
at equilibrium by $R=\partial_V \mean{I}=\beta e^2 L^{(1)}=\beta e^2 K_2(0)/2$.

\subsection{Multiple nonequilibrium constraints}\label{MultNEC}

When multiple nonequilibrium constrains are applied to the system, the
FT can be used to find important symmetries of the
response coefficients \cite{AndrieuxGaspard07b,AndrieuxGaspard04}.
In case of $N$ nonequilibrium constraints, the cumulant GF reads
\begin{eqnarray}
\label{eq-2Gen}
\cz(\{\lambda_{\gamma}\},\{{\cal A}_{\gamma}\}) = \ln \left(\sum_{\{k_{\gamma}\}}
e^{\ic \bar{\lambda} \cdot \bar{k}} p(\{k_{\gamma}\},\{{\cal A}_{\gamma}\})\right)\;,
\end{eqnarray}
where $ \bar{\lambda} \cdot \bar{k} = \sum_{\gamma=1}^N k_{\gamma} \lambda_{\gamma}$.
We assume that it satisfies the FT symmetry
\begin{eqnarray}
\label{eq-1Gen}
\cz(\{\lambda_{\gamma}\},\{{\cal A}_{\gamma}\}) =
\cz(\{\ic {\cal A}_{\gamma}-\lambda_{\gamma}\},\{{\cal A}_{\gamma}\}) \;.
\end{eqnarray}
Proceeding as in section \ref{SingleNEC}, we find that (\ref{eq-6}) generalizes to
\begin{eqnarray}
\label{eq-6Gen}
&&\frac{\partial}{\partial {\cal A}_{\beta}} \left[
\cz(\{\lambda_{\gamma}\},\{0\})-\cz(\{-\lambda_{\gamma}\},\{0\})\right]  \\
&&\hspace{4cm}=-\ic \frac{\partial}{\partial \lambda_{\beta}}
\cz(\{\lambda_{\gamma}\},\{0\}) \;.\nonumber
\end{eqnarray}
The cumulant GF can be expressed as
\begin{eqnarray}
&&\cz(\{\lambda_{\gamma}\},\{{\cal A}_{\gamma}\}) \label{eq-7Gen}\\
&&\hspace{1cm} =\sum_{\{m_{\gamma}\}=1}^{\infty} \bigg( \prod_{\gamma=1}^N
\frac{(\ic \lambda_{\gamma})^{m_{\gamma}}}{m_{\gamma}!} \bigg)
K_{\{m_{\gamma}\}}(\{{\cal A}_{\gamma}\}) \;. \nonumber
\end{eqnarray}
where the cumulants read
\begin{eqnarray}
K_{\{m_{\gamma}\}}(\{{\cal A}_{\gamma}\}) =
\bigg( \prod_{j=1}^{\gamma} (-\ic)^{m_j} \frac{\partial^{m_j}}{\partial \lambda_{j}^{m_j}} \bigg)
\cz(\{0\},\{{\cal A}_{\gamma}\}) \;. \nonumber \\
\label{CumulantGen}
\end{eqnarray}
The generalisation of (\ref{eq-8}) is found using (\ref{eq-7Gen})
in (\ref{eq-6Gen}), so that at a given order in the $\lambda$'s
\begin{eqnarray}
\bigg( 1-\prod_{\gamma=1}^N (- 1)^{m_{\gamma}} \bigg)
\frac{\partial K_{\{m_{\gamma}\}}(\{0\})}{\partial {\cal A}_{\beta}}
= K_{\{m_{\gamma}+\delta_{\gamma \beta}\}}(\{0\}) \;.\nonumber \\
\end{eqnarray}
If we choose $\{m_{\gamma}\}=\{\delta_{\gamma \alpha}\}$, we get that
\begin{eqnarray}
\frac{\partial K_{\{\delta_{\gamma \alpha}\}}(\{0\})}{\partial {\cal A}_{\beta}}
= K_{\{\delta_{\gamma \alpha}+\delta_{\gamma \beta}\}}(\{0\}) \;. \label{PourOnsag}
\end{eqnarray}
Close to equilibrium, the average processes can be expanded
in term of the nonequilibrium constraints as
\begin{eqnarray}
K_{\{\delta_{\gamma \alpha}\}}(\{{\cal A}_{\gamma}\})
= \sum_{\gamma} L_{\alpha \gamma} {\cal A}_{\gamma} + \sum_{\gamma,\gamma'}
L_{\alpha \gamma \gamma'} {\cal A}_{\gamma} {\cal A}_{\gamma'} + \cdots \;.
\nonumber \\ \label{CumulantExp}
\end{eqnarray}
Since the (Onsager) linear response coefficients are given by
\begin{eqnarray}
L_{\alpha \beta}
=\frac{\partial K_{\{\delta_{\gamma \alpha}\}}(\{0\})}{\partial {\cal A}_{\beta}} \;,
\end{eqnarray}
using (\ref{PourOnsag}), we find the Onsager reciprocity relation
\begin{eqnarray}
L_{\alpha \beta}=L_{\beta \alpha} \;. \label{Onsager}
\end{eqnarray}
The generalisation of (\ref{eq-9c}) to multiple nonequilibrium constraints reads
\begin{eqnarray}
\label{FurtherA}
&&\hspace{-0.2cm}\frac{\partial^2}{\partial {\cal A}_{\alpha} \partial {\cal A}_{\beta}} \left[
\cz(\{\lambda_{\gamma}\},\{0\})-\cz(\{-\lambda_{\gamma}\},\{0\})\right] = \\
&&\hspace{-0.2cm}-\ic \left[
\frac{\partial^2}{\partial \lambda_{\alpha} \partial {\cal A}_{\beta}}
\cz(\{\lambda_{\gamma}\},\{0\}) +
\frac{\partial^2}{\partial \lambda_{\beta} \partial {\cal A}_{\alpha}}
\cz(\{-\lambda_{\gamma}\},\{0\}) \right]\;.\nonumber
\end{eqnarray}
This implies that
\begin{eqnarray}
&&\hspace{-0.6cm}\bigg( 1-\prod_{\gamma=1}^N (-1)^{m_{\gamma}} \bigg)
\frac{\partial^2 K_{\{m_{\gamma}\}}(\{0\})}{\partial {\cal A}_{\alpha} \partial {\cal A}_{\beta}}
= \label{FurtherB} \\
&&\hspace{-0.4cm}- \ic \left[ \frac{\partial K_{\{m_{\gamma}+\delta_{\gamma \beta}\}}(\{0\})} {\partial {\cal A}_{\alpha}}
\right. \nonumber\\ &&\hspace{0.4cm} \left. + \bigg( 1-\prod_{\gamma=1}^N
(-1)^{m_{\gamma}+\delta_{\gamma \beta}} \bigg) \frac{\partial K_{\{m_{\gamma}+\delta_{\gamma \alpha}\}}
(\{0\})}{\partial {\cal A}_{\beta}} \right] \;. \nonumber
\end{eqnarray}
For $\{m_{\gamma}\}=\{\delta_{\gamma \theta}\}$, we get
\begin{eqnarray}
\label{FurtherC}
L_{\theta \alpha \beta}&=&
\frac{\partial^2 K_{\{\delta_{\gamma \theta}\}}(\{0\})}{\partial {\cal A}_{\alpha} \partial {\cal A}_{\beta}}\\
&&\hspace{-0.8cm} -\ic \bigg( \frac{\partial K_{\{\delta_{\gamma \theta}+\delta_{\gamma \beta}\}}(\{0\})}
{\partial {\cal A}_{\alpha}} + \frac{\partial K_{\{\delta_{\gamma \theta}+\delta_{\gamma \alpha}\}}(\{0\})}
{\partial {\cal A}_{\beta}} \bigg) \nonumber \;,
\end{eqnarray}
which implies the expected symmetry $L_{\theta \alpha \beta}=L_{\theta \beta \alpha}$.

\section{Conclusions and perspectives}\label{ConcPersp}

The approach to quantum statistics adopted in this
review is based on a two-point projective measurement.
This, together with considerations about the symmetry
between the forward and the time-reversed quantum dynamics,
allow to recover from a simple and unified perspective
all previously derived fluctuation theorems (FTs) for
quantum systems (transient as well as steady-state FTs).
This was the object of section \ref{2pointFT} and \ref{FT}.\\

A generalized quantum master equation (GQME) is presented in section
\ref{MEGF} for a quantum system weakly coupled to reservoirs.
It describes the evolution of the generating function (GF)
associated with the system density matrix conditional to the
outcome from a two-point measurement (of energy or number of
particles) on the reservoir.
When summed over all the possible outcomes, the quantum master
equation (QME) for the system reduced density matrix is recovered.
This formalism has been applied to various model systems and used
to directly demonstrate the validity of steady-state FTs.\\

The GQME formalism circumvents the unraveling of the QME, used to
calculate the quantum statistics of particles or energy, and
originally developed in quantum optics \cite{Gardiner00,Breuer02,
Wiseman93a,Wiseman93b,PlenioKnight98,Brun00,Brun02}.
Since the unraveling of a QME is not unique, a continuous time
measurement on the reservoir is assumed in order to connect
the resulting quantum trajectories to measurable quantities.
This procedure is only possible for Markovian QME which preserve
complete positivity [in the rotating wave approximation (RWA)].
In this regime, the GQME formalism predicts the same
statistics as the unraveling formalism.
This equivalence between the two types of measurements in the
weak coupling limit was first found in Ref. \cite{DeRoeck07,Maes07}.
This results from the fact that the reservoirs are assumed to
always remain described by the same canonical or grand canonical
equilibrium density matrix \cite{EspositoGaspard,EspositoGaspardb} 
and are therefore not affected by the measurement.
The net number of particles or the net amount of
energy transferred during a given time interval is
then the same if the reservoir is continuously monitored
or only measured twice at the beginning and at the end.
The unraveling of non-Markovian QME has been an active field of research
during this last decade \cite{Strunz96,Strunz98,GaspardNagaoka99,Strunz99},
but the connection between the resulting quantum trajectories and
measurable quantities is not straightforward
\cite{Wiseman02,Wiseman03,Wiseman03a,Wiseman03b,Breuer04,Diosi08}.
In the GQME formalism, the connection to measurable quantities
in the non-Markovian regime is unambiguous.
Exploring non-Markovian effects on the particle or energy
statistics could be an important future application. \\


In order to go beyond the approximations used in the GQME formalism
(i.e. initially factorized density matrix, weak coupling), we presented
an alternative approach based on superoperator non-equilibrium Green's
functions (SNGF) in section \ref{GreenFun}.
This Liouville space formalism provides a powerful tool for
calculating the particle statistics in many body quantum systems.
Using this formalism, we showed that initial coherences in the basis of the
measured observable do not affect the steady-state counting statistics and the FT.
This is to be expected since at steady-state, the long time
limit destroys the information about the initial condition.
We showed it using a non-interacting electron model for both
direct and indirect (transport) tunneling between two reservoirs.
However, for transient FTs such as the Crooks relation, the assumption
that the system density matrix is initially diagonal in the basis of
the measured observable seems unavoidable for the FT to be satisfied.
We applied the SNGF formalism to compute the counting statistics in
some simple models and discussed the limit in which the statistics
predicted by the QME is recovered.
The Levitov-Lesovik formula for electron tunneling between two reservoirs,
which goes beyond the weak coupling limit of the QME, was also recovered.
We discussed the approximations required to recover the Levitov-Lesovik
expression from a more general result expressed in terms of the
SNGF for the tunneling region.
In particular, we showed that when several energy channels are available
to tunneling electrons, the Levitov-Lesovik approach does not capture the
quantum coherence between different channels.
This amounts to ignoring the off-diagonal elements of
the self-energy in the eigenbasis of the system.\\

Transient FTs (valid for arbitrary time) have been presented in \ref{WorkFTsection}.
The work FT derived for isolated driven system in section
\ref{WorkFTsectionIsol} is always valid since, besides an
initial canonical density matrix, no assumptions have been made.
The work FT for open driven system derived in section \ref{WorkFTsectionRes}
assumes an initially factorized canonical density matrix between the system
and the reservoir and a definition of work which is only consistent for
weak system-reservoir interaction.
The transient FT for direct heat and matter transfer between two finite
systems and derived in section \ref{TransientFTHM} assumes that the
systems are each initially at equilibrium and weakly interacting.
The steady-state FTs (only valid for long time) presented in section
\ref{SSFT2point} and derived more systematically in section \ref{MEGF}
assumes a weak system-reservoir coupling and the RWA.
However, the FT has been recently shown (numerically) to hold for
QME without RWA \cite{Welack08} and the Levitov-Lesovik formula presented in
section \ref{LevLes} is obtained nonperturbatively and satisfies the FT.
FTs seem therefore to characterize universal feature of nonequilibrium
fluctuations in quantum as well as in classical systems. \\

We now discuss some future perspectives.\\



We mentioned in the introduction and in section \ref{2pointFT} that
an alternative approach to counting statistics, where the GF used
is an influence functional following from a path integral
description of the system-detector interaction, has been
developed during the last decade.
It is only in a semi-classical limit that the two-point measurement
approach predicts the same statistics as this approach.
Determining the region of applicability of both prescriptions is an open
problem that could lead to a better understanding of quantum measurements.\\

Various numerical methods have been developed for using the Jarzynski
relation to efficiently calculate equilibrium free energies of classical
systems \cite{Jarzynski08,LechnerDellago07,DellagoCo06}.
Extending these methods to quantum systems will be of interest.\\

Finally, we note that in this review we have focused on systems maintained
in a steady-state distribution by a single non-equilibrium constraint.
Investigating systems subjected to multiple nonequilibrium
constraints could reveal interesting features.
\section*{Acknowledgments}

The support of the National Science Foundation (Grant No.
CHE-0745891) and NIRT (Grant No. EEC 0303389) is gratefully
acknowledged.
M. E. is funded by the FNRS Belgium (charg\'e de recherche) and by
the Government of Luxembourg (bourse de formation-recherche).\\

\appendix
\section{Time-reversed evolution}\label{appA}

We explain why (\ref{Aaaaf}) corresponds to the time-reversed expression
of the two-point probability (\ref{Aaaaa}) and discuss how to physically
implement a time-reversed evolution. The effect of a static magnetic field
is also discussed.\\

In order to implement the time-reversal operation in quantum mechanics,
it is necessary to introduce the antilinear operator $\Theta$
($\Theta \ic=-\ic \Theta$) which satisfies $\Theta^2=1$
(i.e. $\Theta^{-1}=\Theta$) \cite{Merzbacher,Wigner}.
An arbitrary observable $\hat{A}$ can be even or
odd with respect to the time-reversal operation, i.e
\begin{eqnarray}
\Theta \hat{A} \Theta = \epsilon_A \hat{A} \;,
\end{eqnarray}
where $\epsilon_A=\pm 1$.
For example, the position operator $\hat{R}$ is even ($\epsilon_R=1$) while the
momentum $\hat{P}$ or angular momentum $\hat{L}$ are odd ($\epsilon_{P,L}=-1$).
It can be verified that the Heisenberg commutation relations
are preserved under the time-reversal operation.
When acting on a time dependent Hamiltonian $\hat{H}(t;B)$
that depends on a static magnetic field $B$, we get
\begin{eqnarray}
\Theta \hat{H}(t;B) \Theta = \hat{H}(t;-B) \;.
\label{TimeReversedHamiltonian}
\end{eqnarray}
If a forward evolution operator [as in (\ref{unitary_evolution})
but with a static magnetic field] evolves according to
\begin{eqnarray}
\frac{d}{dt} \hat{U}(t,0;B) = -\frac{\ic}{\hbar} \hat{H}(t;B) \hat{U}(t,0;B) \;,
\label{UnitOperEvol}
\end{eqnarray}
with the initial condition $\hat{U}(0,0;B)=\hat{1}$, than the time-reversed
evolution operator is defined by \cite{GaspardAndrieux08}
\begin{eqnarray}
\hat{U}_{\rm tr}(t,0;-B) &\equiv& \Theta \hat{U}(T-t,0;B) \hat{U}^{\dagger}(T,0;B)
\Theta \nonumber\\ &=& \Theta \hat{U}(T-t,T;B) \Theta \;, \label{UnitOperBack}
\end{eqnarray}
and its evolution is given by
\begin{eqnarray}
\frac{d}{dt} \hat{U}_{\rm tr}(t,0;-B) = -\frac{\ic}{\hbar} \hat{H}(T-t;-B)
\hat{U}_{\rm tr}(t,0;-B) \;,  \label{UnitOperEvolBack}
\end{eqnarray}
with the initial condition $\hat{U}_{\rm tr}(0,0;B)=\hat{1}$.
This can be verified using the change of variable $t \to T-t$ in (\ref{UnitOperEvol}),
then multiplying the resulting equation by $\Theta$ from the left and by
$\hat{U}^{\dagger}(T,0;B) \Theta$ from the right and then using
(\ref{TimeReversedHamiltonian}) and (\ref{UnitOperBack}).\\

From now on we choose $t=T$ (the time at which the time
reversal operation is performed is $t$), and we define
\begin{eqnarray}
\hat{\rho}(t) &\equiv& \hat{U}(t,0;B) \hat{\rho}_0 \hat{U}^{\dagger}(t,0;B) \label{nouvDefPR1} \\
\Theta \hat{\rho}^{{\rm tr}}(t) \Theta &\equiv& \hat{U}_{\rm tr}(t,0;-B)
\Theta \hat{\rho}^{{\rm tr}}_0 \Theta \hat{U}^{\dagger}_{\rm tr}(t,0;-B) \;.\label{nouvDefPR2}
\end{eqnarray}
We note that by multiplying (\ref{nouvDefPR2}) by $\Theta$ from the left
and from the right, we get
\begin{eqnarray}
\hat{\rho}^{{\rm tr}}(t) = \hat{U}^{\dagger}(t,0;B)
\hat{\rho}^{{\rm tr}}_0 \hat{U}(t,0;B) \;.\label{nouvDefPR3}
\end{eqnarray}
We verify that if $\hat{\rho}^{{\rm tr}}_0=
\hat{\rho}(t)$, then $\hat{\rho}^{{\rm tr}}(t)=\hat{\rho}_0$.
This means that, as for classical systems, if a system initially described by $\hat{\rho}_0$
evolves according to the forward evolution between $0$ and $t$, then the time-reversal operation
is applied and the resulting density matrix is evolved according to the backward evolution
during a time $t$ and finally the time-reversal operation is again applied, the
resulting density matrix is the initial condition $\hat{\rho}_0$.
It follows from this discussion that if the two-point probability
(\ref{Aaaaa}) [with a static magnetic field $B$] is defined as
\begin{eqnarray}
&&\hspace{0cm}P[a_t,a_0] \equiv \label{TwoPointProbMagn}\\
&&\hspace{0.5cm}\trace \left\{ \hat{P}_{a_t} \hat{U}(t,0;B) \hat{P}_{a_0} \hat{\rho}_0
\hat{P}_{a_0} \hat{U}^{\dagger}(t,0;B) \hat{P}_{a_t} \right\} \;,
\nonumber
\end{eqnarray}
the time-reversed expression of this two-point probability has to be defined as
\begin{eqnarray}
&&\hspace{-0.2cm}P^{{\rm tr}}[a_0,a_t] \equiv \label{TwoPointProbMagnTR}\\ &&\hspace{0.2cm}
\trace \left\{ \hat{P}_{a_0} \hat{U}_{\rm tr}(t,0;-B) \hat{P}_{a_t} \Theta \hat{\rho}^{{\rm tr}}_0
\Theta \hat{P}_{a_t} \hat{U}_{\rm tr}^{\dagger}(t,0;-B) \hat{P}_{a_0} \right\} \;. \nonumber
\end{eqnarray}
We note that we could have included the final time-reversal operation in
the definition, but it has no effect anyway due to the trace invariance.
By inserting $\Theta^2$ in between all the operators in (\ref{TwoPointProbMagnTR}),
and using (\ref{UnitOperBack}) with $T=t$, we find that
\begin{eqnarray}
&&P^{{\rm tr}}[a_0,a_t] \equiv \label{TwoPointProbMagnTRbis} \\
&&\hspace{0.5cm}\trace \left\{ \hat{P}_{a_0} \hat{U}^{\dagger}(t,0;B) \hat{P}_{a_t}
\hat{\rho}^{{\rm tr}}_0 \hat{P}_{a_t} \hat{U}(t,0;B) \hat{P}_{a_0} \right\} \nonumber \;,
\end{eqnarray}
which is identical to the definition used in (\ref{Aaaaf}).
It is convenient to use (\ref{TwoPointProbMagnTRbis}) as a starting point
because it allows to avoid mentioning the presence of a static magnetic Field.
However, it is important to keep in mind that the physical evolution corresponding
to the time-reversed dynamics associated to a forward dynamics with an Hamiltonian
$\hat{H}(t;B)$ is an evolution with an Hamiltonian where the driving protocol is
time-reversed, where the sign of the static magnetic field is changed $\hat{H}(T-t;-B)$
and where the initial condition is $\Theta \hat{\rho}^{{\rm tr}}_0 \Theta$.

\section{Fluctuation theorem for coarse-grained dynamics} \label{CoarsegrainedFT}

Here, we show that using a coarse-graining of the initial density matrices,
$R$ defined in section \ref{FTgen} becomes a measurable quantity and
$\mean{R}$ a difference of Gibbs-von Neumann entropy.
We follows closely Refs. \cite{Maes04c,Maes05}.\\

We define
\begin{eqnarray}
R[a_t,a_0] \equiv \ln \frac{P[a_t,a_0]}{P^{{\rm tr}}[a_0,a_t]}
\equiv - R^{{\rm tr}}[a_0,a_t] \label{defR}
\end{eqnarray}
and
\begin{eqnarray}
p(R) &\equiv& \sum_{a_t,a_0} P[a_t,a_0] \delta(R-R[a_t,a_0]) \nonumber\\
p^{{\rm tr}}(R)&\equiv& \sum_{a_t,a_0} P^{{\rm tr}}[a_0,a_t]
\delta(R-R^{{\rm tr}}[a_0,a_t]) \;. \label{detailedFTdef}
\end{eqnarray}
Note that (\ref{defR}), in contrast to (\ref{defRtil}), is expressed
exclusively in terms of measurable quantities (eigenvalues of $A(t)$).
An integral FT follows
\begin{eqnarray}
\mean{\e^{-R}} \equiv \sum_{a_t,a_0} P[a_t,a_0] \e^{-R[a_t,a_0]} = 1
\;, \label{integralFT}
\end{eqnarray}
which implies $\mean{R} \geq 0$, as well as a detailed FT
\begin{eqnarray}
\frac{p(R)}{p^{{\rm tr}}(-R)} = \e^{R}  \;. \label{detailedFT}
\end{eqnarray}
The coarse-graining of a density matrix $\hat{\rho}$ within its
non-measured part reads
\begin{eqnarray}
\tilde{\hat{\rho}} = \sum_{a} \frac{p_a}{d_a} \hat{P}_{a} \;,
\label{maxentdensitymat}
\end{eqnarray}
where $p_a= \trace \hat{\rho} \hat{P}_{a}$ is the probability to measures $a$,
and $d_a$ is the number of states with the value $a$.
When, as in \cite{Maes04b}, such a procedure is applied to $\hat{\rho}^{\rm tr}_0$
and $\hat{\rho}_0$, $\mean{R}$ can be related to an entropy change.
In this case
\begin{eqnarray}
P[a_t,a_0] &=& \trace \{ \hat{U}^{\dagger}(t,0) \hat{P}_{a_t} \hat{U}(t,0) \hat{P}_{a_0} \}
\frac{p_{a_0}}{d_{a_0}}  \nonumber \\
P^{{\rm tr}}[a_0,a_t] &=& \trace \{ \hat{U}^{\dagger}(t,0) \hat{P}_{a_t} \hat{U}(t,0) \hat{P}_{a_0} \}
\frac{p_{a_t}^{{\rm tr}}}{d_{a_t}} \label{the2probentropy} \;.
\end{eqnarray}
Therefore, using (\ref{the2probentropy}) in (\ref{defR}), we get
\begin{eqnarray}
R[a_t,a_0] = s_{a_t}^{{\rm tr}}- s_{a_0} \;, \label{Rdifftrajent}
\end{eqnarray}
where
\begin{eqnarray}
s_{a_t}^{{\rm tr}} \equiv - \ln \frac{p_{a_t}^{{\rm tr}}}{d_{a_t}} \  \  \;,
\  \ s_{a_0} \equiv - \ln \frac{p_{a_0}}{d_{a_0}} \;.
\end{eqnarray}
The average of $R$ now reads
\begin{eqnarray}
\mean{R} = \sum_{a_t,a_0} R[a_t,a_0] P[a_t,a_0] = S^{{\rm tr}}-S \;,
\label{trajentropy}
\end{eqnarray}
where
\begin{eqnarray}
S^{{\rm tr}} \equiv \sum_{a_t} s_{a_t}^{{\rm tr}} p_{a_t}^{{\rm tr}}
\  \  \;, \  \ S \equiv \sum_{a_0} s_{a_0} p_{a_0}  \label{entropy}
\end{eqnarray}
are the Gibbs-von Neumann entropies associated to the
coarse-grained density matrix $\hat{\rho}^{\rm tr}_0$ and $\hat{\rho}_0$.
Indeed, if the coarse-grained density matrix $\tilde{\hat{\rho}}$ is used in
the expression for the Gibbs-von Neumann entropy $S = \trace \tilde{\hat{\rho}}
\ln \tilde{\hat{\rho}}$, we get $S = \sum_a s_a p_{a}$.

\section{Large deviation and fluctuation theorem} \label{largedev}

Below, we briefly describe large deviation theory
and show that a symmetry of the long time limit of the cumulant
GF such as (\ref{23april-3}) or (\ref{23april-6b}) translates
into a steady-state FT for the probabilities.\\

We consider a probability distribution $p(t,k)$, where $k$ is a counting
variable associated to a continuous time random walk (we assume that the
waiting time distributions have a finite first and second moment).
For fixed time, the central limit theorem is only valid up to a given
accuracy in a central region of the probability distribution hows width
does not converge uniformly with time.
Large deviation goes beyond the central limit theorem and allows to
describe the behaviour of the tail of the distribution \cite{Sornette,Touchette}.
It relies on the assumption that the probability $\tilde{p}(t,\xi)$
that $\xi=k/t$ takes a value in the interval $[\xi,\xi+d\xi]$ behaves as
\begin{eqnarray}
\tilde{p}(t,\xi) = C(\xi,t) e^{R(\xi) t} \;,
\label{FTaaaaac}
\end{eqnarray}
where the large deviation function (LDF) is defined by
\begin{eqnarray}
R(\xi) \equiv \lim_{t \to \infty} \frac{1}{t} \ln \tilde{p}(t,\xi)
\label{FTaaaaab}
\end{eqnarray}
and where
\begin{eqnarray}
\lim_{t \to \infty} \frac{1}{t} \ln C(\xi,t) = 0 \;.
\label{FTaaaaaca}
\end{eqnarray}
We will show that the LDF is determined by the
long time limit of the cumulant GF given by
\begin{eqnarray}
{\cal S}(\lambda) \equiv \lim_{t \to \infty}
\frac{1}{t} \ln G(t,\lambda) \;, \label{FTaaaaae}
\end{eqnarray}
where the moment GF is defined as
\begin{eqnarray}
G(t,\lambda) \equiv \sum_k p(t,k) e^{- \lambda k}\;.
\label{FTaaaaaabis}
\end{eqnarray}
Note that for convenience, we have absorbed a factor $-\ic$ in the definition of
$\lambda$ compared to the standard definition of the moment GF used in the main text.
The GF can be rewritten in terms of $\tilde{p}(t,\xi)$ as
\begin{eqnarray}
G(t,\lambda) = \int d\xi \tilde{p}(t,\xi) e^{- \lambda \xi t} \;.
\label{FTaaaaaa}
\end{eqnarray}
We can then rewrite (\ref{FTaaaaaa}) as
\begin{eqnarray}
G(t,\lambda)= \int d\xi C(\xi,t) e^{(R(\xi)-\lambda \xi) t}
\;. \label{FTaaaaad}
\end{eqnarray}
At long times, the main contribution to this integral comes
from the value of $\xi$, $\xi^*$, that maximizes
the argument of the exponential.
$\xi^*$ is therefore the value of $\xi$ such that
$\lambda=\frac{dR}{d\xi} \vert_{\xi=\xi^*}$.
At long times, using steepest descent integration,
(\ref{FTaaaaad}) becomes
\begin{eqnarray}
&&\hspace{-0.3cm}G(t,\lambda) \label{FTaaaaadbis}\\
&&\hspace{0.3cm}\approx e^{(R(\xi^*)-\lambda \xi^*) t} \int d\xi
C(\xi,t) e^{- \frac{1}{2} \big\vert \frac{d^2 R(\xi)}{d \xi^2}
\vert_{\xi=\xi^*} \big\vert (\xi-\xi^*)^2 t} \nonumber \\
&&\hspace{0.3cm}\approx e^{(R(\xi^*)-\lambda \xi^*) t} C(\xi^*,t)
\bigg( \bigg\vert \frac{d^2 R(\xi)}{d \xi^2}\vert_{\xi=\xi^*} \bigg\vert
\frac{t}{2\pi} \bigg)^{-\frac{1}{2}} \;.\nonumber
\end{eqnarray}
We assumed $R(\xi)$ concave to have a maximum.
Substituting (\ref{FTaaaaadbis}) in (\ref{FTaaaaae}) gives
\begin{eqnarray}
{\cal S}(\lambda) = R(\xi) - \lambda \xi \;, \label{FTaaaaaebis}
\end{eqnarray}
where
\begin{eqnarray}
\lambda=\frac{dR(\xi)}{d\xi} \;. \label{FTaaaaaetris}
\end{eqnarray}
This shows that ${\cal S}(\lambda)$ is the inverse Legendre transform of the LDF.
By taking the derivative of (\ref{FTaaaaaebis}) with respect to $\lambda$, we get
\begin{eqnarray}
\frac{d {\cal S}(\lambda)}{d \lambda} = \frac{d R(\xi)}{d \xi}
\frac{d \xi}{d \lambda} -\lambda \frac{d \xi}{d \lambda}- \xi \;,
\end{eqnarray}
which using (\ref{FTaaaaaetris}) leads to
\begin{eqnarray}
\xi= -\frac{d {\cal S}(\lambda)}{d \lambda} \;. \label{FTaaaaafbis}
\end{eqnarray}
This shows that the LDF is given by the Legendre transform of ${\cal S}(\lambda)$
\begin{eqnarray}
R(\xi) = {\cal S}(\lambda) + \lambda \xi \label{FTaaaaaf}
\end{eqnarray}
By taking the derivative of (\ref{FTaaaaafbis}) with respect to $\lambda$
and using the derivative of (\ref{FTaaaaaetris}) with respect to $\xi$,
we can confirm that $R(\xi)$ is concave because ${\cal S}(\lambda)$ is convex.\\

We now assume that the cumulant GF satisfies the symmetry
\begin{eqnarray}
{\cal S}(\lambda) = {\cal S}(A-\lambda) \;. \label{Caaaaib}
\end{eqnarray}
We note that the symmetry (\ref{Caaaaib}) with the standard definition
of the moment GF would read ${\cal S}(\lambda) = {\cal S}(\ic A-\lambda)$].
Using the symmetry (\ref{Caaaaib}), Eq. (\ref{FTaaaaaf}) implies that
$R(-\xi) = {\cal S}(A - \lambda) - (A - \lambda) \xi$, so that
\begin{eqnarray}
R(\xi) - R(-\xi) = A \xi \;.
\label{FTaaaaag}
\end{eqnarray}
Using Eq. (\ref{FTaaaaac}), we get
\begin{eqnarray}
\ln \frac{\tilde{p}(t,\xi)}{\tilde{p}(t,-\xi)}
=  A \xi t + \ln \frac{C(\xi,t)}{C(-\xi,t)} \;.
\label{FTaaaaaha}
\end{eqnarray}
Using (\ref{FTaaaaaca}), this gives the steady-state FT
\begin{eqnarray}
\lim_{t \to \infty} \frac{1}{t} \ln \frac{p(t,k)}{p(t,-k)} = A \xi \;,
\label{FTaaaaahbis}
\end{eqnarray}
which is often written as
\begin{eqnarray}
\frac{p(t,k)}{p(t,-k)} \stackrel{t \to \infty}{=} e^{A k}
\;. \label{FTaaaaah}
\end{eqnarray}
Eqs. (\ref{4april-7}) and (\ref{7may-1}) are of this form.

\section{Derivation of the generalized quantum master equation}\label{GFMEweak}

Eq. (\ref{GFtot}) satisfies the equation of motion
\begin{eqnarray}
\label{totalevolution} \dot{\hat{\rho}}(\lambda,t) &=& \shat{L}_{\lambda}
\hat{\rho}(\lambda,t)
= -\frac{i}{\hbar} \big( \hat{H}_{\lambda} \hat{\rho}(\lambda,t) - \hat{\rho}(\lambda,t) \hat{H}_{-\lambda} \big) \\
&=& (\shat{L}_{0} + v \shat{L}'_{\lambda}) \hat{\rho}(\lambda,t) \nonumber \\
&=&-\frac{i}{\hbar} [ \hat{H}_{0}, \hat{\rho}(\lambda,t)] - v
\frac{i}{\hbar} \big( \hat{V}_{\lambda} \hat{\rho}(\lambda,t) -\hat{\rho}(\lambda,t)
\hat{V}_{-\lambda} \big) \nonumber \; ,
\end{eqnarray}
where we multiplied $\hat{V}$ by a scalar $v$ to keep track
of the order in the perturbation expansion below.
Superoperators are denoted by a breve [see appendix \ref{superoperator}].
In the interaction representation where
\begin{eqnarray}
\hat{\rho}_I(\lambda,t) &=& \e^{-\shat{L}_{0} t} \hat{\rho}(\lambda,t)
= \e^{\frac{i}{\hbar} \hat{H}_{0} t} \hat{\rho}(\lambda,t) \e^{-\frac{i}{\hbar} \hat{H}_{0} t} \; , \\
\shat{L}'_{\lambda}(t) &=& \e^{-\shat{L}_{0} t} \shat{L}'_{\lambda}
\e^{\shat{L}_{0} t} \; , \label{interactionB}
\end{eqnarray}
(\ref{totalevolution}) takes the simple form
\begin{eqnarray}
\dot{\hat{\rho}}_I(\lambda,t) = v \shat{L}'_{\lambda}(t)
\hat{\rho}_I(\lambda,t) \;. \label{A1111e}
\end{eqnarray}
By integrating Eq. (\ref{A1111e}) and truncating it to order
$v^2$, we get the perturbative expansion
\begin{eqnarray}
\hat{\rho}_I(\lambda,t)&=&\shat{W}(\lambda,t) \hat{\rho}(0)
=\e^{-\shat{L}_{0} t} \e^{\shat{L}_{\lambda} t} \hat{\rho}(0) \label{A1111f} \\
&=&\left[ \shat{W}_0(\lambda,t) + v \shat{W}_1(\lambda,t) \nonumber \right. \\
&&\hspace{1cm} \left. + v^2 \shat{W}_2(\lambda,t) + {\cal O}(v^3) \right]
\hat{\rho}(0) \; , \nonumber
\end{eqnarray}
where
\begin{eqnarray}
\shat{W}_0(\lambda,t) &=& \shat{1} \; ; \label{A1111g}\\
\; \shat{W}_1(\lambda,t)&=&\int_{0}^{t} dT \, \shat{L}'_{\lambda}(T) \; ; \nonumber\\
\shat{W}_2(\lambda,t)&=&\int_{0}^{t}dT \int_{0}^{T} d\tau \,
\shat{L}'_{\lambda}(T) \shat{L}'_{\lambda}(T-\tau)\; . \nonumber
\end{eqnarray}
The inverse of $\shat{W}(t)$ reads
\begin{eqnarray}
\shat{W}^{-1}(\lambda,t)&=&\shat{W}_0(\lambda,t) -
v \shat{W}_1(\lambda,t) \label{A1111h}\\
&&+ v^2 \left[\shat{W}_1^2(\lambda,t)-\shat{W}_2(\lambda,t)
\right] + {\cal O}(v^3) \; . \nonumber
\end{eqnarray}
Indeed, one can check that $\shat{W}(\lambda,t)
\shat{W}^{-1}(\lambda,t)= \shat{1} + {\cal O}(v^3)$. For later use, we also
notice that
\begin{eqnarray}
\dot{\shat{W}}(\lambda,t) \shat{A} \shat{W}^{-1}(\lambda,t)&=&
v \dot{\shat{W}}_1(\lambda,t) \shat{A} \label{A1111i} \\
&&\hspace{-3cm}+ v^2 \left[\dot{\shat{W}}_2(\lambda,t) \shat{A}
- \dot{\shat{W}}_1(\lambda,t) \shat{A} \shat{W}_{1}(\lambda,t) \right]
+ {\cal O}(v^3) \; . \nonumber
\end{eqnarray}
We define the projection superoperator (acting in reservoir space)
\begin{eqnarray}
\shat{P} &=& \sum_{r} \sket{\rho^{eq}_R} \sbra{rr}
\label{A1111pred}\; ,
\end{eqnarray}
where $\hat{\rho}^{eq}_R$ is the equilibrium density matrix of the reservoir.
We used the Liouville space notation [see appendix \ref{superoperator}].
$\shat{P}$ satisfies the usual properties of projection
superoperators $\shat{P} + \shat{Q} = \shat{1}, \shat{P}^2 = \shat{P},
\shat{Q}^2 = \shat{Q}$ and $\shat{P}\shat{Q} =\shat{Q}\shat{P} = 0$.
When acting on the density matrix $\hat{\rho}(t)$, the projection operator gives
\begin{eqnarray}
\shat{P} \sket{\rho(\lambda,t)} &=& \sket{\rho_S(\lambda,t)} \otimes
\sket{\rho^{eq}_R} \label{A1111pbisB} \; .
\end{eqnarray}

We now let $\shat{P}$ and $\shat{Q}$ act on the density matrix of
the total system in the interaction picture (\ref{A1111f}) and find
\begin{eqnarray}
\shat{P} \sket{\rho_I(\lambda,t)} &=& \shat{P} \shat{W}(t)
(\shat{P}+\shat{Q}) \sket{\rho_I(0)} \label{A1111qa} \\
\shat{Q} \sket{\rho_I(\lambda,t)} &=& \shat{Q} \shat{W}(t)
(\shat{P}+\shat{Q}) \sket{\rho_I(0)} \label{A1111qb} \; .
\end{eqnarray}
Hereafter, we consider initial conditions such that $\shat{Q}
\sket{\rho(0)}=0$. This means that the reservoir part of the
initial condition is diagonal in the reservoir eigenbasis and is
thus invariant under the evolution when $v = 0$. Taking the
time derivative of Eq. (\ref{A1111qa}) and Eq. (\ref{A1111qb}) and
using $\sket{\rho_I(0)}=\shat{W}^{-1}(\lambda,t)
\sket{\rho_I(\lambda,t)}$, we get
\begin{eqnarray}
\shat{P} \sket{\dot{\rho}_I(\lambda,t)} &=& \shat{P} \dot{\shat{W}}
(\lambda,t) \shat{P} \shat{W}^{-1}(\lambda,t)
\shat{P} \sket{\rho_I(\lambda,t)} \label{A1111ra} \\
&&+\shat{P} \dot{\shat{W}}(\lambda,t) \shat{P} \shat{W}^{-1}(\lambda,t)
\shat{Q} \sket{\rho_I(\lambda,t)} \nonumber \\
\shat{Q} \sket{\dot{\rho}_I(\lambda,t)} &=& \shat{Q} \dot{\shat{W}}
(\lambda,t) \shat{P} \shat{W}^{-1}(\lambda,t)
\shat{P} \sket{\rho_I(\lambda,t)} \label{A1111rb} \\
&&+\shat{Q} \dot{\shat{W}}(\lambda,t) \shat{P} \shat{W}^{-1}(\lambda,t)
\shat{Q} \sket{\rho_I(\lambda,t)} \; .\nonumber
\end{eqnarray}
So far these equations are exact. If we restrict ourselves to
second-order perturbation theory in $v$, we can obtain the important result
that the $\shat{P}$ projected density matrix evolution is decoupled
from the $\shat{Q}$ projected part. Indeed, with the help of Eq.
(\ref{A1111i}), we have
\begin{eqnarray}
\shat{P} \dot{\shat{W}}(\lambda,t) \shat{P} \shat{W}^{-1}(\lambda,t)
\shat{Q} &=& v \shat{P} \dot{\shat{W}}_1(\lambda,t) \shat{P}
\shat{Q}
\label{A1111s} \\
&&\hspace{-4cm} +v^2 \shat{P} \dot{\shat{W}}_2(\lambda,t) \shat{P}
\shat{Q} - v^2 \shat{P} \dot{\shat{W}}_1(\lambda,t) \shat{P}
\shat{W}_1(\lambda,t) \shat{Q} + {\cal O}(v^3) \; . \nonumber
\end{eqnarray}
The first two terms of the right-hand side are zero because
$\shat{P} \shat{Q} = 0$ and the third one also because
\begin{eqnarray}
\hspace{-0.2cm} \shat{P} \dot{\shat{W}}_1(\lambda,t) \shat{P}
= \sum_{r,r'} \sket{\rho^{eq}_R} \sbra{rr}
\shat{L}_I'(\lambda,t) \sket{\rho^{eq}_R} \sbra{r'r'} \label{A1111t}
\end{eqnarray}
vanishes since $\hat{\rho}^{eq}_R$ commutes with $\hat{H}_R$.\\

Having shown that the relevant projected density matrix evolves in
an autonomous way, we will now evaluate the generator of its
evolution using second-order perturbation theory. Again using Eq.
(\ref{A1111i}), we find that
\begin{eqnarray}
\shat{P} \dot{\shat{W}}(\lambda,t) \shat{P} \shat{W}^{-1}(\lambda,t)
\shat{P} &=& v \shat{P} \dot{\shat{W}}_1(\lambda,t) \shat{P}
\label{A1111u} \\
&&\hspace{-4cm} +v^2 \shat{P} \dot{\shat{W}}_2(\lambda,t) \shat{P}
- v^2 \shat{P} \dot{\shat{W}}_1(\lambda,t) \shat{P} \shat{W}_1(\lambda,t)
\shat{P} + {\cal O}(v^3) \; . \nonumber
\end{eqnarray}
The only term of right-hand side which is not zero is the second one
[see Eq. (\ref{A1111t})] whereupon we get
\begin{eqnarray}
\shat{P} \sket{\dot{\rho}_I(\lambda,t)} &=& v^2 \shat{P}
\int_{0}^{t} d\tau \shat{L}'_{\lambda}(t) \shat{L}'_{\lambda}(t-\tau)
\shat{P} \sket{\rho_I(\lambda,t)} \nonumber \\
&&+ {\cal O}(v^3) \; .\label{A1111v}
\end{eqnarray}
Now leaving the interaction representation and using the fact that
$\shat{P} \e^{-\shat{L}_0 t} = \e^{-\shat{L}_S t} \shat{P}$, we
obtain
\begin{eqnarray}
\shat{P} \sket{\dot{\rho}(\lambda,t)} &=&
\shat{L}_S \shat{P} \sket{\rho(\lambda,t)} \label{A1111w} \\
&&\hspace{-2cm}+ v^2 \e^{\shat{L}_S t} \shat{P} \int_{0}^{t}
d\tau \shat{L}'_{\lambda}(t) \shat{L}'_{\lambda}(t-\tau)
\e^{-\shat{L}_S t} \shat{P} \sket{\rho(\lambda,t)} \nonumber \; .\nonumber
\end{eqnarray}
By taking the trace of (\ref{A1111w}) we get
\begin{eqnarray}
\dot{\hat{\rho}}_S(\lambda,t) &=& \shat{L}_S \hat{\rho}_S(\lambda,t)
+ v^2 \sum_{r} \int_{0}^{t} d\tau \label{A1111y} \\
&&\hspace{-1cm}  \times \; \e^{\shat{L}_S t} \sbra{rr}
\shat{L}_{\lambda}'(t) \shat{L}_{\lambda}'(t-\tau) \sket{\rho^{eq}_R}
\e^{-\shat{L}_S t}\hat{\rho}_S(\lambda,t)  \; . \nonumber
\end{eqnarray}
Explicit evaluation leads to Eq. (\ref{GFevolofRedGen}).

\section{Bidirectional Poisson statistics}\label{bidirpoisson}

The GF of section \ref{IsolTunnelJunct} corresponds to
a bidirectional Poisson process. We give here some basic
properties of this process.\\

The GF
of the probability distribution $p(k)$
can be expanded in terms of moments $\mean{k^{n}}$
as
\begin{eqnarray}
G(\lambda) = \sum_k \e^{\ic \lambda k} p(k)
= \sum_{n=1}^{\infty} \mean{k^{n}} \frac{(\ic \lambda)^n}{n!}
\;.
\end{eqnarray}
The Poisson distribution and its GF are given by
\begin{eqnarray}
p(k)= \frac{\mu^k \e^{-\mu}}{k!} \ \ \;, \ \
G(\lambda)=\exp{\{\mu (\e^{\ic \lambda}-1)\}} \;.
\end{eqnarray}
Note that $\mu=\mean{k}$.
If $k=k_1-k_2$ where $p(k_1,k_2) = p_1(k_1) p_1(k_2)$
and $p_1(k)$ and $p_2(k)$ are Poissonian, we get that
\begin{eqnarray}
G(\lambda) &=& G_1(\lambda_1=\lambda)G_2(\lambda_2=-\lambda) \nonumber\\
&=& \exp{\{\mu_1 (\e^{\ic \lambda}-1) + \mu_2 (\e^{-\ic \lambda}-1)\}}
\;. \label{GFdiffpoisson}
\end{eqnarray}
If the average of the positive process is related to the
average of the negative one by $\mu_1=\mu_2 \exp(-A)$, we find
that the GF displays the FT symmetry $G(\lambda)=G(A-\lambda)$.
By inverting Eq. (\ref{GFdiffpoisson}), we get
\begin{eqnarray}
p(k)=\e^{-(\mu_1+\mu_2)} \e^{A k/2} I_k
\left[-\frac{(\mu_1-\mu_2)}{\sinh (A/2)}\right] \;,
\end{eqnarray}
where $I_k$ is the modified Bessel function of order $k$.

\section{Liouville space and superoperator algebra}\label{superoperator}

In Liouville space, a $N \times N$ Hilbert space operators $\hat{\rho}$ is mapped
into a $N^2$ vector $\sket{\rho}$ and a superoperator $\shat{A}$ (linear map)
acting on an operator $\hat{\rho}$ becomes a $N^2 \times N^2$ matrix acting
on the vector $\sket{\rho}$:
$\shat{A} \hat{\rho} \leftrightarrow \shat{A} \sket{\rho}$ \cite{fano,zwanzig,reuven,up-shaul-JCP,HarbolaPhysRep08}.
We recall some basic definitions
\begin{eqnarray}
&&{\rm scalar \; product:} \ \ \sbraket{A}{B} \equiv \trace \hat{A}^{\dagger} \hat{B} \; ,
\label{A1111k}\\
&&{\rm identity:} \ \ \shat{1} \equiv \sum_{n,n'} \sket{nn'} \sbra{nn'} \; ,
\label{A1111l}\\
&&\hspace*{0cm} \sket{nn'} \leftrightarrow \ket{n} \bra{n'} \; , \;
\sbra{nn'} \leftrightarrow \ket{n'} \bra{n} \; .
\label{A1111m}
\end{eqnarray}
Useful consequences of these definitions are
\begin{eqnarray}
\sbraket{nn'}{\bar{n}\bar{n}'}&=&\delta_{n\bar{n}} \delta_{n'\bar{n}'}
\label{A1111n}\\
\sbraket{nn'}{A}&=&\bra{n} A \ket{n'} \label{A1111o} \\
\sbraket{1}{A}&=& \trace \hat{A} \label{19may-1}
\end{eqnarray}
We define left and right Liouville space operators as
\begin{eqnarray}
\label{liouv-operator}
\breve{A}_L |X\gg \leftrightarrow \hat{A}\hat{X},~~~
\breve{A}_R |X\gg \leftrightarrow \hat{X} \hat{A} \;.
\end{eqnarray}
We also define
\begin{eqnarray}
\label{trans}
\breve{A}_+ \equiv \frac{1}{\sqrt{2}}(\breve{A}_L+\breve{A}_R),~~~
\breve{A}_- \equiv \frac{1}{\sqrt{2}}(\breve{A}_L-\breve{A}_R).
\end{eqnarray}
This linear transformation is symmetric.
The inverse transformation can be obtained by simply
interchanging $+$ and $-$ with $L$ and $R$, respectively.
Thus most of the expressions in the following are symmetric
and the indices used to represent superoperators can take
both $+,-$ and $L,R$ values without any other change.
The advantage of the $+,-$ representation is that a single operation
$A_-$ in Liouville space represents the commutation with $A$ in
Hilbert space. Thus all the intertwined commutations, that appear in
perturbation expansions in Hilbert space transform to a compact
notation that is more easy to interpret in terms of the double sided
Fynmann diagrams \cite{MukamelB}. Similarly a single operation of
$A_+$ in Liouville space corresponds to an anticommutator in
Hilbert space.
\begin{eqnarray}
\label{transform}
\breve{A}_- |X\gg &\leftrightarrow& \frac{1}{\sqrt{2}}(\hat{A}\hat{X}-\hat{X}\hat{A})\\
\breve{A}_+ |X\gg &\leftrightarrow& \frac{1}{\sqrt{2}}(\hat{A}\hat{X}+\hat{X}\hat{A})
\end{eqnarray}

For any product of operators in Hilbert space, we can define
corresponding superoperators in Liouville space using the following
identities.
\begin{eqnarray}
\label{super-identity}
(\hat{A}_i\hat{A}_j\cdots \hat{A}_k)_L&=&
\breve{A}_{iL}\breve{A}_{jL}\cdots\breve{A}_{kL}\nonumber\\
(\hat{A}_i\hat{A}_j\cdots \hat{A}_k)_R&=&\breve{A}_{kR}\cdots
\breve{A}_{jR}\breve{A}_{iR}.
\end{eqnarray}
Applying this immediately gives,
\begin{eqnarray}
(\hat{A}_i\hat{A}_j)_- &=&
\frac{1}{2\sqrt{2}}\left[[\breve{A}_{i+},\breve{A}_{j+}]+
[\breve{A}_{i-},\breve{A}_{j-}] \right.\nonumber\\
&+&\left. \{\breve{A}_{i+},\breve{A}_{j-}\}
+\{\breve{A}_{i-},\breve{A}_{j+}\}
\right]\label{pmm} \\
(\hat{A}_i\hat{A}_j)_+ &=&
\frac{1}{2\sqrt{2}}\left[\{\breve{A}_{i+},\breve{A}_{j+}\}+
\{\breve{A}_{i-},\breve{A}_{j-}\} \right.\nonumber\\
&+&\left. [\breve{A}_{i+},\breve{A}_{j-}]
+[\breve{A}_{i-},\breve{A}_{j+}]\right]\label{pmmm}.
\end{eqnarray}
Equations (\ref{super-identity})-(\ref{pmmm}) are useful for recasting
functions of Hilbert space operators, such as Hamiltonian, in terms
of the superoperators in Liouville space.

Another useful quantity in Liouville space is the time ordering
operator $\breve{T}$; when acting on a product of superoperators
(each at different times), it rearranges them in increasing order of
time from right to left.
\begin{eqnarray}
\label{time-ordering}
\breve{T}\breve{A}_{i\alpha}(t)\breve{A}_{j\beta}(\tp)
&=& \breve{A}_{j\beta}(\tp)\breve{A}_{i\alpha}(t),~~~~~~ t < \tp \nonumber\\
&=& \breve{A}_{i\alpha}(t)\breve{A}_{j\beta}(\tp),~~~~~~  \tp < t.
\end{eqnarray}
where $\alpha,\beta=L,R,+,-$. Note that, unlike the Hilbert space
where we have two time ordering operators describing the evolution
in opposite (forward and backward) directions, a Liouville space
operator $\breve{T}$ always acts to its right and therefore all
processes are given in terms of forward times alone. This makes it
easier to give physical interpretation to various algebraic
expressions commonly obtained in perturbation expansions which can
be converted readily in terms of different Liouville space diagrams.

We finally note that using (\ref{19may-1}) and (\ref{liouv-operator})
we get for $\alpha=L,R$ that
\begin{eqnarray}
\label{19may-2}
\ll I|\breve{A}_\alpha |\rho\gg = \langle \hat{A} \rangle =
\mbox{Tr} \{\hat{A}\hat{\rho}\}
\end{eqnarray}
and using (\ref{liouv-operator}), (\ref{trans}) and (\ref{19may-2}), we get
\begin{eqnarray}
\label{liu-hil}
 \ll I |\breve{A}_-|\rho\gg  = 0 ~~,~~
\ll I |\breve{A}_+|\rho\gg = \sqrt{2} \langle \hat{A}\rangle  \;.
\end{eqnarray}

\section{Probability distribution for electron transfers}
\label{probdistrib}

In the model considered in Sec. \ref{GeneralFormulation},
we consider electron transfer between system $A$ and $B$.
We measure the number of electron in system $A$ at time $0$ and time $t$.
The number operator for system $A$ is defined as
$\hat{N}= \sum_{i\in A} \hat{c}_{i}^{\dagger} \hat{c}_{i}$,
where $\hat{c}^\dag(\hat{c})$ are creation (annihilation) operators.
Only the coupling $\hat{V}$ can induce electron transfer:
$[\hat{H}_{A}+\hat{H}_{B}, \hat{N}] = 0$

The total density matrix follows a unitary dynamics in
Liouville space
\begin{eqnarray}
\sket{\rho(t)} = \shat{{U}}(t,0) \sket{\rho(0)}
= \shat{{U}}_L(t,0) \shat{{U}}_R^{\dagger}(t,0) \sket{\rho(0)} \;,
\label{aaaaaa}
\end{eqnarray}
where
\begin{eqnarray}
\label{aaaaab}
\shat{{U}}(t,0) &=& \exp{\{- \ic \sqrt{2} \shat{{H}}_{-}t\}}
\end{eqnarray}
with $\shat{{H}}_{-}$ is the superoperator corresponding to the
total Hamiltonian, $\sqrt{2}\shat{{H}}_{-}=\shat{{H}}_{L}-\shat{{H}}_{R}$ and
\begin{eqnarray}
\shat{{U}}_{\alpha}(t,0) &=& \exp{\{- \ic \sqrt{2}\shat{{H}}_{\alpha}t\}},
~~~~~\alpha= L,R \label{aaaaac} \;.
\end{eqnarray}
By measuring the number of electrons in $A$, when the system right
before the measurement is described by $\sket{\rho(0)}$, we get the
outcome $n$ with a probability $\sbra{I} \shat{P}_n
\sket{\rho(0)}$ and the density matrix of the system after the
measurement becomes $\shat{P}_n  \sket{\rho(0)}$, where the
projection operator in Liouville space is defined as
\begin{eqnarray}
\shat{P}_n &=& \delta_K(n-\shat{{N}}_{L})
\delta_K(n-\shat{{N}}_{R}) \label{prob-basic} \\
&=&\int_0^{2\pi} \frac{d\lambda d\lambda'}{(2\pi)^2} \e^{-\ic
\lambda(n-\shat{{N}}_{L})} \e^{-\ic\lambda'(n-\shat{{N}}_{R})} \nonumber \;.
\end{eqnarray}
$\delta_K$ is the Kronecker delta and $\shat{{N}}_{\alpha}$ are
the left and right superoperators corresponding to the number
operator in $A$. We have $\shat{{P}}_n \shat{{P}}_{n'}= \delta_K(n-n') \shat{P}_n$
and
\begin{eqnarray}
\label{identity} \exp{\{\ic \lambda \shat{{N}}_{\alpha}\}}
\shat{{P}}_n = \exp{\{\ic \lambda n\}} \shat{{P}}_n\;.
\end{eqnarray}

The net number of electrons $k$ transferred between $A$ and $B$
during time $t$ is a fluctuating quantity. The probability for measuring
$k$ electrons during this time interval is given by
\begin{eqnarray}
p(k,t)= \sum_n \sbra{I} \shat{{P}}_{n-k} \shat{{U}}(t,0)
\shat{{P}}_n\sket{\rho(0)}\;. \label{prob-1}
\end{eqnarray}

Substituting (\ref{aaaaaa}) and (\ref{prob-basic}) in
(\ref{prob-1}) and using (\ref{identity}) with the fact that left and
right superoperators commute, we get
\begin{eqnarray}
\label{prob-2bisbis}
p(k,t) &=& \int_0^{2\pi} \frac{d\lambda_1
d\lambda_2}{(2\pi)^2}
\e^{\ic (\lambda_1 + \lambda_2) k} \\
&&\hspace{-1.5cm} \sbra{I} \e^{\ic \lambda_1 \shat{{N}}_{L}}
\shat{{U}}_L(t,0) \e^{-\ic \lambda_1 \shat{{N}}_{L}}
\nonumber \\ &&\hspace{0.5cm} \times \e^{\ic \lambda_2 \shat{{N}}_{R}}
\shat{{U}}_{R}^{\dagger}(t,0) \e^{-\ic \lambda_2
\shat{{N}}_{R}} \sket{\rho(0)} \;.\nonumber
\end{eqnarray}
Making the change of variables, $\lambda_1=-\Lambda-\lambda/2$ and
$\lambda_2=\Lambda-\lambda/2$, we get
\begin{eqnarray}
\label{prob-3} p(k,t) = \int_0^{2\pi} \frac{d\lambda}{2\pi} \e^{- \ic
\lambda k} G(\lambda,t) \;,
\end{eqnarray}
where the GF reads
\begin{eqnarray}
\label{prob-3bis} G(\lambda,t) = \int_0^{2\pi}
\frac{d\Lambda}{2\pi} G(\lambda,\Lambda,t)
\end{eqnarray}
and
\begin{eqnarray}
G(\lambda,\Lambda,t) &=& \sbra{I} \e^{-\ic (
\Lambda+\lambda_X/2) \shat{{N}}_L} \shat{{U}}_L(t,0)
\e^{\ic ( \Lambda+\lambda_X/2) \shat{{N}}_L} \nonumber\\
&&\hspace{-1.8cm}  \times \e^{\ic ( \Lambda-\lambda_X/2) \shat{{N}}_R}
\shat{{U}}_R^{\dagger}(t,0) \e^{-\ic (
\Lambda-\lambda_X/2) \shat{{N}}_R} \sket{\rho(0)} \;.
\label{prob-3tris}
\end{eqnarray}
Equation (\ref{prob-3bis}) is identical to the trace of $\hat{\rho}(\lambda,t)$
defined in (\ref{Aaaacx2BIS}).

The density matrix right before the first measurement ($t=0$ can be constructed
by switching the interaction $V$ adiabatically
from the remote past, $t\to-\infty$. This gives
\begin{eqnarray}
\label{GFcomplete}
G(\lambda,\Lambda,t) = \sbra{I}\shat{{U}}_0(t,0)\shat{{U}}_I(\gamma(t),t,-\infty)\sket{\rho(-\infty)}\;,
\end{eqnarray}
where
\begin{eqnarray}
\label{prob-3quad}
\shat{{U}}_I(\gamma(t),t,-\infty)&=& \exp_+{\left\{- \ic
\int_{-\infty}^{t}d\tau \sqrt{2}\shat{{V}}_{-}(\gamma(\tau),\tau)
\right\}}\nonumber\\
\shat{{U}}_0(t,0) &=& \theta(t)\exp\{-\ic\sqrt{2}({H}_{0-}t)\}.
\end{eqnarray}
We define
\begin{eqnarray}
\label{prob-3quadbis}
\sqrt{2}\shat{{V}}_{-}(\gamma(\tau),\tau) =
\shat{{V}}_{L}(\gamma_L(\tau),\tau) - \shat{{V}}_{R}(\gamma_R(\tau),\tau)
\end{eqnarray}
where
\begin{eqnarray}
&&\hspace{-0.8cm} \shat{{V}}_L(\gamma_L(\tau),\tau)
=\e^{-\ic \gamma_L(\tau) \shat{{N}}_L} \left(\shat{{V}}_L(\tau)\right)
\e^{\ic \gamma_L(\tau) \shat{{N}}_L} \nonumber\\
&&\hspace{-0.8cm} \shat{{V}}_R(\gamma_R(\tau),\tau)
=\e^{\ic \gamma_R(\tau) \shat{{N}}_R} \left(\shat{{V}}_R(\tau)\right)
\e^{-\ic \gamma_R(\tau) \shat{{N}}_R}\label{new-1}
\end{eqnarray}
with $\shat{{V}}_\alpha=\shat{{J}}_\alpha+\shat{{J}}_\alpha^\dag$ and
\begin{eqnarray}
\label{glr-1}
\gamma_L(t)&=&\theta(t)(\Lambda+\lambda/2)\nonumber\\
\gamma_R(t)&=&\theta(t)(\Lambda-\lambda/2).
\end{eqnarray}
The time dependence of
operators in (\ref{prob-3quad}) is in the interaction picture
with respect to $\hat{H}_0$.
\begin{eqnarray}
\shat{{V}}_\alpha(t)= e^{\ic\sqrt{2}\breve{H}_{0-} t}
\shat{{V}}_\alpha e^{-\ic\sqrt{2} \breve{H}_{0-} t}.
\end{eqnarray}

Equation (\ref{GFcomplete}) is the GF used in Eq. (\ref{chi-y}).

\section{Path-integral evaluation of the generating function for fermion transport}
\label{sec-path-int}

The fermion coherent states $|\psi\rangle$ are defined through the
eigenvalue equation for the Fermi destruction operators,
$\hat{c}_x|\psi_i\rangle=\psi_{xi}|\psi\rangle$ and $\langle
\psi_i|\hat{c}^\dag_x=\langle \psi|\bar{\psi}_{xi}$, where $\psi$ and
$\bar{\psi}$ are independent Grassmann variables (see Appendix \ref{Grassmann}) which satisfy
anticommutation relations similar to the Fermi operators \cite{negle}.

It is convenient to introduce coherent states in Liouville space
corresponding to the superoperator $\breve{c}_{x\alpha}$, $x=a,b,s$, as
\begin{eqnarray}
\label{coherent-state}
\breve{c}_{xL}|\psi\rangle\rangle = \psi_{xL}|\psi\rangle\rangle\nonumber\\
\breve{c}_{xR}^\dag |\psi\rangle\rangle =
\psi_{xR}|\psi\rangle\rangle
\end{eqnarray}
The state $|\psi\rangle\rangle$ can be expressed in terms of the vacuum state
\begin{eqnarray}
\label{oct-81}
|\psi\rangle\rangle = \mbox{e}^{\sum_x(-\psi_{xL}c_{xL}^\dag-\psi_{xR}c_{xR})}|0\rangle\rangle
\end{eqnarray}
and
\begin{eqnarray}
\label{oct-82}
\langle\langle\psi| = \langle\langle 0| \mbox{e}^{{\sum_x(\bar{\psi}_{xL}c_{xL}-\bar{\psi}_{xR}c^\dag_{xR})}}.
\end{eqnarray}
Note that $c_R^\dag$ is {\em not} the hermitian conjugate of $c_R$ \cite{HarbolaPhysRep08}.
Grassmann variables $\psi_\alpha$ and $\bar{\psi}_\beta$ anticommute between themselves and with the creation and annihilation operators. Note that, unlike usual fermion case, we now have four generators for the Grassmann algebra, two corresponding to each index $\alpha$. Using (\ref{coherent-state})-(\ref{oct-82}), it can be shown that
\begin{eqnarray}
\langle\langle\psi^\prime|c_L^\dag c_R|\psi\rangle\rangle &=&
\bar{\psi}_R^\prime \bar{\psi}^\prime_L\langle\langle\psi^\prime|\psi\rangle\rangle\label{oct-83-1}\\
\langle\langle\psi^\prime|c_R^\dag c_L|\psi\rangle\rangle &=&
\psi_R\psi_L \langle\langle\psi^\prime|\psi\rangle\rangle\label{oct-83-2}\\
\langle\langle\psi^\prime|c_L^\dag c_L|\psi\rangle\rangle &=&
\bar{\psi}_L^\prime\psi_L \langle\langle\psi^\prime|\psi\rangle\rangle\label{oct-83-3}\\
\langle\langle\psi^\prime|c_R c_R^\dag|\psi\rangle\rangle &=&
{\psi}_R \bar{\psi}_R^\prime\langle\langle\psi^\prime|\psi\rangle\rangle\label{oct-83-4}.
\end{eqnarray}
These matrix elements will be useful in the path-integral formulation below.
The scaler product of two coherent states is
\begin{eqnarray}
 \langle\langle\psi^\prime|\psi\rangle\rangle=\mbox{e}^{\sum_\alpha\bar{\psi}_\alpha^\prime\psi_\alpha}.
\end{eqnarray}

Grassmann variables satisfy the closure relation,
\begin{eqnarray}
\label{closure} 1=\int {\cal D}(\bar{\psi}\psi)
\mbox{e}^{-\sum_{\alpha}\bar{\psi}_{ix\alpha}\psi_{ix\alpha}}|\psi_i\rangle\rangle\langle\langle\psi_i|
\end{eqnarray}
where ${\cal D}(\bar{\psi}\psi)=\Pi_{i,x,\alpha}(d\bar{\psi}_{ix\alpha})(d\psi_{ix\alpha})$.

We next switch to $+,-$ notation\cite{up-shaul-JCP}, represented by the index
$\nu=+,-$. Superoperators $H_{0-}$ and ${\cal V}_{-}(\gamma)$ are
\begin{eqnarray}
\label{pm}
\sqrt{2}H_{0-}&=&\sum_{x\nu} \epsilon_x(\breve{c}_{x+}^\dag \breve{c}_{x+}-\breve{c}_{x-} \breve{c}_{x-}^\dag) 
\nonumber\\
\sqrt{2}{\cal V}_{-}(\gamma(t),t)&=& \sum_{x\neq\xp}
\sum_{\nu\nu^\prime}J_{x\xp}^{\nu\nu^\prime}
(\gamma(t))\breve{c}_{x\nu}^\dag(t)\breve{c}_{x\nu^\prime}(t)
\end{eqnarray}
where $J_{x\xp}^{\nu\nu^\prime}(\gamma)(=J_{\xp x}^{\nu\nu^\prime\dag})$ is $2\times 2$ matrices for $\nu,\nu^\prime=+,-$ with elements
\begin{eqnarray}
J_{x\xp}^{++}(\gamma)&=& J_{x\xp}^{--}(\gamma)
= J_{x\xp}(\mbox{e}^{i\gamma_L}+\mbox{e}^{i\gamma_R})/2\nonumber\\
J_{x\xp}^{+-}(\gamma)&=& J_{x\xp}^{-+}(\gamma)
=J_{x\xp}(\mbox{e}^{i\gamma_L}-\mbox{e}^{i\gamma_R})/2
\end{eqnarray}
while for $x,\xp=b,s$, $J_{x\xp}^{++}(\gamma) = J_{x\xp}^{--}(\gamma)= J_{x\xp}$ and
$J_{x\xp}^{+-}(\gamma)=$ $J_{x\xp}^{-+}(\gamma)= 0$.

We will encounter the matrix element of an exponential operator of the type
\begin{eqnarray}
 \langle\langle\psi^\prime|\mbox{e}^{\sqrt{2}\breve{H}_{0-}}|\psi\rangle\rangle&=&
\langle\langle\psi^\prime|\mbox{e}^{\breve{H}_{0L}-\breve{H}_{0R}}|\psi\rangle\rangle\nonumber\\
&=&\langle\langle\psi^\prime|\mbox{e}^{\sum_x\epsilon_x(\breve{c}_{xL}^\dag\breve{c}_{xL}-
\breve{c}_{xR}\breve{c}_{xR}^\dag)}|\psi\rangle\rangle\nonumber\\
&=& \mbox{e}^{\sum_x\epsilon_x(\bar{\psi}^\prime_{xL}{\psi}_{xL}+\bar{\psi}^\prime_{xR}\psi_{xR})}
\langle\langle\psi^\prime|\psi\rangle\rangle\nonumber
\end{eqnarray}
where in going from second to the third line we used (\ref{oct-83-3}) and (\ref{oct-83-4}). We can now make the linear transformation
from $L/R$ variables to the $+/-$ variables. In Hilbert space this corresponds to the Keldysh rotation \cite{kamenev}. Using this transformation we can
write above matrix element as 
\begin{eqnarray}
 \label{matrix-element-1}
 \langle\langle\psi^\prime|\mbox{e}^{\sqrt{2}\breve{H}_{0-}}|\psi\rangle\rangle&=& 
\mbox{e}^{\sum_{x\nu\nu^\prime}\epsilon_x^{\nu\nu^\prime}\bar{\psi}^\prime_{x\nu}{\psi}_{x\nu^\prime}}
\langle\langle\psi^\prime|\psi\rangle\rangle
\end{eqnarray}
where $\epsilon_x^{++}=\epsilon^{--}_x=\epsilon_x$ and $\epsilon^{+-}_x=\epsilon^{-+}_x=0$. This matrix element (\ref{matrix-element-1}) can also 
be obtained directly by formally defining the Grassmann variables corresponding to $+,-$ operators, $\breve{c}_{x\nu}$
and $\breve{c}_{x\nu}^\dag$, by $\psi_{x\nu}$ and $\bar{\psi}_{x\nu}$, respectively, and using (\ref{pm}). We shall use the $+/-$
formulation in the rest of the section. The advantage of using this notation is that we directly work with the retarded and advanced functions
which are naturally linked to the observables (when $\lambda=0$).

We can express the trace in Eq. (\ref{GFcomplete}) in terms
of the coherent states basis,
\begin{eqnarray}
&&\label{trace} G(\lambda,\Lambda,t)= \nonumber\\
&&\int{\cal
D}(\bar{\psi}\psi)\mbox{e}^{-\bar{\psi}\psi}\langle\langle \psi|
\shat{{U}}_0(t,0)\shat{{U}}_I(\gamma(t),t,-\infty) | \rho(-\infty)\rangle\rangle.\nonumber\\
\end{eqnarray}

We next divide the time from $0$ to $t$ in Eq. (\ref{trace}) into $N$
equal segments of length $\delta t$ and introduce the closure
relation (\ref{closure}) after each time interval. We then get,
\begin{eqnarray}
\label{trace-1}
&&G(\lambda,\Lambda,t)=\nonumber\\
&&\int{\cal D}(\bar{\psi}\psi)
\langle\langle\psi_0|\rho(-\infty)\rangle\rangle
\langle\langle\psi_1|\shat{{U}}_I(\gamma(t),t,-\infty)|\psi_0\rangle\rangle\nonumber\\
&&\prod_{i=2}^N \langle\langle \psi_i|\shat{{U}}_0(\delta
t_i)|\psi_{i-1}\rangle\rangle.
\end{eqnarray}
Here the index $i$ on $\psi_i$ carries time index so that
$\psi_{i+1}$ is at $\delta t$ time ahead of $\psi_i$.

$\breve{U}_I$ can be formally evaluated by dividing the time interval from
the initial time $-t_0$ (at the end we can put $t_0\to \infty$ ) to
$t$ in $N^\prime$ number of equal time steps. We then get
\begin{eqnarray}
&&\langle\langle\psi_1|\shat{{U}}_I(\gamma(t),t,t_0)|\psi_0\rangle\rangle
= \\&&\hspace{1cm} \langle\langle\psi_1|e^{-i\sum_i^{N^\prime}\sqrt{2}
\shat{{V}}_{-}(\gamma(t_i),t_i){\delta t}_i}|\psi_0\rangle\rangle.
\nonumber
\end{eqnarray}
Here $\delta t > 0$ is small enough so that only the linear order
term contributes. The exponential can then be factorized into
products of exponentials. By inserting the identity between
exponentials, we obtain (repeated indices are summed over),
\begin{eqnarray}
\label{new-eq-1} &&\langle\langle\psi_{i+1}|e^{-\ic\sqrt{2}\shat{{\cal
V}}_{-}(i\gamma(t_i),t_i)\bar{\delta t}_i}|\psi_i\rangle\rangle \\
&&\hspace{2.4cm}\approx e^{-iV_{x\xp}^{\nu\nu^\prime}(\gamma(t_i))
\bar{\psi}_{ix\nu}\psi_{i\xp\nu^\prime}{\delta t}_i}
\langle\langle\psi_{i+1}|\psi_i\rangle\rangle. \nonumber
\end{eqnarray}

The second matrix element of the evolution operator
$\shat{{U}}_0(t)$ between two coherent states is
\begin{eqnarray}
\label{matrix-element} \langle\langle \psi_{i+1}
|\shat{{U}}_0(\delta t_i) |\psi_i\rangle\rangle\approx
\mbox{e}^{-i\epsilon_x^{\nu\nu^\prime} \bar{\psi}_{ix\nu}\psi_{i x\nu^\prime}{\delta t}_i}
\langle\langle \psi_{i+1}|\psi_i\rangle\rangle.
\end{eqnarray}

Using Eqs. (\ref{new-eq-1}) and (\ref{matrix-element}) in Eq.
(\ref{trace}), we obtain for the GF,
\begin{eqnarray}
\label{gf-2} &&G(\lambda,\Lambda,t)= \int {\cal
D}[\bar{\psi}\psi]\langle\langle \psi_{0}|\rho(-\infty)\rangle\rangle
\prod_{i=1}^{M=N+N^\prime}\nonumber\\
&&\mbox{exp}\left\{i\bar{\psi}_{ix\nu}
\left(i\frac{\psi_{ix\nu}-\psi_{i-1x\nu}}{\delta
t_i}-\epsilon_x\psi_{ix\nu}\right){\delta t}_i\right\}\nonumber\\
&&\mbox{exp}\left\{-i\bar{\psi}_{ix\nu}V_{x\xp}^{\nu\nu^\prime}(\gamma(t_i))
\psi_{i\xp\nu^\prime}{\delta t}_i\right\}.
\end{eqnarray}
Here $\gamma(t_i)=0$ for $i< M-N$.

Setting $M\to\infty$, $t_0\to\infty$ and $\delta t_i\to 0$, we get
\begin{eqnarray}
\label{gf-3} &&G(\lambda,\Lambda,t)= \int {\cal D}[\bar{\psi}\psi]
e^{iS(\bar{\psi},\psi)}
\end{eqnarray}
where in the continuous time notation ${\cal D}[\bar{\psi}\psi]\equiv \prod_\tau d\bar{\psi}(\tau)
d\psi(\tau)$ and the action $S(\bar{\psi},\psi)$ is defined as
\begin{eqnarray}
\label{action} S(\bar{\psi},\psi)&=&\int d\tau
\left(\frac{}{}\bar{\psi}_{x\nu}(\tau){g_{x\xp}^{\nu\nu^\prime}}^{-1}(\tau)
\psi_{\xp\nu^\prime}(\tau) \right.\nonumber\\
&-&\left.\bar{\psi}_{x\nu}(\tau)V_{x\xp}^{\nu\nu^\prime}
(\gamma(\tau))\psi_{\xp\nu^\prime}(\tau)\frac{}{}\right).
\end{eqnarray}
$g_{x\xp}^{\nu\nu^\prime}$ is a $2\times 2$ matrix corresponding to
$\nu,\nu^\prime=+,-$ which satisfies
\begin{eqnarray}
\left(i\frac{\partial}{\partial
t}-\epsilon_x\right)g_{x\xp}^{\nu\nu^\prime}(t,\tp)=
\delta(t-\tp)\delta_{x,\xp}\delta_{\nu\nu^\prime}.
\end{eqnarray}

Using the integral identity for independent Grassmann variables
$\bar{\eta},\eta, \bar{\kappa}$ and $\kappa$
\begin{eqnarray}
\label{integral-identity} \int {\cal
D}(\bar{\eta},\eta)e^{-\bar{\eta}_iA_{ij}\eta_j}e^{{\bar
\kappa}_i\eta_i+\bar{\eta}_i\kappa_i} =Det[A]e^{{\bar
\kappa}_i[A]^{-1}_{ij}\kappa_j}
\end{eqnarray}
we can trace out the leads' degrees of freedom to obtain,
\begin{eqnarray}
\label{traced-gf} G(\lambda,\Lambda,t) = \int {\cal
D}[\bar{\psi}\psi]e^{iS(\bar{\psi}\psi)}
\end{eqnarray}
with
\begin{eqnarray}
S(\bar{\psi}\psi)&=&\int d\tau d\tau^\prime \bar{\psi}_{s\nu}(\tau)
\left[{g_{s\sp}^{\nu\nu^\prime}}^{-1}(\tau,\tau^\prime)\right.\nonumber\\
&&\quad\quad\quad-\left.\Sigma_{s\sp}^{\nu\nu^\prime}(\tau,\tau^\prime,\gamma)\right]
\psi_{\sp\nu^\prime}(\tau^\prime).
\end{eqnarray}
The self energy $\Sigma(\gamma)$ is
\begin{eqnarray}
&&\label{se} \Sigma_{s\sp}^{\nu\nu^\prime}(t,\tp,\gamma)=  \\
&&\hspace{0.5cm}\sum_{xx^\prime\in A,B}\sum_{\nu_1\nu_2}V_{sx}^{\nu\nu_1}
(\gamma(t))g_{xx^\prime}^{\nu_1\nu_2}(t,\tp)
V_{x^\prime\sp}^{\nu_2,\nu^\prime}(\gamma(\tp)) \nonumber,
\end{eqnarray}
where repeated arguments are summed over and $g_{xx^\prime}$ are the
Greens functions for the non-interacting leads. The counting
parameter appears in the self-energy only through coupling terms
$V_{xs}^{\nu\nu^\prime}(\gamma)$. Finally, using Eq. (\ref{integral-identity}) we can
perform the Gaussian integral in Eq. (\ref{traced-gf}) to obtain,
\begin{eqnarray}
\label{gf-final} G(\lambda,\Lambda,t) = \exp[{\cal Z}(\lambda,\Lambda,t)]
\end{eqnarray}
where
\begin{eqnarray}
{\cal Z}(\lambda,\Lambda,t)=\int_{-\infty}^t d\tau
\mbox{ln}\mbox{Det}[g^{-1}(\tau=0) -\Sigma(\tau,\tau,\gamma(\tau))].
\nonumber\\
\end{eqnarray}
Here $g(\tau,\tau^\prime)$ and $\Sigma(\tau,\tau^\prime,\gamma(\tau))$
are matrices in $+,-$ superoperator indices and defined in the system space.
This result for the GF was used in (\ref{gf-final-10-a}).

\section{Grassmann Algebra}\label{Grassmann}

Here we briefly review come properties of the Grassmann algebra used in Appendix
\ref{sec-path-int}. Fermion coherent states $|\eta\rangle$ are defined in terms of the
vacuum state $|0\rangle$ \cite{negle}.
\begin{eqnarray}
\label{1}
|\eta\rangle&=&e^{c^\dag \eta}|0\rangle=|0\rangle+c^\dag \eta|0\rangle\\
\langle \eta |&=&\langle 0|e^{\eta^*c}=\langle 0|+\langle0|\eta^* c
\end{eqnarray}
where $\eta$ and $\eta^*$ are two independent complex numbers. Here
we consider a single degree of freedom. This can be generalized
easily for several degrees of freedom for which,
$|\eta\rangle=e^{\sum_i\hat{c}_i^\dag \eta_i}|0\rangle$.

Since coherent states are the eigenstates of the annihilation
operator, $c|\eta\rangle=\eta|\eta\rangle$, from Eq. (\ref{1}), we
have
\begin{eqnarray}
\label{2} (\eta)^2=(\eta^*)^2=0
\end{eqnarray}
which is a consequence of $c^2=(c^*)^2=0$. Also since $c$, $c^\dag$
anticommute, it can be shown from the eigenvalue equations
\begin{eqnarray}
\label{3} \eta\eta^*+\eta^*\eta=0.
\end{eqnarray}
The independent variables $\eta$ and $\eta^*$ which satisfy Eqs.
(\ref{1}) and (\ref{2}) are called Grassmann variables. Thus
elements of the Grassmann algebra can be second order polynomials at
the most.
\begin{eqnarray}
\label{3b} f(\eta,\eta^*)=A+B\eta+C\eta^*+D\eta\eta^*
\end{eqnarray}
and the complex conjugate of a product of two elements is equal to
the product of the conjugates written in the reverse order.

Using Eqs. (\ref{1}) and (\ref{2}), we can write the
overlap between the two coherent states as
\begin{eqnarray}
\label{4} \langle \eta|\eta\rangle=1+\eta^*\eta=e^{\eta^*\eta}.
\end{eqnarray}
Integration of the Grassmann variables is defined by,
\begin{eqnarray}
\int d\eta&=&\int d\eta^*=0\label{5}\\
\int d\eta \eta &=& \int d\eta^* \eta^*=1\label{5a}.
\end{eqnarray}
The differential elements $d\eta$ and $d\eta^*$ anticommute with
each other. Using Eqs. (\ref{1}), (\ref{4}), (\ref{5}) and
(\ref{5a}) it is straightforward to show that
\begin{eqnarray}
\label{6} \int d\eta d\eta^* e^{\eta^*\eta}|\eta\rangle\langle\eta|
= 1
\end{eqnarray}
which is the closure relation for coherent states.

Differentials of the Grassmann variables are defined as,
\begin{eqnarray}
\label{7a} \frac{\partial}{\partial \eta}f(\eta,\eta^*)=B+D\eta^*,~~~
\frac{\partial}{\partial \eta^*}f(\eta,\eta^*)=C-D\eta.
\end{eqnarray}
This implies that
\begin{eqnarray}
\label{7} \frac{\partial}{\partial \eta}\frac{\partial}{\partial
\eta^*}=-\frac{\partial}{\partial \eta^*} \frac{\partial}{\partial
\eta}.
\end{eqnarray}
Taking integral of $f(\eta,\eta^*)$ with respect to $\eta$ or
$\eta^*$ and comparing with Eqs. (\ref{7}), we obtain the operator
identities
\begin{eqnarray}
\label{n-1-1} \int d\eta = \frac{\partial}{\partial \eta},~~~ \int
d\eta^* = \frac{\partial}{\partial \eta^*}.
\end{eqnarray}

Using Eqs. (\ref{5}), (\ref{5a}) and (\ref{n-1-1}), it is
straightforward to see that for any $N\times N$ matrix $A$,
\begin{eqnarray}
\int {\cal D}(\eta^*\eta)\mbox{e}^{\sum_{ij}\eta_i^*A_{ij}\eta_j} =
\mbox{Det}[A]
\end{eqnarray}
where ${\cal D}(\eta^*\eta)= \prod_i d\eta^*_i d\eta_i$.


\end{document}